%% file: main.tex
\documentclass[prd,aps,showpacs,superscriptaddress,floatfix,twocolumn]{revtex4-1}

\usepackage{amssymb}                             %
\usepackage{grffile} %
\usepackage{comment}
\usepackage{mathrsfs}
\usepackage{amsmath}
\usepackage{epsfig}
\usepackage{epsf}
\usepackage{epsfig}
\usepackage{color}
\usepackage{ifpdf}
\usepackage{graphics}
\usepackage{slashed}
\usepackage{float}
\usepackage[T1]{fontenc}
\usepackage{longtable}
\usepackage{booktabs}
\usepackage{afterpage}
\usepackage{hyperref}
\usepackage{lmodern}
\usepackage[utf8]{inputenc}
\usepackage{parskip}
\newcommand{\nclearpage}{}

\def\simge{%
    \mathrel{\rlap{\raise 0.511ex
        \hbox{$>$}}{\lower 0.511ex \hbox{$\sim$}}}}
\def\simle{%
    \mathrel{\rlap{\raise 0.511ex 
        \hbox{$<$}}{\lower 0.511ex \hbox{$\sim$}}}}

\def\be{\begin{equation}}
\def\ee{\end{equation}}
\def\bea{\begin{eqnarray}}
\def\eea{\end{eqnarray}}

\graphicspath{suppl/}

\newcommand{\changes}{}

\begin{abstract}
    We report the first result for the hadronic light-by-light scattering contribution to the muon anomalous magnetic moment with all errors systematically controlled. Several ensembles using 2+1 flavors of physical mass M\"obius domain-wall fermions, generated by the RBC/UKQCD collaborations, are employed to take the continuum and infinite volume limits of finite volume lattice QED+QCD.
    We find
    $a_\mu^{\rm HLbL} = 7.87(3.06)_\text{stat}(1.77)_\text{sys}\times 10^{-10}$.
    Our value is consistent with previous model results and leaves little room for this notoriously difficult hadronic contribution to explain the difference between the Standard Model and the BNL experiment.
\end{abstract}

\begin{document}
\title{The hadronic light-by-light scattering contribution to the muon anomalous magnetic moment from lattice QCD}

\author{Thomas~Blum}
\address{Physics Department, University of Connecticut, 2152 Hillside Road, Storrs, CT, 06269-3046, USA}
\address{RIKEN BNL Research Center, Brookhaven National Laboratory, Upton, New York 11973, USA}
\author{Norman~Christ}
\address{Physics Department, Columbia University, New York, New York 10027, USA}
\author{Masashi~Hayakawa}
\address{Department of Physics, Nagoya University, Nagoya 464-8602, Japan}
\address{Nishina Center, RIKEN, Wako, Saitama 351-0198, Japan}
\author{Taku~Izubuchi}
\address{Physics Department, Brookhaven National Laboratory, Upton, New York 11973, USA}
\address{RIKEN BNL Research Center, Brookhaven National Laboratory, Upton, New York 11973, USA}
\author{Luchang~Jin}
\email{ljin.luchang@gmail.com}
\address{Physics Department, University of Connecticut, 2152 Hillside Road, Storrs, CT, 06269-3046, USA}
\address{RIKEN BNL Research Center, Brookhaven National Laboratory, Upton, New York 11973, USA}
\author{Chulwoo~Jung}
\address{Physics Department, Brookhaven National Laboratory, Upton, New York 11973, USA}
\author{Christoph~Lehner}
\address{Universit\"at Regensburg, Fakult\"at f\"ur Physik, 93040, Regensburg, Germany}
\address{Physics Department, Brookhaven National Laboratory, Upton, New York 11973, USA}

\maketitle

\section{Introduction}

The anomalous magnetic moment of the muon is providing an important test of the Standard Model. The current discrepancy between experiment and theory stands between three and four standard deviations. An ongoing experiment at Fermilab (E989) and one planned at J-PARC (E34) aim to reduce the uncertainty of the BNL E821 value~\cite{Bennett:2006fi} by a factor of four, and \href{http://www.int.washington.edu/PROGRAMS/19-74W/}{similar efforts} are underway on the theory side~\cite{%
Lehner:2019wvv,%
Blum:2018mom,%
Davier:2019can,%
Keshavarzi:2018mgv,%
Blum:2015you,%
Davies:2019efs,%
Giusti:2019xct,%
Aubin:2019usy,%
Colangelo:2017qdm,%
Colangelo:2019lpu,%
Colangelo:2019uex,%
Colangelo:2017fiz,%
Colangelo:2015ama,%
Colangelo:2014pva,%
Hoferichter:2018kwz,%
Hoferichter:2019gzf,%
Hoferichter:2018dmo,%
Gerardin:2016cqj,%
Gerardin:2019rua,%
Hagelstein:2017obr,%
Masjuan:2017tvw,%
Shintani:2019wai,%
Colangelo:2018mtw,%
Giusti:2018mdh,%
Chakraborty:2018iyb,%
Izubuchi:2018tdd,%
Borsanyi:2017zdw,%
Chakraborty:2017tqp%
}.
A key part of the latter is to compute the hadronic light-by-light (HLbL) contribution from first principles using lattice QCD~\cite{
Blum:2014oka,%
Blum:2015gfa,%
Blum:2016lnc,%
Green:2015sra,%
Asmussen:2016lse,%
Gerardin:2017ryf,%
Meyer:2018til%
}.
Such a calculation, with all errors under control, is crucial to interpret the anticipated improved experimental
results~\cite{%
Davoudiasl:2018fbb,%
Crivellin:2018qmi%
}.

The magnetic moment is an intrinsic property of a spin-1/2 particle, and is defined through its interaction with an external magnetic field $\boldsymbol{B}$,
$H_\mathrm{int} = - \boldsymbol{\mu} \cdot \boldsymbol{B}$. Here
\begin{eqnarray}
\boldsymbol{\mu} &=& - g \frac{e}{2 m} \bf S,
\end{eqnarray}
where $\bf S$ is the particle's spin, $q$ and $m$ are the electric charge and mass, respectively, and $g$ is the Land\'e g-factor. The Dirac equation predicts that $g=2$, exactly, so any difference from 2 must arise from interactions.
Lorentz and gauge symmetries tightly constrain the form of the interactions,
\begin{gather}
\langle \mu ({\bf p^\prime}) | J_\nu(0) |\mu({\bf p})\rangle =\nonumber\\
-e \bar u({\bf p^\prime})\left(F_1(q^2)\gamma_\nu+i\frac{F_2(q^2)}{4 m}[\gamma_\nu,\gamma_\rho] q_\rho\right)u({\bf p}),
\label{eq:ff}
\end{gather}
where $J_\nu$ is the electromagnetic current, and $F_1$ and $F_2$ are form factors, giving the charge and magnetic moment at zero momentum transfer ({\changes $q^2=(p^\prime-p)^2=0$}), or static limit. ${u}(\bf{p})$ and $\bar u({\bf p})$ are Dirac spinors. The anomalous part of the magnetic moment is given by $F_2(0)$ alone, and is known as the anomaly,
\begin{eqnarray}
a_\mu &\equiv& (g-2)/2 = F_2(0).
\end{eqnarray}

\begin{figure}
    \centering
    \includegraphics[width=0.15\textwidth]{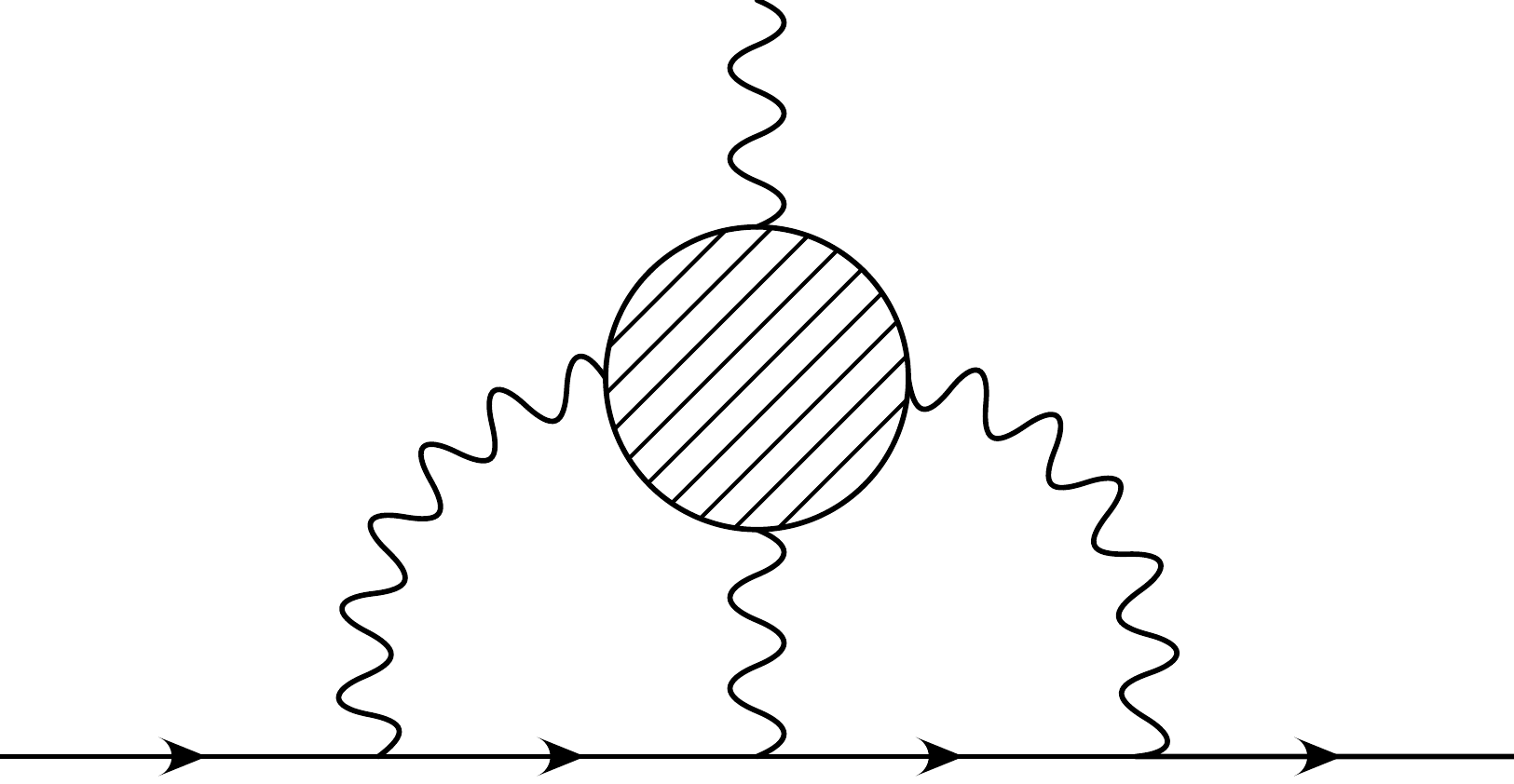}~~+~~\includegraphics[width=0.15\textwidth]{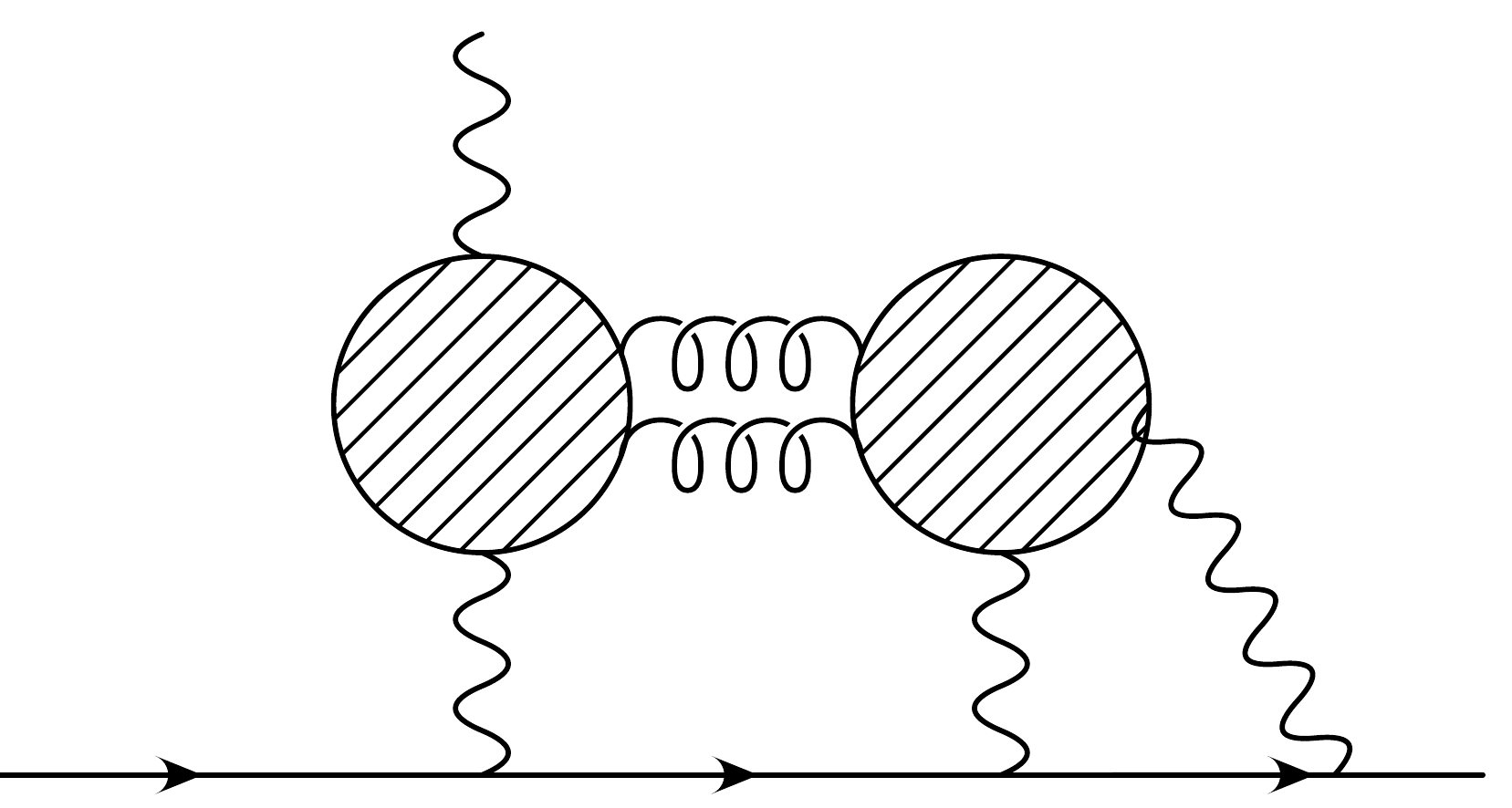}~~$+~~\cdots$
    \caption{Leading contributions from hadronic light-by-light scattering to the muon anomaly. The shaded circles represent quark loops containing QCD interactions to all orders. Horizontal lines represent muons. Quark- connected (left) and disconnected (right) diagrams are shown. Ellipsis denote diagrams obtained by permuting the photon contractions with the muons and diagrams with three and four quark loops with photon couplings (See Fig.~\ref{fig:disco diags}).}
    \label{fig:hlbl feyn diags}
\end{figure}

The desired matrix element in (\ref{eq:ff}) is extracted in quantum field theory from a correlation function of fields as depicted in the Feynman diagrams shown in Fig.~\ref{fig:hlbl feyn diags}. Here we work in coordinate (Euclidean) space and use Lattice QCD for the hadronic part which is intrinsically non-perturbative.
QED is treated using the same discrete, finite, lattice as used for the hadronic part,
while we remove the spatial zero modes of the photon propagator.
This method is called QED$_L$~\cite{Hayakawa:2008an}. It is perturbative with respect to QED, $i.e$, only diagrams where the hadronic part is connected to the muon by three photons enter the calculation.

\section{QED$_L$ Method}

\begin{figure}
    \centering
    \includegraphics[width=0.45\textwidth]{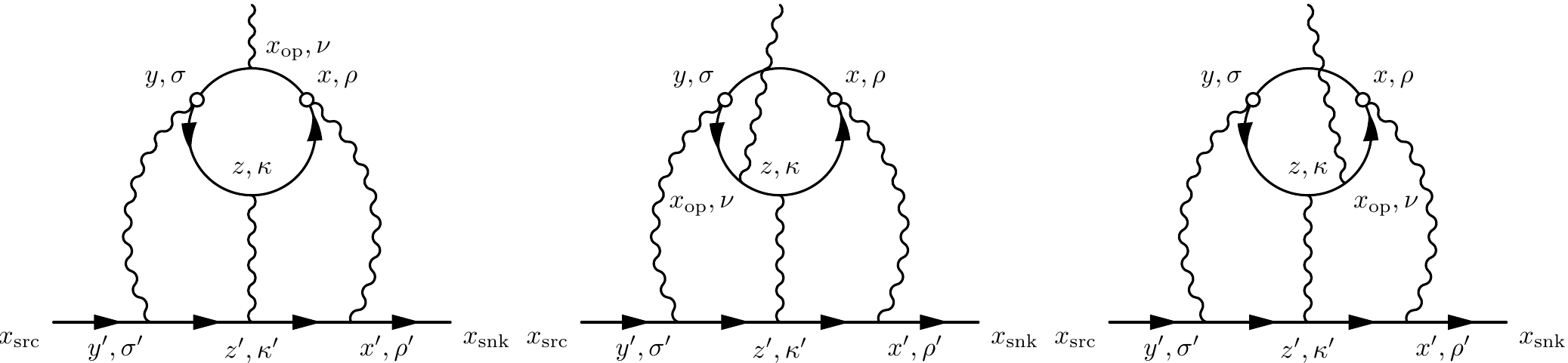}
    \caption{Connected diagrams. Sums over $x$ and $y$ are computed stochastically. The third internal vertex $z$ and the external vertex $x_{\rm op}$ are summed over exactly. The sums on the muon line are done exactly using FFT's. Strong interactions to all orders are not shown.}
    \label{fig:point src method}
\end{figure}

\begin{figure}
    \centering
\includegraphics[width=0.15\textwidth]{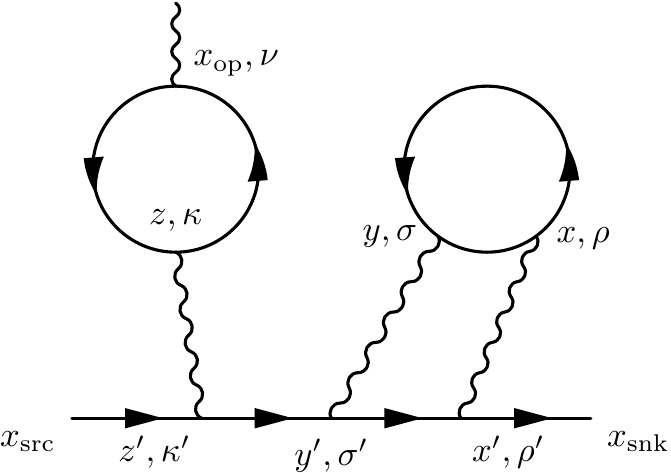}
\includegraphics[width=0.15\textwidth]{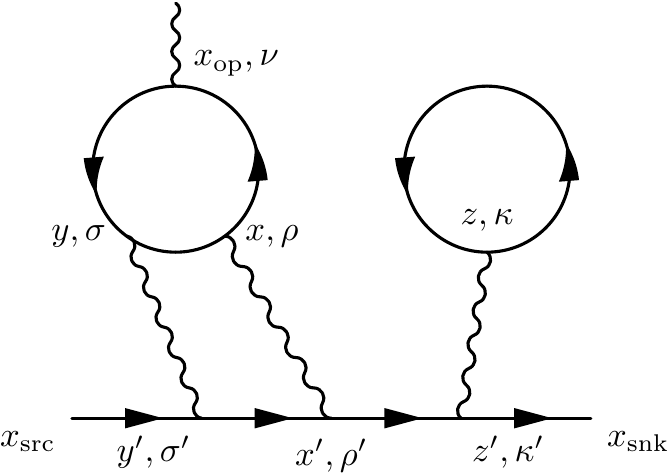}
\includegraphics[width=0.15\textwidth]{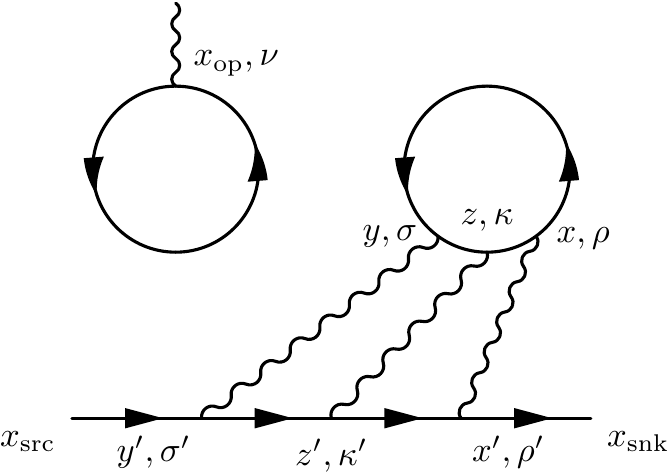}\\
\includegraphics[width=0.15\textwidth]{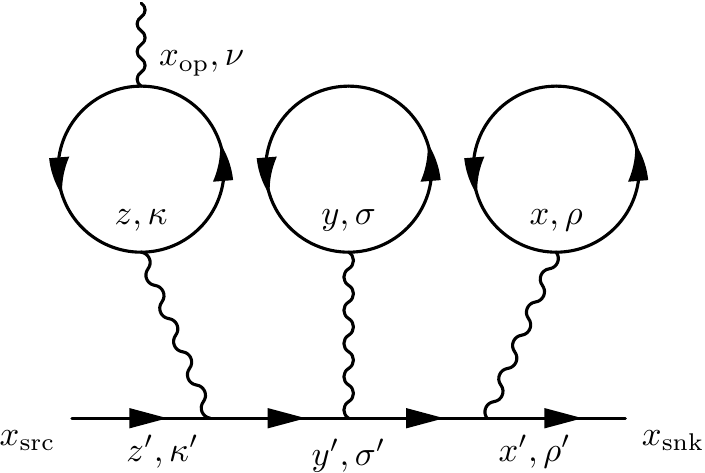}
\includegraphics[width=0.15\textwidth]{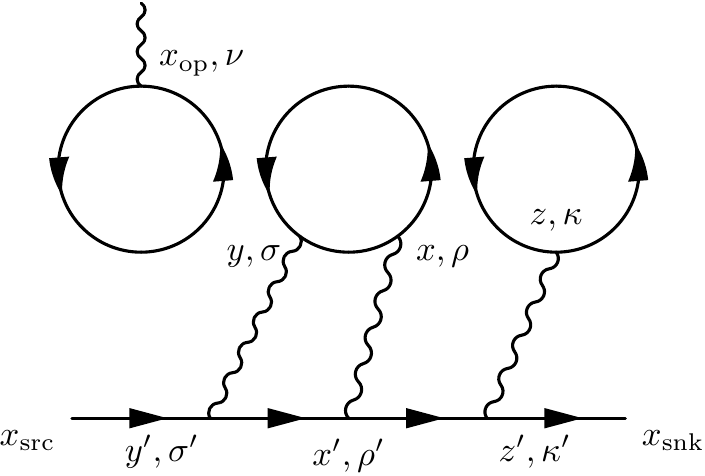}
\includegraphics[width=0.2\textwidth]{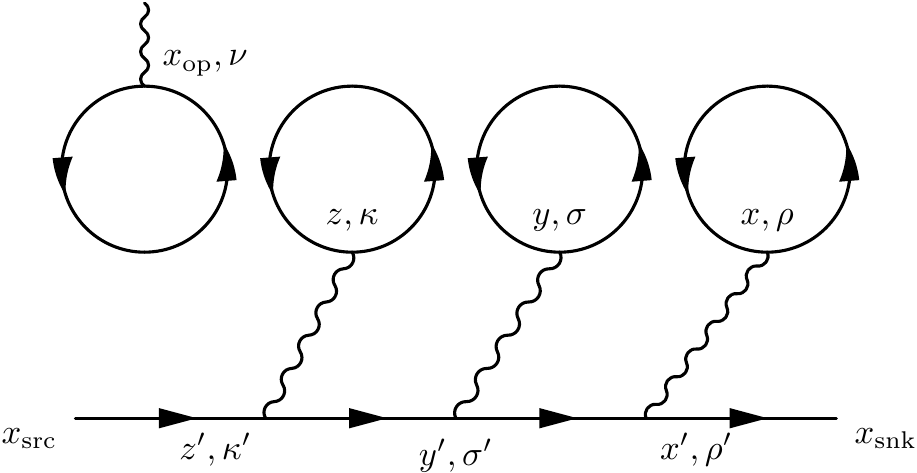}
    \caption{Disconnected diagrams contributing to the muon anomaly. The top leftmost is the leading one, and does not vanish in the $SU(3)$ flavor limit. Strong interactions to all orders, including gluons connecting the quark loops, are not shown.}
    \label{fig:disco diags}
\end{figure}

Here the muon, photons, quarks, and gluons are treated on a finite, discrete lattice. The method is described in detail in Ref.~\cite{Blum:2015gfa}, and the diagrams to be computed are shown in Figs.~\ref{fig:point src method} and~\ref{fig:disco diags}. It is still not possible to do all of the sums over coordinate space vertices exactly with currently available compute resources. Therefore we resort to a hybrid method where two of the vertices on the hadronic loop(s) are summed stochastically: point source propagators from coordinates $x$ and $y$ are computed, and their sink points are contracted at the third internal vertex $z$ and the external vertex $x_{\rm op}$. Since the propagators are calculated to all sink points, $z$ and $x_{\rm op}$ can be summed over the entire volume. The sums over vertices $x$ and $y$ are then done stochastically by computing many ($O(1000)$) random pairs of point source propagators. To do the sampling efficiently, the pairs are chosen with an empirical distribution \footnote{\changes We continue to use the distribution as described in our previous work \cite{Blum:2016lnc}. The same distribution is used for all our ensembles.} designed to frequently probe the summand where it is large, less frequently where it is small.
Since QCD has a mass-gap, we know the hadronic loop is exponentially suppressed according to the distance between any pair of vertices, including $|x-y|$. As we will see, the main contribution comes from distances less than about 1 fm.
The muon line and photons are computed efficiently using FFT's; however, because they must be calculated many times, the cost is not negligible.

Two additional, but related, parts of the method bear mentioning. First, the form dictated by the right hand side of Eq.~\ref{eq:ff} suggests the limit $q\to0$ is unhelpful since the desired $F_2$ term is multiplied by 0. Second, in our Monte Carlo lattice QCD calculation the error on the $F_2$ contribution blows up in this limit. The former is avoided by evaluating the first moment with respect to $\bf{x}_{\mathrm{op}}$ at the external vertex and noticing that an induced extra term vanishes exponentially in the infinite volume limit~\cite{Blum:2015gfa}. This moment method allows the direct calculation of the correlation function at $q=0$, and hence $F_2(0)$.
To deal with the second issue, we first recall that it is the Ward identity that guarantees the unwanted term to vanish in the moment method. We thus enforce the Ward identity exactly on a configuration-by-configuration basis~\cite{Blum:2015gfa}. i.e., before averaging over gauge fields by inserting the external photon at all possible locations on the quark loop in Fig.~\ref{fig:point src method}. This makes the factor of $q$ in Eq.~(\ref{eq:ff}) exact for each measurement and not just in the average and reduces the error on $F_2(0)$ significantly. 
Implementing the above techniques produces an order $O(1000)$ fold improvement in the statistical error over the original non-perturbative QED method used to compute the hadronic light-by-light scattering contribution~\cite{Blum:2014oka}.

The quark-disconnected diagrams that occur at $O(\alpha^3)$ are shown Fig.~\ref{fig:disco diags}. All but the upper-leftmost diagram vanish in the $SU(3)$ flavor limit and are suppressed by powers of $m_{u,d} - m_s$, depending on the number of quark loops with a single photon attached. For now we ignore them and concentrate on the leading disconnected diagram which is computed with a method~\cite{Blum:2016lnc} similar to the one described in the {\changes first part of this} section.
To ensure the loops are connected by gluons, explicit vacuum subtraction is required. However, in the leading diagram the moment at $x_{\rm op}$ implies the left-hand loop in Fig.~\ref{fig:disco diags} vanishes due to parity symmetry, and the vacuum subtraction is done to reduce noise. 

{\changes
As for the connected case, two point sources
(at $y$ and $z$ in Fig.~\ref{fig:disco diags})
are chosen randomly, and the sink points
(at $x$ and $x_\text{op}$ in Fig.~\ref{fig:disco diags}) are summed over.
We compute $M$ (usually $M=1024$) point source propagators for each configuration.
All $M^2$ combinations are used to perform the stochastic sum over $y-z$.
This ``$M^2$ trick'' \cite{Blum:2015gfa,Blum:2016lnc} is crucial to bring the statistical fluctuations of the disconnected diagram under control (see Sec. B of the supplemental material for more details).
}

\section{Lattice Setup}

The simulation parameters are given in Tab.~\ref{tab:ensembles}. All particles have their physical masses (isospin breaking for the up and down quark masses is not included). The discrete Dirac operator is known as the (M\"obius) domain wall fermion ((M)DWF)) operator. Similarly the discrete gluon action is given by the plaquette plus rectangle Iwasaki gauge action. Additionally, three ensembles with larger lattice spacing employ the dislocation-suppressing-determinant-ratio (DSDR) to soften explicit chiral symmetry breaking effects for MDWFs~\cite{Renfrew:2009wu}. 
We use All Mode Averaging (AMA)~\cite{Shintani:2014vja} and
Multi-grid Lanczos~\cite{Clark:2017wom} techniques
to speed up the fermion propagator generation.

The muons and photons take discrete free-field forms. The muons are DWFs with infinite size in the extra fifth dimension, and the photons are non-compact in the Feynman gauge. In the latter all modes with ${\bf q}=0$ are dropped, a finite volume formulation of QED known as QED$_L$~\cite{Hayakawa:2008an}.

\begin{table}[htp]
\begin{center}
\begin{tabular}{|c|c|c|c|c|c|c|}
\hline
& 48I & 64I & 24D & 32D & 48D & 32Dfine  \\
\hline
$a^{-1}$ (GeV) & 1.730 & 2.359 & 1.015 & 1.015 & 1.015 & 1.378\\
$a$ (fm) & 0.114 &0.084 & 0.194 & 0.194  & 0.194 & 0.143 \\
$L$ (fm)  & 5.47 & 5.38 & 4.67 & 6.22 & 9.33 & 4.58 \\
$L_s$ & 48 & 64 & 24 & 24  & 24 & 32\\
$m_\pi$ (MeV) & 139 & 135 & 142 & 142 & 142 & 144 \\
$m_\mu$ (MeV) & 106 & 106 & 106 & 106 & 106 & 106 \\
\# meas con & 65 & 43 & 157 & 70 & 8 & 75\\
\# meas discon & 124 & 105 & 156 & 69 & 0 & 69\\
\hline
\end{tabular}
\caption{2+1 flavors of MDWF gauge field ensembles generated by the RBC/UKQCD collaborations~\protect\cite{Blum:2014tka}. 
The lattice spacing $a$, spatial extent $L$, extra fifth dimension size $L_s$, muon pion mass $m_\pi$, and number of QCD configuration used for the connected and the disconnected diagrams.
}
\label{tab:ensembles}
\end{center}
\end{table}

\section{Results}

\label{sec:results}
 
\begin{figure}
     \centering
     \includegraphics[width=0.45\textwidth]{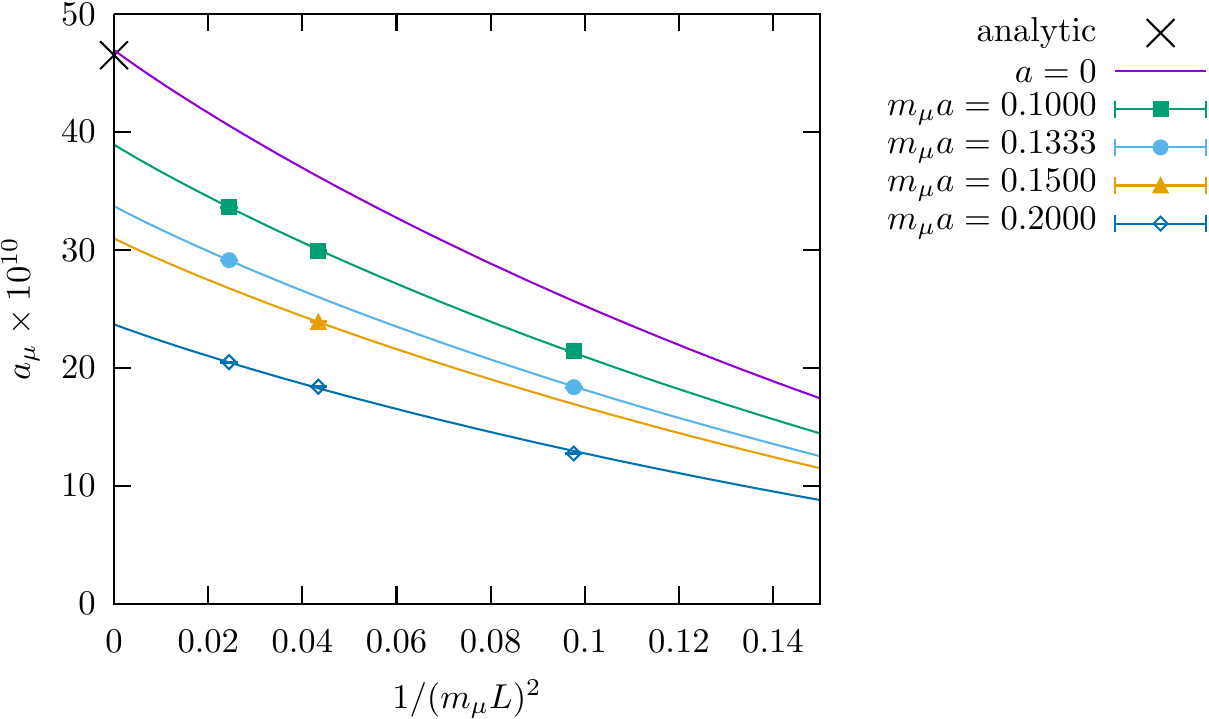}
     \caption{QED light-by-light scattering contribution
     from the muon loop to the muon anomaly.  
     The lattice spacing decreases from bottom to top. Solid lines are from a fit using Eq.~(\ref{eq:qed_extrap}).}
     \label{fig:qed test}
 \end{figure}

 Before moving to the hadronic case, the method was tested in pure QED~\cite{Blum:2015gfa}. Results for several lattice spacings and box sizes are shown in Fig.~\ref{fig:qed test}. The systematic uncertainties are large, but under control. Note that the finite volume errors are polynomial in $1/L$ and not exponential, due to the photons which interact over a long range. The data are well fit to the form
\begin{eqnarray}
\label{eq:qed_extrap}
a_\mu(L,a)&=&a_\mu
\Big(1-\frac{b_2}{(m_\mu L)^2}+\frac{b_3}{(m_\mu L)^3}\Big)
\\
&&\qquad\times\Big(1-c_1 (m_\mu a)^2+c_2 (m_\mu a)^4\Big).
\nonumber
\end{eqnarray}
The continuum and infinite volume limit is $F_2(0)=46.9(2)_\text{stat} \times 10^{-10}$ for the case where the lepton mass in the loop is the same as the muon mass, which is quite consistent with the well known perturbative value~\cite{Laporta:1991zw}, $46.5\times10^{-10}$.

\begin{figure}
    \centering
    \includegraphics[width=0.4\textwidth]{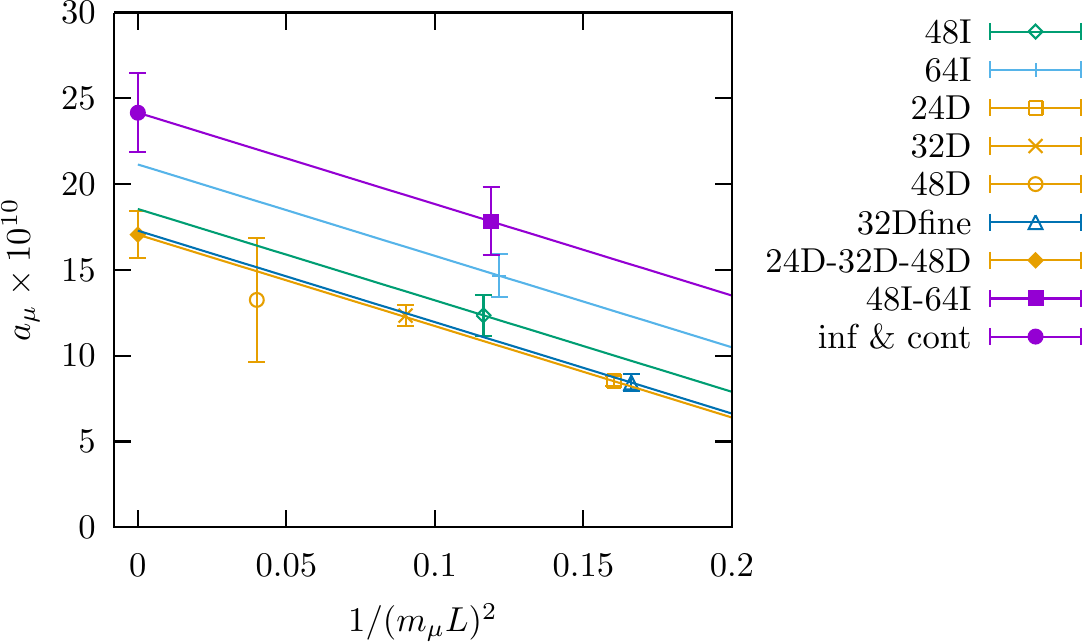}
    \includegraphics[width=0.4\textwidth]{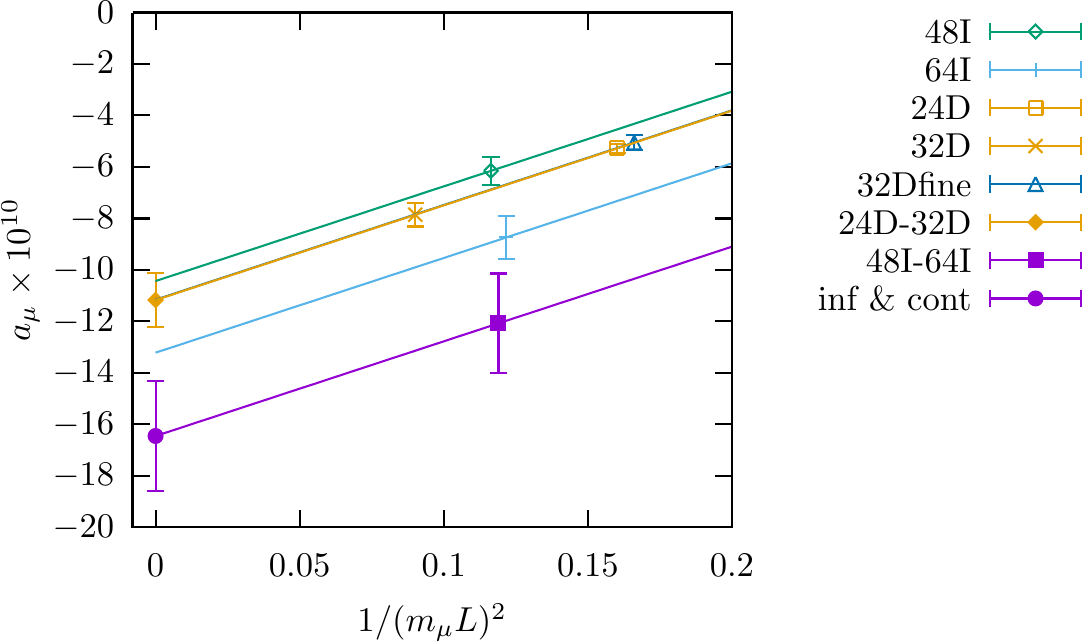}
    \includegraphics[width=0.4\textwidth]{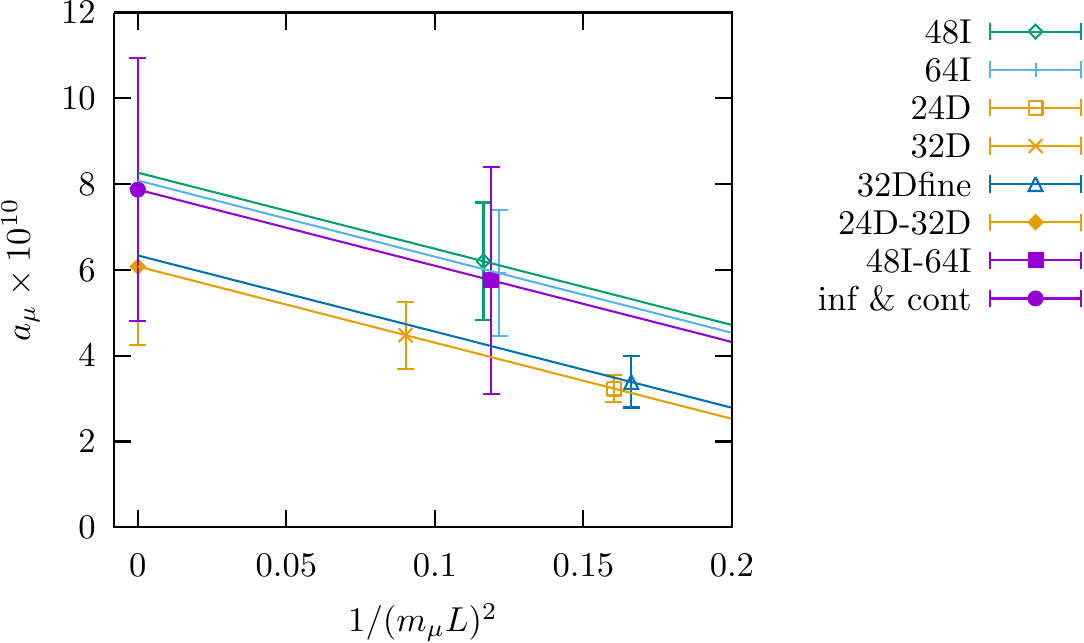}
    \caption{Infinite volume extrapolation. Connected (top), disconnected (middle), and total (bottom).
             We have use the hybrid method to calculate the continuum limit for the connected contribution.}
    \label{fig:qedl-extrap-plot}
\end{figure}

Our physical point calculation~\cite{Blum:2016lnc} started on the $48^3$, $a^{-1}=1.730$ GeV, Iwasaki ensemble listed in the first column of Tab.~\ref{tab:ensembles}, for which we found 
$a_\mu^{\rm con} = 11.60(0.96)_\text{stat}\times 10^{-10}$,
$a_\mu^{\rm discon} = -6.25(0.80)_\text{stat}\times 10^{-10}$,
 and $a_\mu^{\rm tot} = 5.35(1.35)_\text{stat}\times 10^{-10}$ for the connected, leading disconnected, and total HLbL contributions to the muon anomaly, respectively. The errors quoted are purely statistical.
We have since improved the statistics on the leading disconnected diagram with measurements on 59 additional configurations, and the contribution becomes $-6.15 (55)\times 10^{-10}$. Since then we have computed on several additional ensembles in order to take the continuum and infinite volume limits (see Tab.~\ref{tab:ensembles}).
The results are displayed in Fig.~\ref{fig:qedl-extrap-plot} along with curves obtained with the following equation:
\begin{eqnarray}
\label{eq:fit-plus-form-2L-2a-2ad-4ad}
a_\mu(L,a^\text{I},a^\text{D})
&=&
a_\mu
\Big(
1
- \frac{b_2}{(m_\mu L)^2}
\\
&&
\hspace{-1.5cm}
- c_1^\text{I} (a^\text{I}~\mathrm{GeV})^2
- c_1^\text{D} (a^\text{D}~\mathrm{GeV})^2
+ c_2^\text{D} (a^\text{D}~\mathrm{GeV})^4
\Big)
\nonumber
\end{eqnarray}
where $a^\text{I}$, $a^\text{D}$ represent the lattice spacings for the
Iwasaki and I-DSDR ensembles respectively.
For the Iwasaki ensembles, we define the variable $a^\text{D}$ to be zero and vice versa.
Therefore the lattice spacing is always equal to $a=a^\text{I} + a^\text{D}$.
We allow different $a^2$ coefficients for the Iwasaki and I-DSDR ensembles as the gauge
actions are different.
The lattice spacings for the I-DSDR ensembles are not small enough to allow us
to ignore the $a^4$ effects, and therefore we include them in the fit.
As we only have two lattice spacings for the I-DSDR ensembles,
with both $a^2$ and $a^4$ effects unknown,
we cannot extrapolate to the continuum just with the I-DSDR ensembles.
Therefore, based on this fit form,
the continuum limit is obtained from the two Iwasaki ensembles,
and
the I-DSDR ensembles are used to obtain the volume dependence only.
In particular, the 32Dfine ensemble does not affect
the fitted $a_\mu$ at all.
It only helps to determine the parameter $c_2^\text{D}$,
which provides evidence for the size of
the potential $\mathcal{O}(a^4)$ systematic errors.
We find for the connected, disconnected, and total contributions,
$a_\mu^{\rm con} = 23.76(3.96)_\text{stat}(4.88)_\text{sys}\times 10^{-10}$,
$a_\mu^{\rm discon} = -16.45(2.13)_\text{stat}(3.99)_\text{sys}\times 10^{-10}$,
$a_\mu^{\rm tot} = 7.47(4.24)_\text{stat}(1.64)_\text{sys}\times 10^{-10}$, respectively.
For the total contribution, we fit the total contribution for each ensemble,
which is slightly different from the sum of the fitted results
from the connected and the disconnected parts.
Notice there is a large cancellation between the connected and disconnected diagrams that persists for $a\to0$ and $L\to\infty$, so even though the individual contributions are relatively well resolved, the total is not. The cancellation is expected since hadronic light-by-light scattering at long distance is dominated by the $\pi^0$ which contributes to both diagrams, but with opposite sign~\cite{Bijnens:2016hgx,Jin:2016rmu,Gerardin:2017ryf}. Notice also that the $a^2$ and $1/L^2$ corrections are individually large but also tend to cancel in the sum. 

The systematic errors mostly result from the higher order
discretization and finite volume effects which are not
included in the fitting formula Eq.~(\ref{eq:fit-plus-form-2L-2a-2ad-4ad}).
We therefore estimate the errors through
the change of the results after
adding a corresponding term in the fitting formula.
For $\mathcal{O}(1/L^3)$,
we add another $1/(m_\mu L)^3$ term with the same coefficient
as the $1/(m_\mu L)^2$ term.
For $\mathcal{O}(a^4)$ effects,
we add an $a^4$ term
also for the Iwasaki ensembles with coefficient similar to
the I-DSDR ensembles.
For $\mathcal{O}(a^2\log(a^2))$ effects,
we multiply the discretization effect terms
in Eq.~(\ref{eq:fit-plus-form-2L-2a-2ad-4ad}) by
$(1-(\alpha_S / \pi) \log(a^2~\mathrm{GeV}))$.
For $\mathcal{O}(a^2/L)$,
we multiply the discretization effect terms
in Eq.~(\ref{eq:fit-plus-form-2L-2a-2ad-4ad}) by
$(1-1/(m_\mu L))$.
In addition, for the only two contributions which we have not
included in the present HLbL calculation:
(a) strange quark contribution to the connected diagrams;
(b) sub-leading disconnected diagrams' contribution.
We have performed lattice calculations
with the QED$_\infty$ approach~\cite{Blum:2017cer} on the
24D ensemble to estimate the systematic errors.
These systematic errors are added in quadrature and summarized in Tab.~\ref{tab:sys-err-nohybrid}.
In the supplementary materials, these systematic
errors are discussed in more detail.

\begin{table}
\begin{tabular}{cccc}
\toprule
& con & discon & tot\tabularnewline
\midrule
\(a_\mu\) & 23.76(3.96) & -16.45(2.13) & 7.47(4.24)\tabularnewline
sys \(\mathcal{O}(1/L^3)\) & 2.34(0.41) & 1.72(0.32) & 0.83(0.56)\tabularnewline
sys \(\mathcal{O}(a^4)\) & 0.83(0.53) & 0.71(0.28) & 0.96(0.94)\tabularnewline
sys \(\mathcal{O}(a^2\log(a^2))\) & 0.21(0.18) & 0.25(0.09) & 0.03(0.17)\tabularnewline
sys \(\mathcal{O}(a^2/L)\) & 4.18(2.37) & 3.49(1.37) & 0.86(2.20)\tabularnewline
sys strange con & 0.30 & 0 & 0.30\tabularnewline
sys sub-discon & 0 & 0.50 & 0.50\tabularnewline
sys all & 4.88(2.17) & 3.99(1.29) & 1.64(1.15)\tabularnewline
\bottomrule
\end{tabular}
\caption{Central value and various systematic errors.
Numbers in parentheses are statistical error for the corresponding
values.}
\label{tab:sys-err-nohybrid}
\end{table}

\begin{table}
\begin{tabular}{cccc}
\toprule
& con & discon & tot\tabularnewline
\midrule
\(a_\mu\) & 24.16(2.30) & -16.45(2.13) & 7.87(3.06)\tabularnewline
sys hybrid \(\mathcal{O}(a^2)\) & 0.20(0.45) & 0 & 0.20(0.45)\tabularnewline
sys \(\mathcal{O}(1/L^3)\) & 2.34(0.41) & 1.72(0.32) & 0.83(0.56)\tabularnewline
sys \(\mathcal{O}(a^4)\) & 0.88(0.31) & 0.71(0.28) & 0.95(0.92)\tabularnewline
sys \(\mathcal{O}(a^2\log(a^2))\) & 0.23(0.08) & {\changes 0.25(0.09)} & 0.02(0.11)\tabularnewline
sys \(\mathcal{O}(a^2/L)\) & 4.43(1.38) & 3.49(1.37) & 1.08(1.57)\tabularnewline
sys strange con & 0.30 & 0 & 0.30\tabularnewline
sys sub-discon & 0 & 0.50 & 0.50\tabularnewline
sys all & 5.11(1.32) & 3.99(1.29) & 1.77(1.13)\tabularnewline
\bottomrule
\end{tabular}
\caption{Central value and various systematic errors,
use the hybrid continuum limit for the connected diagrams.
Numbers in parentheses are statistical error for the corresponding
values.}
\label{tab:sys-err-hybrid}
\end{table}

\begin{figure}[hbt]
    \centering
    \includegraphics[width=0.45\textwidth]{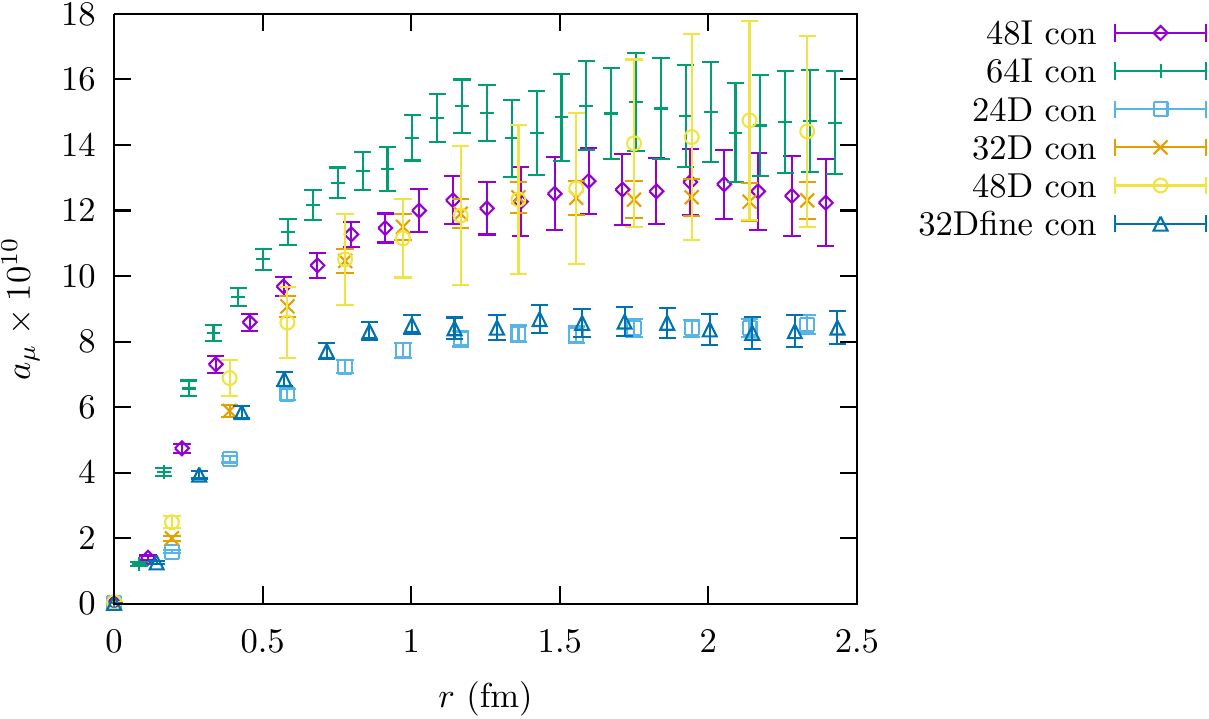}
    \includegraphics[width=0.45\textwidth]{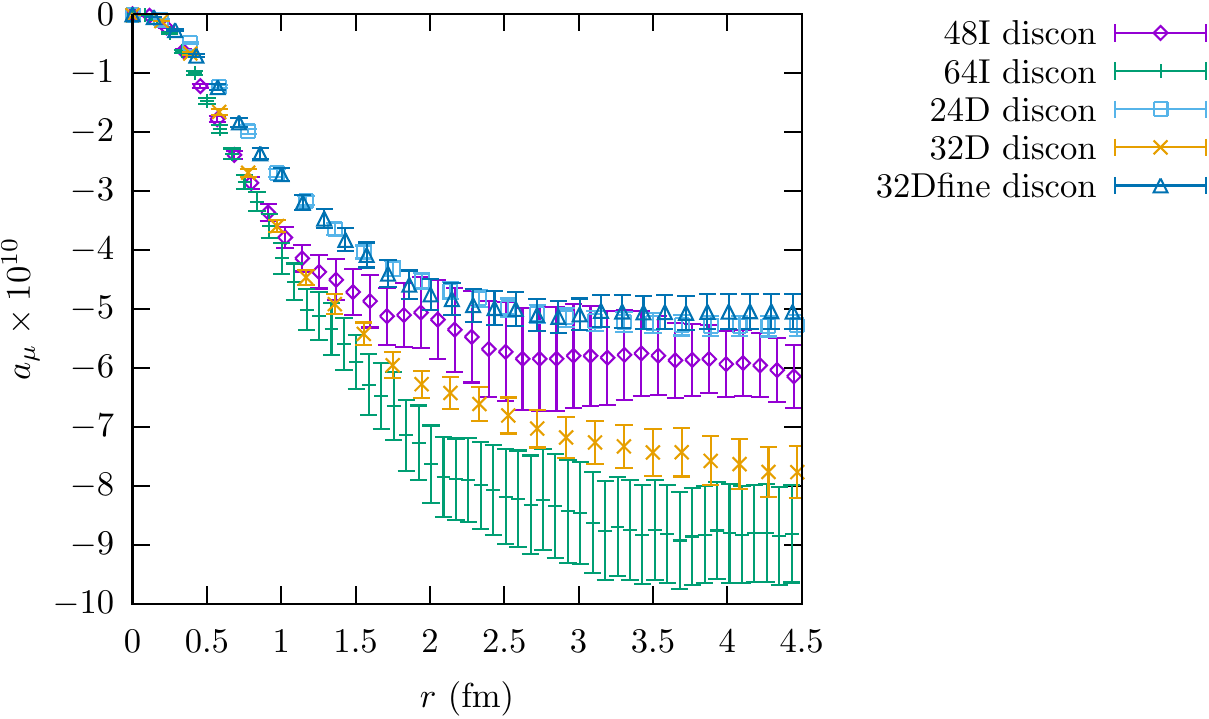}    
    \caption{Cumulative contributions to the muon anomaly, connected (upper) and disconnected (lower). $r$ is the distance between the two sampled currents in the hadronic loop (the other two currents are summed exactly)
    and the horizontal axis is the cumulative contributions from $r$ and below.
    $24^3$ IDSDR (squares), $24^3$ IDSDR (squares), $32^3$ IDSDR (crosses), $48^3$ Iwasaki (diamonds), and $64^3$ Iwasaki (plusses).
    }
    \label{fig:cumulative}
\end{figure}

While the large relative error on the total is a bit unsatisfactory, we emphasize that our result represents an important estimate on the hadronic light-by-light scattering contribution to the muon anomaly, with all systematic errors controlled.
It appears that this contribution cannot bring the Standard Model and the E821 experiment in agreement.

In fact we can do even a bit better with the data on hand. As seen in Fig.~\ref{fig:cumulative}, which shows the cumulative sum of all contributions up to a given separation of the two sampled currents in the hadronic loop, the total connected contribution saturates at a distance of about 1 fm for all ensembles. This suggests the region $r\simge 1$ fm adds mostly noise and little signal, and the situation gets worse in the limits. A more accurate estimate can be obtained by taking the continuum limit for the sum up to $r=1$ fm, and above that by taking the contribution from the relatively precise $48^3$ ensemble. We include a systematic error on this long distance part since it is not extrapolated to $a=0$. The infinite volume limit is taken as before. This hybrid procedure yields $a_\mu^{\rm con}=24.16(2.30)_\text{stat}(5.11)_\text{sys}\times 10^{-10}$, with a statistical error that is roughly $2\times$ smaller and the additional $\mathcal{O}(a^2)$ systematic error from the hybrid procedure is only $0.20\times 10^{-10}$. Unfortunately a similar procedure for the disconnected diagram is not reliable, as can be seen in the lower panel of Fig.~\ref{fig:cumulative}.
The cumulative plots do not reach plateaus around 1 fm, but instead tend to fall
significantly up to 2 fm, or more. Once the cut moves beyond 1 fm it is no longer effective. The different behavior between the two stems from the different sampling strategies used for each~\cite{Blum:2015gfa}. Using the improved connected result, we find our final result for QED$_L$, 
\begin{equation}
    a_\mu^{\rm tot} = 7.87(3.06)_\text{stat}(1.77)_\text{sys}\times 10^{-10},
\end{equation}
where the error is mostly statistical. We also include all systematic errors added in quadrature, including the hybrid $\mathcal{O}(a^2)$ error of the connected diagram.
The systematic errors are summarized in Tab.~\ref{tab:sys-err-hybrid}.

\section{Summary and Outlook}

We have presented results for the hadronic light-by-light scattering contribution to the muon $g-2$ from Lattice QCD+QED calculations with all errors under control. Large discretization and finite volume corrections are apparent but under control, and the value in the continuum and infinite volume limits is compatible with previous model and dispersive treatments, albeit with a large statistical error. Despite the large error, which results after a large cancellation between quark- connected and disconnected diagrams, our calculation suggests that light-by-light scattering can not be behind the approximately 3.7 standard deviation discrepancy between the Standard Model and the BNL experiment E821. Future calculations will reduce the error significantly. The calculations presented here strengthen the much anticipated test of the Standard Model from the new experiments at Fermilab and J-PARC, with the former planning to announce first results soon.

\begin{acknowledgments}
We would like to thank our RBC and UKQCD collaborators for helpful discussions
and critical software and hardware support.
This work is partially supported by the US Department of Energy.
T.B and L.C.J are supported by US DOE grant DE-SC0010339.
N.C is supported by US DOE grant DE-SC0011941.
M.H. is supported in by Japan Grants-in-Aid
for Scientific Research, No.16K05317.
T.I and C.L. are supported in part by US DOE Contract DESC0012704(BNL).
T.I. is also supported by JSPS KAKENHI grant numbers JP26400261, JP17H02906.
C.L. is also supported by a DOE Office of Science Early Career Award.
We developed the computational code based on the Columbia Physics System (CPS)
and Grid (\texttt{https://github.com/paboyle/Grid}).
Computations were performed mainly under the ALCC Program of the US DOE
on the Blue Gene/Q Mira computer at the Argonne Leadership
Class Facility,
a DOE Office of Science Facility supported under Contract DE-AC02-06CH11357.
We gratefully acknowledge computer resources at the Oakforest-PACS
supercomputer system at Tokyo University,
partly through the HPCI System Research Project (hp180151, hp190137),
the BNL SDCC computer clusters at Brookhaven National Lab as well as computing resources provided through USQCD at the Brookhaven and Jefferson Labs.
\end{acknowledgments}

\bibliographystyle{apsrev4-1}

\input{src-suppl.tex}

\end{document}

%% file: src-suppl.tex
{\clearpage}
{\newpage}

\section*{Supplementary Material}

\subsection{QED Test with different fitting forms}

We test various fitting formulas from lattice calculation
of the muonic leptonic light-by-light contribution
to muon $g-2$. The analytic result is known to be
\(a_\mu=0.371\times(\alpha/\pi)^3=46.5\times10^{-10}\)
using the conventional perturbative calculation.

In Ref.~\cite{Blum:2015gfa}, we have performed
the lattice calculation
with three different lattice volumes,
and each with three different lattice spacings.
The results are listed in Tab.~\ref{tab:muon-lbl}.
Then, the continuum limit is calculated
for each lattice volume and then extrapolate to infinite volume limit.
In this paper, we adopt a different strategy.
We use one formula which include both
discretization effects and finite volume effects
to fit all the data points.
In the fitting forms listed below,
we studied the size of
$\mathcal{O}(a^4)$,
$\mathcal{O}(1/L^3)$,
and $\mathcal{O}(a^2/L)$,
effects in addition
to the leading $\mathcal{O}(a^2)$, $\mathcal{O}(1/L^2)$ effects.

\begin{table}[H]
\centering
\begin{tabular}{@{}llll@{}}
\toprule
\(L/a\) & \(ma\) & \(mL\) & \(a_\mu\times10^{10}\)\tabularnewline
\midrule
16 & 0.2 & 3.2 & 12.73(1)\tabularnewline
24 & 0.1333 & 3.2 & 18.36(4)\tabularnewline
32 & 0.1 & 3.2 & 21.46(4)\tabularnewline
24 & 0.2 & 4.8 & 18.40(1)\tabularnewline
32 & 0.15 & 4.8 & 23.90(3)\tabularnewline
48 & 0.1 & 4.8 & 29.93(6)\tabularnewline
32 & 0.2 & 6.4 & 20.48(3)\tabularnewline
48 & 0.1333 & 6.4 & 29.13(4)\tabularnewline
64 & 0.1 & 6.4 & 33.59(6)\tabularnewline
\bottomrule
\end{tabular}
\caption{Data for Muonic QED light-by-light.}
\label{tab:muon-lbl}
\end{table}

\begin{itemize}

\item \textit{fit-product-form-3L-4a}

\begin{eqnarray}
a_\mu(a,L)
&=&
a_\mu
\Big(
1
- \frac{b_2}{(m_\mu L)^2}
+ \frac{b_3}{(m_\mu L)^3}
\Big)
\\ &&\qquad\times
\Big(
1
- c_1 (m_\mu a)^2
+ c_2 (m_\mu a)^4
\Big)
\nonumber
\end{eqnarray}
This is the form used in the QED light-by-light calculation
in Section~\ref{sec:results}.
The results of the fit are listed in the following table:
\begin{center}
\begin{tabular}{@{}ccc@{}}
\toprule
Variable & Value & Statistical Error\tabularnewline
\midrule
\(a_\mu\) & 46.93696 & 0.20628\tabularnewline
\(b_2\) & 6.51815 & 0.21910\tabularnewline
\(b_3\) & 6.00380 & 0.63394\tabularnewline
\(c_1/10\) & 1.86523 & 0.01896\tabularnewline
\(c_2/10^2\) & 1.56695 & 0.03898\tabularnewline
\bottomrule
\end{tabular}
\end{center}

\item \textit{fit-product-form-2L-4a}

\begin{eqnarray}
a_\mu(a,L)
&=&
a_\mu
\Big(
1
- \frac{b_2}{(m_\mu L)^2}
\Big)
\\ &&\qquad\times
\Big(
1
- c_1 (m_\mu a)^2
+ c_2 (m_\mu a)^4
\Big)
\nonumber
\end{eqnarray}

The results of the fit is listed in the following table:
\begin{center}
\begin{tabular}{@{}ccc@{}}
\toprule
Variable & Value & Statistical Error\tabularnewline
\midrule
\(a_\mu\) & 45.34372 & 0.13747\tabularnewline
\(b_2\) & 4.44949 & 0.00982\tabularnewline
\(c_1/10\) & 1.88140 & 0.01844\tabularnewline
\(c_2/10^2\) & 1.59916 & 0.03796\tabularnewline
\bottomrule
\end{tabular}
\end{center}

\item \textit{fit-product-form-2L-2a}

\begin{eqnarray}
a_\mu(a,L)
&=&
a_\mu
\Big(
1
- \frac{b_2}{(m_\mu L)^2}
\Big)
\Big(
1
- c_1 (m_\mu a)^2
\Big)
\end{eqnarray}

The results of the fit is listed in the following table:
\begin{center}
\begin{tabular}{@{}ccc@{}}
\toprule
Variable & Value & Statistical Error\tabularnewline
\midrule
\(a_\mu\) & 41.74717 & 0.05870\tabularnewline
\(b_2\) & 4.44563 & 0.00981\tabularnewline
\(c_1/10\) & 1.15123 & 0.00210\tabularnewline
\bottomrule
\end{tabular}
\end{center}

\item \textit{fit-plus-form-3L-4a}

\begin{eqnarray}
a_\mu(a,L)
&=&
a_\mu
\Big(
1
- \frac{b_2}{(m_\mu L)^2}
+ \frac{b_3}{(m_\mu L)^3}
\\ &&\qquad
- c_1 (m_\mu a)^2
+ c_2 (m_\mu a)^4
\Big)
\nonumber
\end{eqnarray}
The results of the fit is listed in the following table:
\begin{center}
\begin{tabular}{@{}ccc@{}}
\toprule
Variable & Value & Statistical Error\tabularnewline
\midrule
\(a_\mu\) & 43.09059 & 0.13226\tabularnewline
\(b_2\) & 4.91796 & 0.14967\tabularnewline
\(b_3\) & 4.61329 & 0.41827\tabularnewline
\(c_1/10\) & 1.51229 & 0.01716\tabularnewline
\(c_2/10^2\) & 1.30464 & 0.03263\tabularnewline
\bottomrule
\end{tabular}
\end{center}

\item \textit{fit-plus-form-3L-4a-cross}

\begin{eqnarray}
a_\mu(a,L)
&=&
a_\mu
\Big(
1
- \frac{b_2}{(m_\mu L)^2}
+ \frac{b_3}{(m_\mu L)^3}
\\ &&\qquad
- c_1 (m_\mu a)^2 \big( 1 - \frac{b_1'}{m_\mu L}\big)
+ c_2 (m_\mu a)^4
\Big)
\nonumber
\end{eqnarray}
The results of the fit is listed in the following table:
\begin{center}
\begin{tabular}{@{}ccc@{}}
\toprule
Variable & Value & Statistical Error\tabularnewline
\midrule
\(a_\mu\) & 46.21446 & 0.15636\tabularnewline
\(b_2\) & 6.54789 & 0.13960\tabularnewline
\(b_3\) & 7.06359 & 0.38837\tabularnewline
\(c_1/10\) & 1.84720 & 0.01728\tabularnewline
\(c_2/10^2\) & 1.19503 & 0.03024\tabularnewline
\(b_1'\) & 1.07316 & 0.01327\tabularnewline
\bottomrule
\end{tabular}
\end{center}

\end{itemize}

\begin{figure*}[hbt]
\centering
\includegraphics[width=0.45\textwidth]{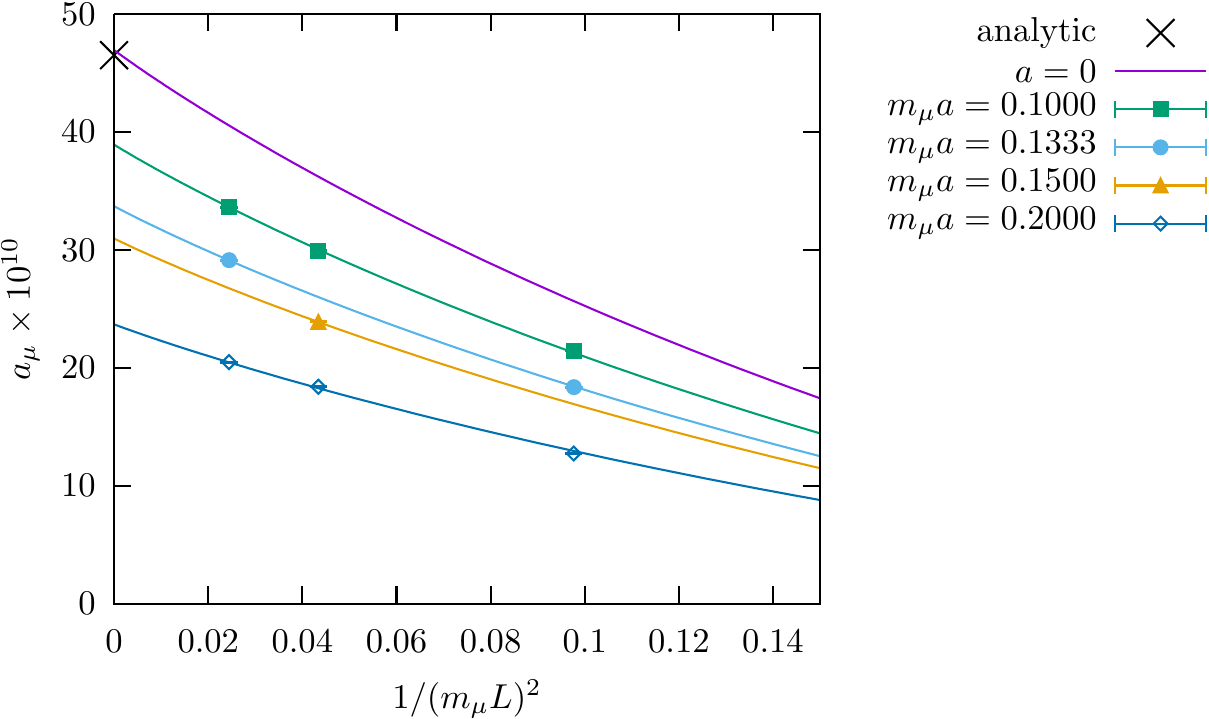}
\hfill
\includegraphics[width=0.45\textwidth]{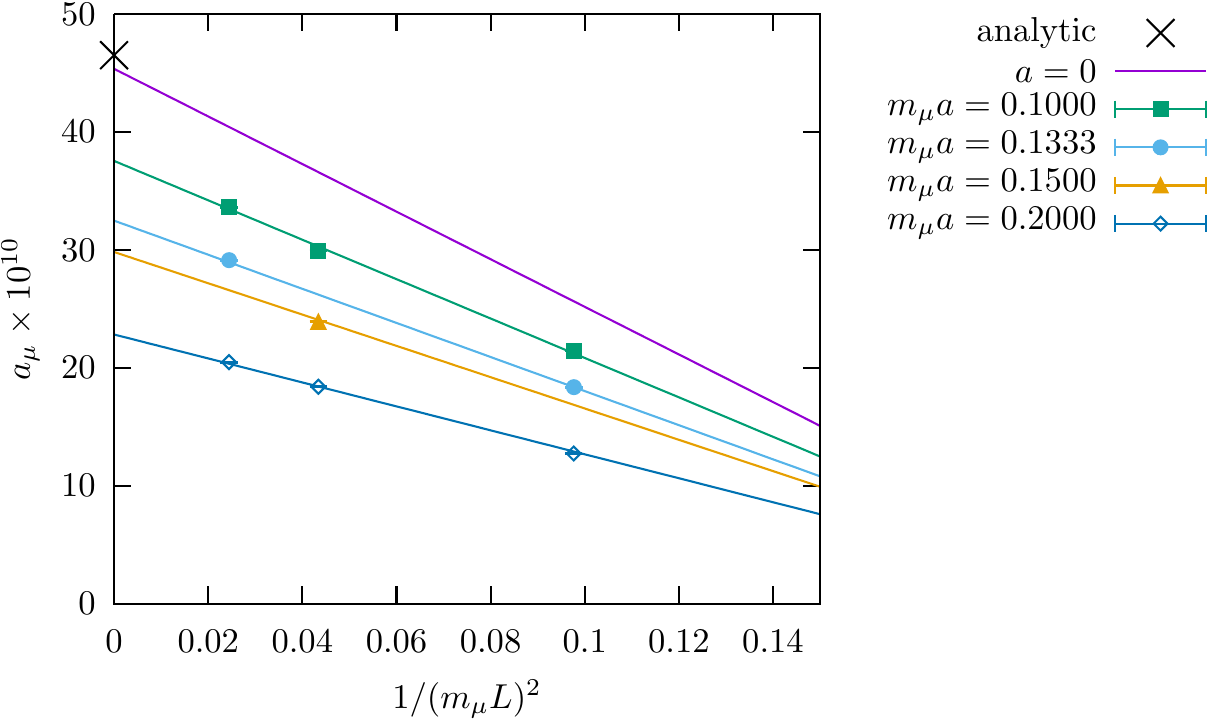}
\\
\includegraphics[width=0.45\textwidth]{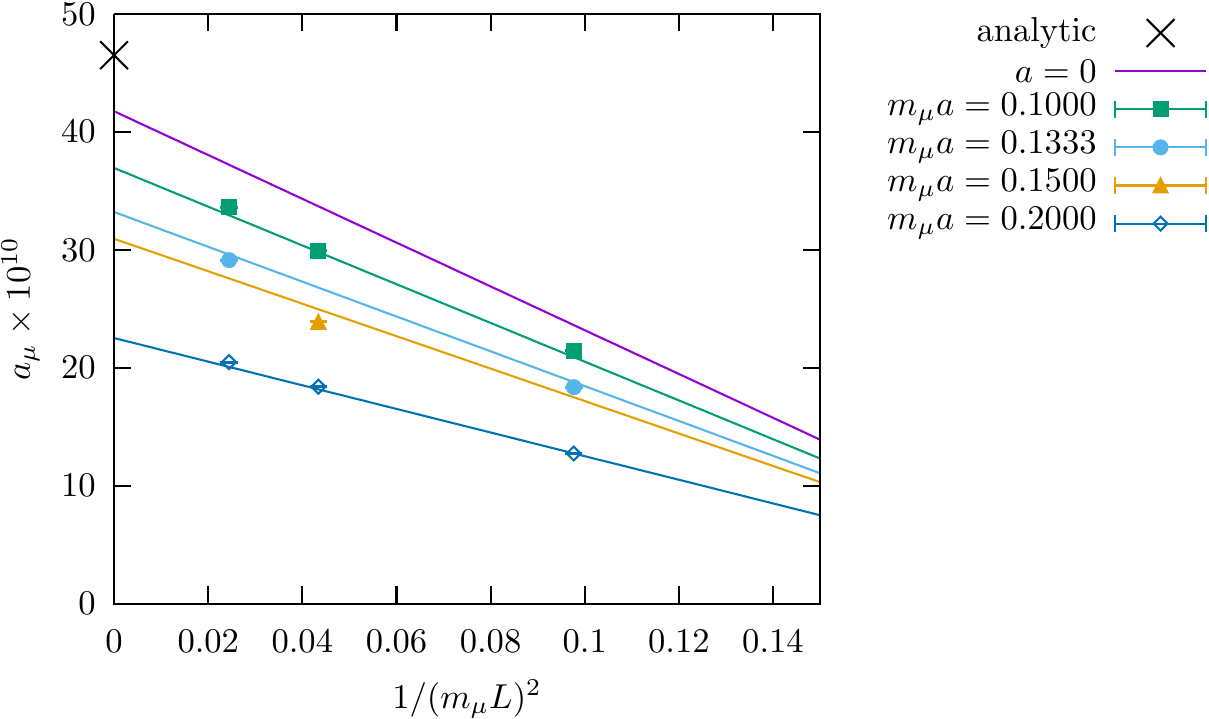}
\hfill
\includegraphics[width=0.45\textwidth]{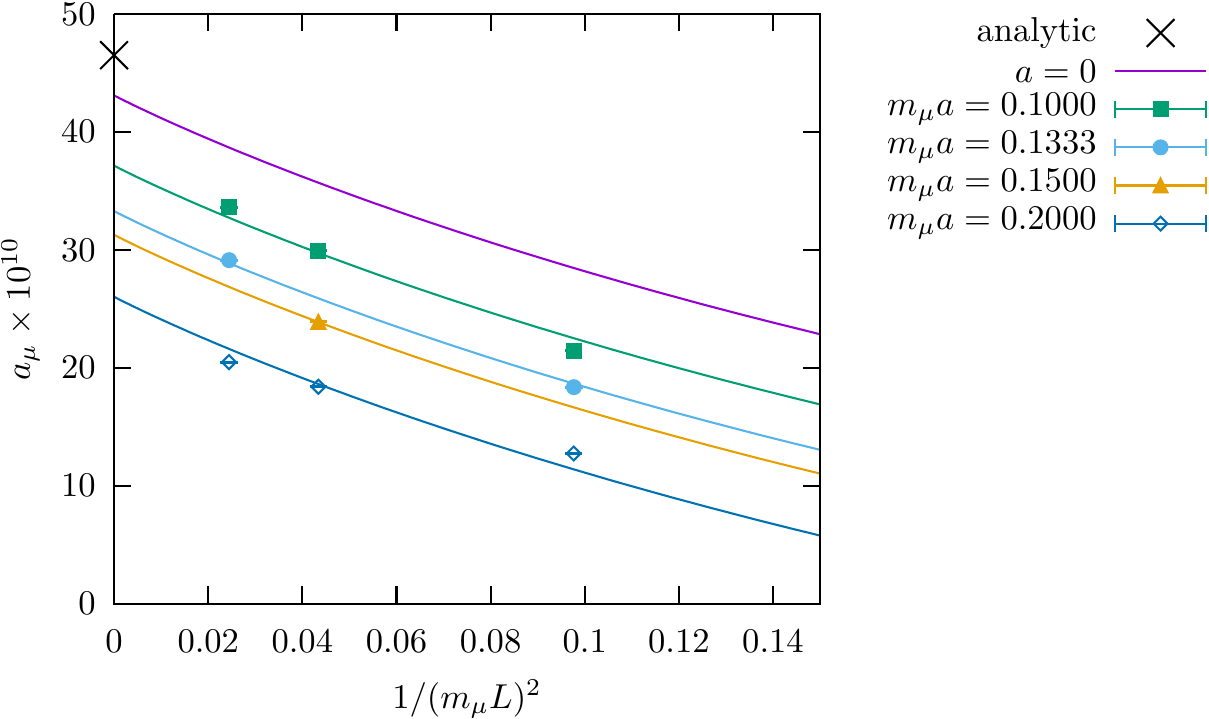}
\\
\includegraphics[width=0.45\textwidth]{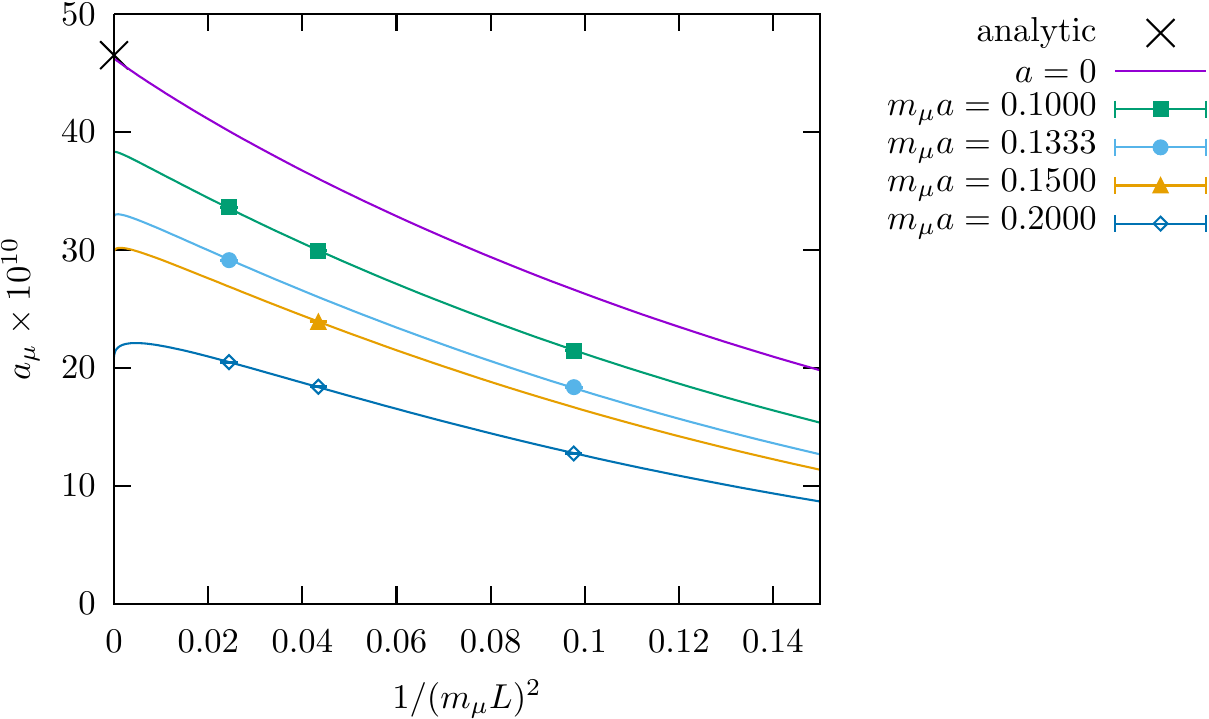}
\caption{QED test with the follwing forms:
\textit{fit-product-form-3L-4a},
\textit{fit-product-form-2L-4a},
\textit{fit-product-form-2L-2a},
\textit{fit-plus-form-3L-4a},
\textit{fit-plus-form-3L-4a-cross},
in this order.
}
\end{figure*}

\subsection{Points sampling}
\label{sec:points-sampling}

{
\changes
How the point source locations are chosen in the connected and the disconnected diagrams calculation was described in the ``Computational details'' section
of our previous work \cite{Blum:2016lnc},
where the strategy was first applied in the 48I ensemble calculation.
We use exactly the same sample strategy in all the ensembles we present in
this paper, except for the 32D and 48D ensembles.
Due to the larger physical sizes, we use the same sampling method
but twice as many points for 32D and eight times as many points for 48D.

In the connected-diagram calculation, we use independently chosen
point-pairs as point $x$ and $y$ in Fig.~\ref{fig:point src method}.
The point-pairs provide various values for $r=x-y$,
and allow a stochastic summation over $r$.
We use importance sampling to enhance the statistics at relatively short
distance, where most of the signal, and relatively little statistical error, arises.
That is, point-pairs are sampled with a non-uniform distribution --
short distances are more likely to be sampled than very long ones.
In particular, all possible $r$ with length less than or equal
to $5$ (in lattice units) are included, up to discrete symmetries.
For distance larger than $5$, we use the probability distribution
$p(r) \propto \exp (- 0.01 | r |) / | r |^4$.
Point source propagators as well as sequential propagators are used to compute the amplitude.

In the leading disconnected-diagram calculation,
we choose $M$ (in most cases $M=1024$) point
source locations distributed roughly uniformly over the entire lattice.
Point source propagators are calculated for each of the $M$ point source locations.
To compute the amplitude for the leading disconnected diagram shown in
Fig.~\ref{fig:disco diags}, we need two stochastically chosen point sources as point $y$ and $z$.
To achieve maximum statistics, all $M^2$ combinations of the point
source propagators are used.
This is referred to as the ``$M^2$ trick'' \cite{Blum:2015gfa,Blum:2016lnc}.

}

\subsection{Systematic error estimation}
\label{sec:systematics}

We discuss how we estimate all systematic errors
presented in our results, including:
hybrid $\mathcal{O}(a^2)$,
$\mathcal{O}(a^4)$,
$\mathcal{O}(a^2\log(a^2))$,
$\mathcal{O}(1/L^3)$,
$\mathcal{O}(a^2/L)$,
strange quark contribution to the connected diagrams,
and sub-leading disconnected diagrams' contribution.
Many of the above systematic errors are resulting from
lacking the corresponding terms
in the fitting formula, Eq.~(\ref{eq:fit-plus-form-2L-2a-2ad-4ad}),
which we shall refer to as the
``\textit{fit-plus-form-2L-2a-2ad-4ad}''.
We therefore estimate the systematic errors
by fitting with different fitting formulas with
these terms included.
We listed all the fitting forms below.

\begin{itemize}

    \item \textit{fit-plus-form-2L-2a-2ad-4ad}
    
\begin{eqnarray}
a_\mu(L,a^\text{I},a^\text{D})
&=&
a_\mu
\Big(
1
- \frac{b_2}{(m_\mu L)^2}
\\ && \hspace{-1.5cm}
- c_1^\text{I} (a^\text{I}~\mathrm{GeV})^2
- c_1^\text{D} (a^\text{D}~\mathrm{GeV})^2
+ c_2^\text{D} (a^\text{D}~\mathrm{GeV})^4
\Big)
\nonumber
\end{eqnarray}

    \item \textit{fit-plus-form-3L-2a-2ad-4ad}
   
\begin{eqnarray}
a_\mu(L,a^\text{I},a^\text{D})
&=&
a_\mu
\Big(
1
- \frac{b_2}{(m_\mu L)^2}
+ \frac{b_2}{(m_\mu L)^3}
\\ && \hspace{-1.5cm}
- c_1^\text{I} (a^\text{D}~\mathrm{GeV})^2
- c_1^\text{D} (a^\text{D}~\mathrm{GeV})^2
+ c_2^\text{D} (a^\text{D}~\mathrm{GeV})^4
\Big)
\nonumber
\end{eqnarray}

    \item \textit{fit-plus-form-2L-2a-4a-2ad}

\begin{eqnarray}
a_\mu(L,a^\text{I},a^\text{D})
&=&
a_\mu
\Big(
1
- \frac{b_2}{(m_\mu L)^2}
\\ && \hspace{-1.0cm}
- c_1^\text{I} (a^\text{I}~\mathrm{GeV})^2
- c_1^\text{D} (a^\text{D}~\mathrm{GeV})^2
+ c_2 (a~\mathrm{GeV})^4
\Big)
\nonumber
\end{eqnarray}

    \item \textit{fit-plus-form-2L-2a-4a-4ad}

\begin{eqnarray}
a_\mu(L,a^\text{I},a^\text{D})
&=&
a_\mu
\Big(
1
- \frac{b_2}{(m_\mu L)^2}
\\ && \hspace{-1.0cm}
- c_1 (a~\mathrm{GeV})^2
+ c_2^\text{I} (a^\text{I}~\mathrm{GeV})^4
+ c_2^\text{D} (a^\text{D}~\mathrm{GeV})^4
\Big)
\nonumber
\end{eqnarray}

    \item \textit{fit-plus-form-2L-2a-2ad-4ad-lna}

\begin{eqnarray}
a_\mu(L,a^\text{I},a^\text{D})
&=&
a_\mu
\Big(
1
- \frac{b_2}{(m_\mu L)^2}
\\
&&
\hspace{-1.5cm}
-\Big(
c_1^\text{I} (a^\text{I}~\mathrm{GeV})^2
+ c_1^\text{D} (a^\text{D}~\mathrm{GeV})^2
- c_2^\text{D} (a^\text{D}~\mathrm{GeV})^4
\Big)
\nonumber
\\
&&
\hspace{0.0cm}
\times
\Big(1 - \frac{\alpha_S}{\pi} \log\big((a~\mathrm{GeV})^2\big)\Big)
\Bigg)
\nonumber
\end{eqnarray}
    
    \item \textit{fit-plus-form-2L-2a-2ad-4ad-cross}

 \begin{eqnarray}
a_\mu(L,a^\text{I},a^\text{D})
&=&
a_\mu
\Bigg(
1
- \frac{b_2}{(m_\mu L)^2}
\\ && \hspace{-1.5cm}
-
\Big(
c_1^\text{I} (a^\text{I}~\mathrm{GeV})^2
+ c_1^\text{D} (a^\text{D}~\mathrm{GeV})^2
- c_2^\text{D} (a^\text{D}~\mathrm{GeV})^4
\Big)
\nonumber
\\ && \hspace{2.0cm}
\times\Big(1 - \frac{1}{m_\mu L}\Big)
\Bigg)
\nonumber
\end{eqnarray}
    
    \item \textit{fit-product-form-2L-2a-2ad-4ad}

\begin{eqnarray}
a_\mu(L,a^\text{I},a^\text{D})
&=&
a_\mu
\Big(
1
- \frac{b_2}{(m_\mu L)^2}
\Big)
\\ && \hspace{-1.0cm}
\times\Big(1
- c_1^\text{I} (a^\text{I}~\mathrm{GeV})^2
- c_1^\text{D} (a^\text{D}~\mathrm{GeV})^2
\nonumber \\ && \hspace{1.0cm}
+ c_2^\text{D} (a^\text{D}~\mathrm{GeV})^4
\Big)
\nonumber
\end{eqnarray}

    \item \textit{fit-product-form-3L-2a-2ad-4ad}

\begin{eqnarray}
a_\mu(L,a^\text{I},a^\text{D})
&=&
a_\mu
\Big(
1
- \frac{b_2}{(m_\mu L)^2}
+ \frac{b_2}{(m_\mu L)^3}
\Big)
\\ && \hspace{-1.0cm}
\times\Big(1
- c_1^\text{I} (a^\text{I}~\mathrm{GeV})^2
- c_1^\text{D} (a^\text{D}~\mathrm{GeV})^2
\nonumber \\ && \hspace{1.0cm}
+ c_2^\text{D} (a^\text{D}~\mathrm{GeV})^4
\Big)
\nonumber
\end{eqnarray}
    
    \item \textit{fit-product-form-2L-2a-4a-2ad}

\begin{eqnarray}
a_\mu(L,a^\text{I},a^\text{D})
&=&
a_\mu
\Big(
1
- \frac{b_2}{(m_\mu L)^2}
\Big)
\\ && \hspace{-1.0cm}
\times\Big(1
- c_1^\text{I} (a^\text{I}~\mathrm{GeV})^2
- c_1^\text{D} (a^\text{D}~\mathrm{GeV})^2
\nonumber \\ && \hspace{1.0cm}
+ c_2 (a~\mathrm{GeV})^4
\Big)
\nonumber
\end{eqnarray}
    
    \item \textit{fit-product-form-2L-2a-4a-4ad}

\begin{eqnarray}
a_\mu(L,a^\text{I},a^\text{D})
&=&
a_\mu
\Big(
1
- \frac{b_2}{(m_\mu L)^2}
\Big)
\\ && \hspace{-1.0cm}
\times\Big(1
- c_1 (a~\mathrm{GeV})^2
\nonumber \\ && \hspace{-0.3cm}
+ c_2^\text{I} (a^\text{I}~\mathrm{GeV})^4
+ c_2^\text{D} (a^\text{D}~\mathrm{GeV})^4
\Big)
\nonumber
\end{eqnarray}

    \item \textit{fit-product-form-2L-2a-2ad-4ad-lna}
 
\begin{eqnarray}
a_\mu(L,a^\text{I},a^\text{D})
&=&
a_\mu
\Big(
1
- \frac{b_2}{(m_\mu L)^2}
\Big)
\\ && \hspace{-2.1cm} \times
\Bigg(
1-
\Big(
c_1^\text{I} (a^\text{I}~\mathrm{GeV})^2
+ c_1^\text{D} (a^\text{D}~\mathrm{GeV})^2
- c_2^\text{D} (a^\text{D}~\mathrm{GeV})^4
\Big)
\nonumber
\\ && \hspace{1.5cm}\times
\Big(1 - \frac{\alpha_S}{\pi} \log\big((a~\mathrm{GeV})^2\big)\Big)
\Bigg)
\nonumber
\end{eqnarray}
    
    \item \textit{fit-product-form-2L-2a-2ad-4ad-cross}

\begin{eqnarray}
a_\mu(L,a^\text{I},a^\text{D})
&=&
a_\mu
\Big(
1
- \frac{b_2}{(m_\mu L)^2}
\Big)
\\ && \hspace{-1.5cm}
\times\Bigg(1
-
\Big(
c_1^\text{I} (a^\text{I}~\mathrm{GeV})^2
+ c_1^\text{D} (a^\text{D}~\mathrm{GeV})^2
\nonumber \\ && \hspace{0.0cm}
- c_2^\text{D} (a^\text{D}~\mathrm{GeV})^4
\Big)
\Big(1 - \frac{1}{m_\mu L}\Big)
\Bigg)
\nonumber
\end{eqnarray}

\end{itemize}

Again, in these formulas, \(a^\text{I}\) is the lattice spacing for
the Iwasaki ensembles,
48I and 64I. We define it to be zero for I-DSDR ensembles.
Similarly, \(a^\text{D}\) is the lattice spacing for I-DSDR ensembles, and
it is zero for Iwasaki ensembles. With this notation, the lattice spacings
for all our ensembles are always equal to $a=a^\text{I}+a^\text{D}$.
In two of the fitting forms, $\alpha_S(a)$ is used to estimate the
size of the $a^2 \log(a^2)$ contribution.
We use the 2-loop $\alpha_S^{\overline{\mathrm{MS}}}$ with
$\Lambda_\text{QCD}=325~\mathrm{MeV}$:

\begin{center}
\begin{tabular}{@{}cc@{}}
\toprule
\(a^{-1} / \mathrm{GeV}\) & \(\alpha_S\)\tabularnewline
\midrule
1.015 & 0.44\tabularnewline
1.378 & 0.34\tabularnewline
1.730 & 0.30\tabularnewline
2.359 & 0.26\tabularnewline
\bottomrule
\end{tabular}
\end{center}

All these fit forms correct the \(\mathcal{O}(a^2)\) lattice artifacts
presented in DWF lattice calculations for Iwasaki ensembles (48I and 64I).
The results obtained has a relative large statistical error
mostly because of the continuum extrapolation.
In Section~\ref{sec:results},
we introduced the hybrid continuum limit
for the connected diagrams' contribution,
where
we calculate the continuum
limit for contributions from the region \(r\leq 1~\mathrm{fm}\) and use
the relatively precise 48I ensemble results without extrapolation for
the \(r>1~\mathrm{fm}\) region.
This hybrid procedure is possible due to the small size of
the contribution from the region \(r>1~\mathrm{fm}\).
This is expected due to the exponential suppression from
QCD mass-gap, and, in addition, for the connected diagrams, we label the
three vertex locations on the quark loop that connect to the internal
photons as \(x\), \(y\), \(z\) and require
\(r=|x-y|\le \min(|x-z|,|y-z|)\).
This is a direct consequence of Eq.~(3) in Ref.~\cite{Blum:2016lnc}.
Operationally, this hybrid continuum limit for the connected diagrams
is implemented by replacing the long distance region (\(r>1~\mathrm{fm}\)) of the 64I
ensemble data with the corresponding 48I ensemble data:

\[a_\mu^\text{64I} \leftarrow a_\mu^\text{64I}(r\le 1~\mathrm{fm}) + a_\mu^\text{48I}(r>1~\mathrm{fm}).\]

We perform the above replacement under the super jackknife procedure
before the global fit together with all the results from other ensembles.

To estimate the hybrid \(\mathcal{O}(a^2)\)
systematic error resulting from the above replacement,
we perform a slightly different replacement:
\[a_\mu^\text{64I}
\leftarrow a_\mu^\text{64I}(r\le 1~\mathrm{fm})
+
\frac
{a_\mu^\text{64I,con}(r\le 1~\mathrm{fm})}
{a_\mu^\text{48I,con}(r\le 1~\mathrm{fm})}
a_\mu^\text{48I}(r>1~\mathrm{fm}).\]
The difference of the results obtained with these two replacements
is used as the estimation of the hybrid $\mathcal{O}(a^2)$ systematic error.

We then estimate systematic errors due to lacking higher
order finite volume or discretization terms in
our default fitting form \textit{fit-plus-form-2L-2a-2ad-4ad}
by comparing the results between the fitting formulas:
\begin{itemize}
    \item  \(\mathcal{O}(1/L^3)\): Difference between
  \textit{fit-plus-form-2L-2a-2ad-4ad} and \textit{fit-plus-form-3L-2a-2ad-4ad};
  
  \item \(\mathcal{O}(a^4)\): The maximum of the following two differences:
  (a) \textit{fit-plus-form-2L-2a-2ad-4ad} and \textit{fit-plus-form-2L-2a-4a-2ad};
  (b) \textit{fit-plus-form-2L-2a-2ad-4ad} and \textit{fit-plus-form-2L-2a-4a-4ad}.

  \item \(\mathcal{O}(a^2\log(a^2))\): Difference between
  \textit{fit-plus-form-2L-2a-2ad-4ad} and \textit{fit-plus-form-2L-2a-2ad-4ad-lna}.

  \item \(\mathcal{O}(a^2/L)\): The maximum of the following two differences:
  (a) \textit{fit-plus-form-2L-2a-2ad-4ad} and
 \textit{fit-plus-form-2L-2a-2ad-4ad-cross};
  (b) \textit{fit-plus-form-2L-2a-2ad-4ad} and
  \textit{fit-product-form-2L-2a-2ad-4ad}.
\end{itemize}

\begin{figure}[hbt]
\centering
\includegraphics[width=0.4\textwidth]{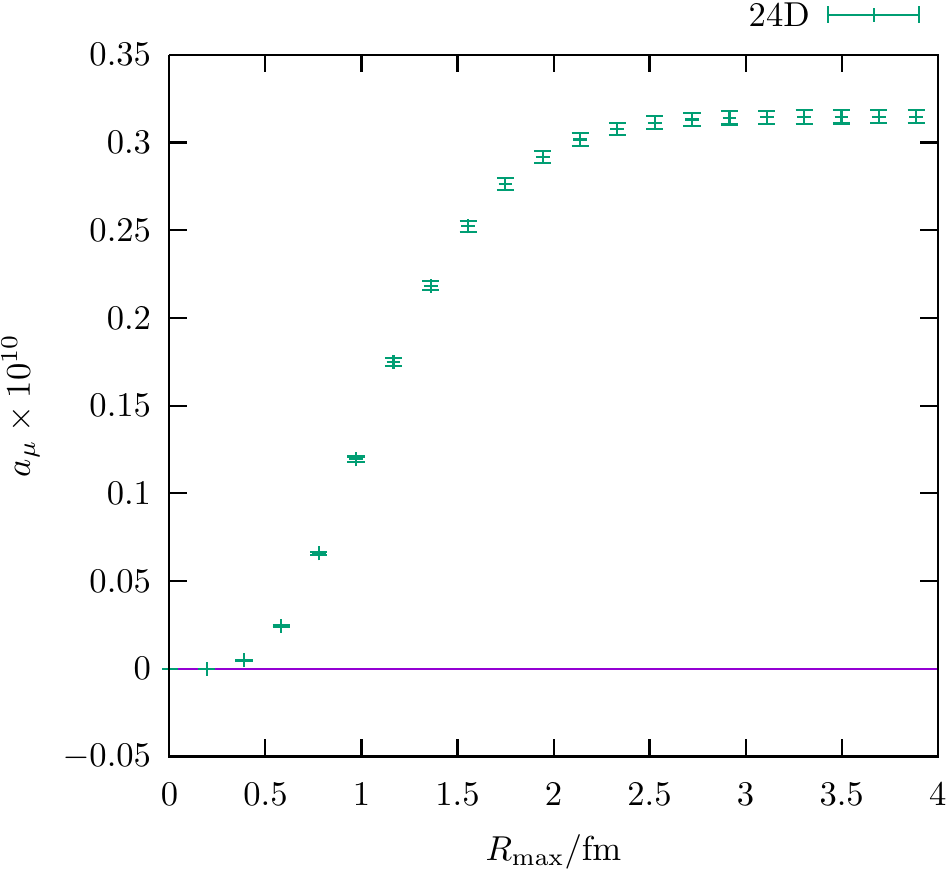}
\caption{Strange quark contribution to the connected diagrams calculated with QED$_\infty$.
The strange quark contribution to the disconnected diagrams is already
included in the present QED$_L$ calculation.}
\label{fig:sys-strange}
\end{figure}

\begin{figure}[hbt]
\centering
\includegraphics[width=0.4\textwidth]{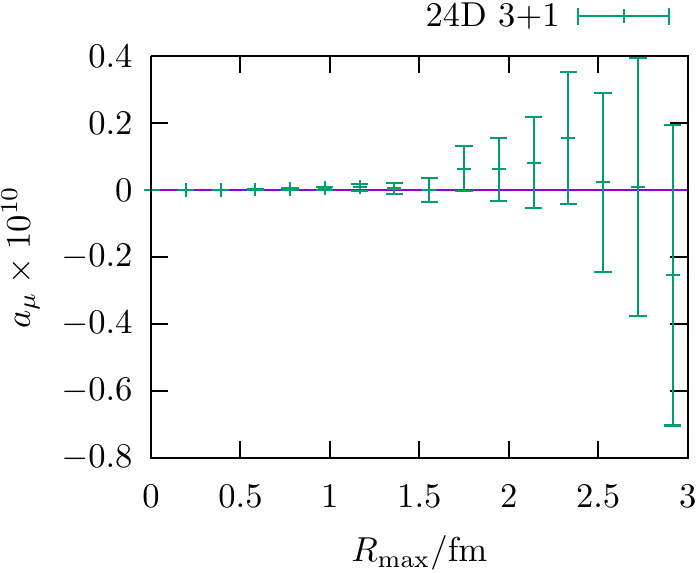}
\caption{Sub-leading disconnected diagrams' contribution.
We only include the second diagram in Fig.~\ref{fig:disco diags}.
Only light quark contribution is calculated except for the tadpole part, where we reuse the calculation for
the disconnected diagrams in HVP calculations described in Ref.~\cite{Blum:2015you} and both light quark and strange quark
contribution is included.
The remaining diagrams are equally or more suppressed.
}
\label{fig:sys-sub-leading}
\end{figure}

For the systematic errors due to not including
(a) the strange quark contribution to the connected diagrams and
(b) sub-leading disconnected diagrams' contributions,
we performed lattice calculation with the 24D ensemble using
the QED\(_\infty\) method~\cite{Blum:2017cer}.
The results are plotted in Fig.~\ref{fig:sys-strange} and Fig.~\ref{fig:sys-sub-leading}.
We estimate the systematic error due to omitting these two contributions
to be: \(\pm 0.3\times 10^{-10}\) and \(\pm 0.5\times 10^{-10}\)
respectively.

We then add all these systematic effects in quadrature as our final
systematic error. We also use these procedure per jackknife sample and
calculate the statistical uncertainty of the systematic error.
This procedure is repeated for each of the ``con'', ``discon'', and ``tot'' contributions.
The ``tot'' contribution is calculated by perform a single fit with
the sum of ``con'' and the ``discon'' contributions for each ensemble.
Since we only calculated the ``con'' contribution for the 48D ensemble,
48D is not included in the fit for ``tot''.
As such, the ``tot'' fit is usually not exactly equal to
the sum of the individual ``con'' and ``discon'' fits, which we label them as ``sum''.
The results are shown in the
Tab.~\ref{tab:hybrid-plus-form-results}.
Statistical errors for central values and systematic errors are
listed in the table.
We use the ``tot'' in this table as our central value.

We did the same exercise for \textit{fit-product-form-2L-2a-2ad-4ad}
as a comparison.
The results are shown in Tab.~\ref{tab:hybrid-product-form-results}.
Based on the experiences from the QED test, result from the
\textit{fit-product-form-2L-2a-2ad-4ad} fit the data better.
However, in QCD calculations,
\textit{fit-plus-form-2L-2a-2ad-4ad} leads to smaller statistical
error.
At present level of accuracy, smaller statistical error is more important
than potentially smaller systematic error.
In particular, the systematic error is largely
cancelled between the connected diagrams and the disconnected diagrams,
while the statistical error does not.
The fitting results with all the fitting forms are also summarized in
Tab~\ref{tab:hybrid-all-fits}.

We have also performed the calculation without the hybrid continuum limit.
The results are shown in
Tabs.~\ref{tab:nohybrid-plus-form-results},
\ref{tab:nohybrid-product-form-results},
\ref{tab:nohybrid-all-fits}.

Finally, the detailed results and plots for each fitting forms
with or without the hybrid continuum limit are listed
in remaining tables and plots.

\begin{table*}
\begin{tabular}{@{}ccccc@{}}
\toprule
& con & discon & sum & tot\tabularnewline
\midrule
\(a_\mu\) & 24.16(2.30) & -16.45(2.13) & 7.71(3.04) &
7.87(3.06)\tabularnewline
sys hybrid \(\mathcal{O}(a^2)\) & 0.20(0.45) & 0.00(0.00) & 0.20(0.45) &
0.20(0.45)\tabularnewline
sys \(\mathcal{O}(1/L^3)\) & 2.34(0.41) & 1.72(0.32) & 0.63(0.54) &
0.83(0.56)\tabularnewline
sys \(\mathcal{O}(a^4)\) & 0.88(0.31) & 0.71(0.28) & 0.99(0.92) &
0.95(0.92)\tabularnewline
sys \(\mathcal{O}(a^2\log(a^2))\) & 0.23(0.08) & 0.25(0.09) & 0.02(0.12)
& 0.02(0.11)\tabularnewline
sys \(\mathcal{O}(a^2/L)\) & 4.43(1.38) & 3.49(1.37) & 0.94(1.87) &
1.08(1.57)\tabularnewline
sys strange con & 0.30 & 0.00 & 0.30 & 0.30\tabularnewline
sys subleading discon & 0.00 & 0.50 & 0.50 & 0.50\tabularnewline
sys all & 5.11(1.32) & 3.99(1.29) & 1.62(1.25) &
1.77(1.13)\tabularnewline
\bottomrule
\end{tabular}
\caption{Plus form fitting results, use the hybrid continuum limit for the connected diagrams.}
\label{tab:hybrid-plus-form-results}
\end{table*}

\begin{table*}
\begin{tabular}{@{}ccccc@{}}
\toprule
& con & discon & sum & tot\tabularnewline
\midrule
\(a_\mu\) & 28.59(3.65) & -19.94(3.49) & 8.65(4.88) &
8.95(4.59)\tabularnewline
sys hybrid \(\mathcal{O}(a^2)\) & 0.33(0.73) & 0.00(0.00) & 0.33(0.73) &
0.31(0.70)\tabularnewline
sys \(\mathcal{O}(1/L^3)\) & 4.18(0.84) & 3.24(0.76) & 0.94(1.13) &
1.28(1.01)\tabularnewline
sys \(\mathcal{O}(a^4)\) & 1.40(0.53) & 1.18(0.47) & 1.48(1.62) &
1.36(1.46)\tabularnewline
sys \(\mathcal{O}(a^2\log(a^2))\) & 0.38(0.13) & 0.41(0.15) & 0.03(0.20)
& 0.02(0.16)\tabularnewline
sys \(\mathcal{O}(a^2/L)\) & 4.43(1.38) & 3.49(1.37) & 1.15(1.74) &
1.08(1.57)\tabularnewline
sys strange con & 0.30 & 0.00 & 0.30 & 0.30\tabularnewline
sys subleading discon & 0.00 & 0.50 & 0.50 & 0.50\tabularnewline
sys all & 6.28(1.57) & 4.95(1.53) & 2.20(1.64) &
2.26(1.51)\tabularnewline
\bottomrule
\end{tabular}
\caption{Product form fitting results, use the hybrid continuum limit for the connected diagrams.}
\label{tab:hybrid-product-form-results}
\end{table*}
\begin{table*}
\begin{tabular}{@{}ccccccc@{}}
\toprule
fit-form & con & discon & sum & sum-sys (\(a^2\)) & tot & tot-sys
(\(a^2\))\tabularnewline
\midrule
fit-plus-form-2L-2a-2ad-4ad & 24.16(2.30) & -16.45(2.13) & 7.71(3.04) &
0.20(0.45) & 7.87(3.06) & 0.20(0.45)\tabularnewline
fit-plus-form-3L-2a-2ad-4ad & 26.51(2.53) & -18.17(2.29) & 8.34(3.34) &
0.20(0.45) & 8.70(3.38) & 0.20(0.45)\tabularnewline
fit-plus-form-2L-2a-4a-2ad & 25.04(2.50) & -17.16(2.35) & 7.88(3.32) &
0.23(0.51) & 8.03(3.33) & 0.23(0.51)\tabularnewline
fit-plus-form-2L-2a-4a-4ad & 24.71(2.43) & -16.01(1.98) & 8.70(3.03) &
0.20(0.44) & 8.82(3.04) & 0.20(0.44)\tabularnewline
fit-plus-form-2L-2a-2ad-4ad-lna & 24.39(2.36) & -16.70(2.21) &
7.69(3.14) & 0.21(0.47) & 7.85(3.16) & 0.21(0.47)\tabularnewline
fit-plus-form-2L-2a-2ad-4ad-cross & 26.18(2.76) & -17.84(2.59) &
8.34(3.65) & 0.26(0.57) & 8.38(3.65) & 0.26(0.57)\tabularnewline
fit-product-form-2L-2a-2ad-4ad & 28.59(3.65) & -19.94(3.49) & 8.65(4.88)
& 0.33(0.73) & 8.95(4.59) & 0.31(0.70)\tabularnewline
fit-product-form-3L-2a-2ad-4ad & 32.78(4.42) & -23.19(4.20) & 9.59(5.91)
& 0.37(0.83) & 10.23(5.51) & 0.36(0.80)\tabularnewline
fit-product-form-2L-2a-4a-2ad & 29.99(4.01) & -21.12(3.88) & 8.87(5.38)
& 0.37(0.82) & 9.19(5.04) & 0.35(0.79)\tabularnewline
fit-product-form-2L-2a-4a-4ad & 29.36(3.89) & -19.23(3.24) & 10.13(4.88)
& 0.32(0.72) & 10.31(4.68) & 0.31(0.69)\tabularnewline
fit-product-form-2L-2a-2ad-4ad-lna & 28.97(3.76) & -20.35(3.64) &
8.62(5.05) & 0.34(0.76) & 8.93(4.74) & 0.33(0.73)\tabularnewline
fit-product-form-2L-2a-2ad-4ad-cross & 31.86(4.75) & -22.06(4.47) &
9.80(6.27) & 0.43(0.97) & 9.69(5.70) & 0.41(0.96)\tabularnewline
\bottomrule
\end{tabular}
\caption{Values for all fitting forms, use the hybrid continuum limit for the connected diagrams.}
\label{tab:hybrid-all-fits}
\end{table*}

\begin{table*}
\begin{tabular}{@{}ccccc@{}}
\toprule
& con & discon & sum & tot\tabularnewline
\midrule
\(a_\mu\) & 23.76(3.96) & -16.45(2.13) & 7.31(4.23) &
7.47(4.24)\tabularnewline
sys hybrid \(\mathcal{O}(a^2)\) & 0.00(0.00) & 0.00(0.00) & 0.00(0.00) &
0.00(0.00)\tabularnewline
sys \(\mathcal{O}(1/L^3)\) & 2.34(0.41) & 1.72(0.32) & 0.63(0.54) &
0.83(0.56)\tabularnewline
sys \(\mathcal{O}(a^4)\) & 0.83(0.53) & 0.71(0.28) & 0.99(0.94) &
0.96(0.94)\tabularnewline
sys \(\mathcal{O}(a^2\log(a^2))\) & 0.21(0.18) & 0.25(0.09) & 0.03(0.17)
& 0.03(0.17)\tabularnewline
sys \(\mathcal{O}(a^2/L)\) & 4.18(2.37) & 3.49(1.37) & 0.69(2.48) &
0.86(2.20)\tabularnewline
sys strange con & 0.30 & 0.00 & 0.30 & 0.30\tabularnewline
sys subleading discon & 0.00 & 0.50 & 0.50 & 0.50\tabularnewline
sys all & 4.88(2.17) & 3.99(1.29) & 1.48(1.20) &
1.64(1.15)\tabularnewline
\bottomrule
\end{tabular}
\caption{Plus form fitting results, do not use the hybrid continuum limit.}
\label{tab:nohybrid-plus-form-results}
\end{table*}
\begin{table*}
\begin{tabular}{@{}ccccc@{}}
\toprule
& con & discon & sum & tot\tabularnewline
\midrule
\(a_\mu\) & 27.94(6.31) & -19.94(3.49) & 8.00(6.75) &
8.33(6.40)\tabularnewline
sys hybrid \(\mathcal{O}(a^2)\) & 0.00(0.00) & 0.00(0.00) & 0.00(0.00) &
0.00(0.00)\tabularnewline
sys \(\mathcal{O}(1/L^3)\) & 4.09(1.12) & 3.24(0.76) & 0.85(1.31) &
1.19(1.16)\tabularnewline
sys \(\mathcal{O}(a^4)\) & 1.31(0.87) & 1.18(0.47) & 1.49(1.65) &
1.37(1.48)\tabularnewline
sys \(\mathcal{O}(a^2\log(a^2))\) & 0.35(0.29) & 0.41(0.15) & 0.06(0.29)
& 0.04(0.26)\tabularnewline
sys \(\mathcal{O}(a^2/L)\) & 4.18(2.37) & 3.49(1.37) & 0.94(2.24) &
0.86(2.20)\tabularnewline
sys strange con & 0.30 & 0.00 & 0.30 & 0.30\tabularnewline
sys subleading discon & 0.00 & 0.50 & 0.50 & 0.50\tabularnewline
sys all & 6.01(2.54) & 4.95(1.53) & 2.04(1.55) &
2.09(1.46)\tabularnewline
\bottomrule
\end{tabular}
\caption{Product form fitting results, do not use the hybrid continuum limit.}
\label{tab:nohybrid-product-form-results}
\end{table*}
\begin{table*}
\begin{tabular}{@{}ccccccc@{}}
\toprule
fit-form & con & discon & sum & sum-sys (\(a^2\)) & tot & tot-sys
(\(a^2\))\tabularnewline
\midrule
fit-plus-form-2L-2a-2ad-4ad & 23.76(3.96) & -16.45(2.13) & 7.31(4.23) &
0.00(0.00) & 7.47(4.24) & -0.00(0.00)\tabularnewline
fit-plus-form-3L-2a-2ad-4ad & 26.10(4.10) & -18.17(2.29) & 7.93(4.44) &
0.00(0.00) & 8.30(4.48) & 0.00(0.00)\tabularnewline
fit-plus-form-2L-2a-4a-2ad & 24.58(4.43) & -17.16(2.35) & 7.42(4.70) &
0.00(0.00) & 7.58(4.71) & -0.00(0.00)\tabularnewline
fit-plus-form-2L-2a-4a-4ad & 24.31(3.69) & -16.01(1.98) & 8.30(3.87) &
0.00(0.00) & 8.43(3.88) & -0.00(0.00)\tabularnewline
fit-plus-form-2L-2a-2ad-4ad-lna & 23.97(4.13) & -16.70(2.21) &
7.27(4.40) & 0.00(0.00) & 7.43(4.41) & 0.00(0.00)\tabularnewline
fit-plus-form-2L-2a-2ad-4ad-cross & 25.67(4.92) & -17.84(2.59) &
7.83(5.20) & 0.00(0.00) & 7.87(5.20) & 0.00(0.00)\tabularnewline
fit-product-form-2L-2a-2ad-4ad & 27.94(6.31) & -19.94(3.49) & 8.00(6.75)
& 0.00(0.00) & 8.33(6.40) & 0.00(0.00)\tabularnewline
fit-product-form-3L-2a-2ad-4ad & 32.03(7.36) & -23.19(4.20) & 8.85(7.95)
& 0.00(0.00) & 9.52(7.47) & 0.00(0.00)\tabularnewline
fit-product-form-2L-2a-4a-2ad & 29.25(7.08) & -21.12(3.88) & 8.13(7.54)
& 0.00(0.00) & 8.48(7.14) & -0.00(0.00)\tabularnewline
fit-product-form-2L-2a-4a-4ad & 28.72(5.91) & -19.23(3.24) & 9.49(6.21)
& 0.00(0.00) & 9.69(5.95) & -0.00(0.00)\tabularnewline
fit-product-form-2L-2a-2ad-4ad-lna & 28.29(6.59) & -20.35(3.64) &
7.94(7.04) & 0.00(0.00) & 8.28(6.67) & 0.00(0.00)\tabularnewline
fit-product-form-2L-2a-2ad-4ad-cross & 31.00(8.24) & -22.06(4.47) &
8.94(8.73) & 0.00(0.00) & 8.88(8.04) & 0.00(0.00)\tabularnewline
\bottomrule
\end{tabular}
\caption{Values for all fitting forms, do not use the hybrid continuum limit.}
\label{tab:nohybrid-all-fits}
\end{table*}

\clearpage

\input{src-hybrid.tex}

\clearpage

\input{src-nohybrid.tex}

%% file: src-hybrid.tex
\begin{table*}
\begin{tabular}{@{}cccccc@{}}
\toprule
 & $a_\mu$ & $b_2$ & $c_1^\text{I}$ & $c_1^\text{D}-c_1^\text{I}$ & $c_2^\text{D}$ \tabularnewline
\midrule
con               & 24.16(2.30) & 2.20(0.33) & 0.69(0.20) & 0.13(0.18) & 0.53(0.16) \tabularnewline
discon            & -16.45(2.13) & 2.24(0.40) & 1.10(0.28) & -0.15(0.14) & 0.63(0.19) \tabularnewline
sum               & 7.71(3.04) \tabularnewline
sum-sys ($a^2$)   & 0.20(0.45) \tabularnewline
tot               & 7.87(3.06) & 2.26(1.33) & -0.15(1.15) & 0.68(0.76) & 0.31(0.69) \tabularnewline
tot-sys ($a^2$)   & 0.20(0.45) \tabularnewline
\bottomrule
\end{tabular}
\caption{Fit results for \textit{fit-plus-form-2L-2a-2ad-4ad} with the hybrid continuum limit.}
\label{tab:fit-plus-form-2L-2a-2ad-4ad_hybrid}
\end{table*}

\begin{figure*}
\centering
\includegraphics[width=0.3\textwidth]{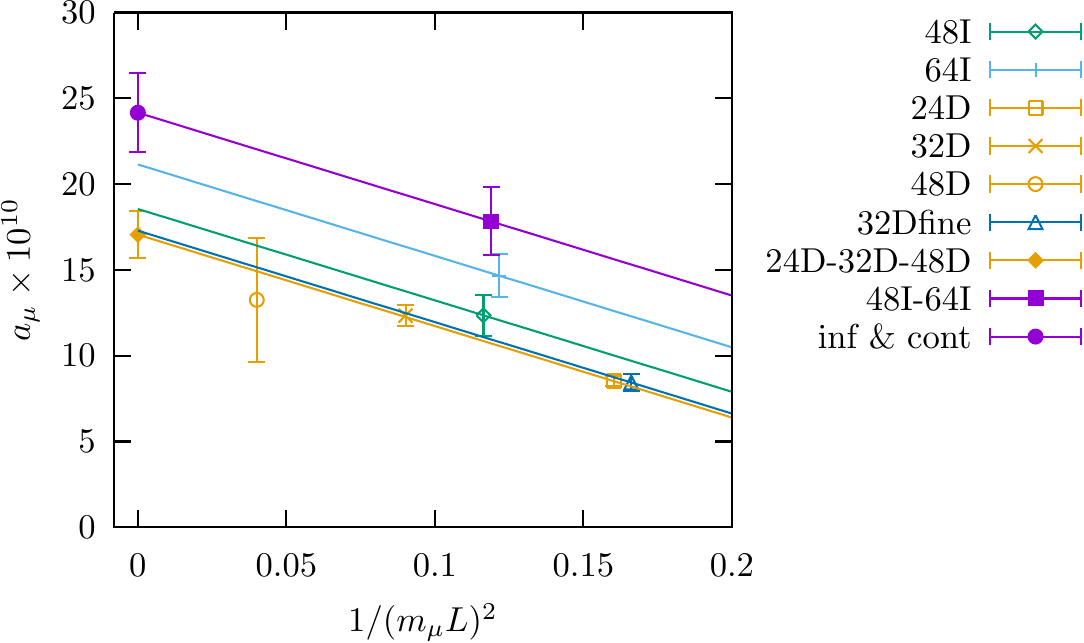}
\includegraphics[width=0.3\textwidth]{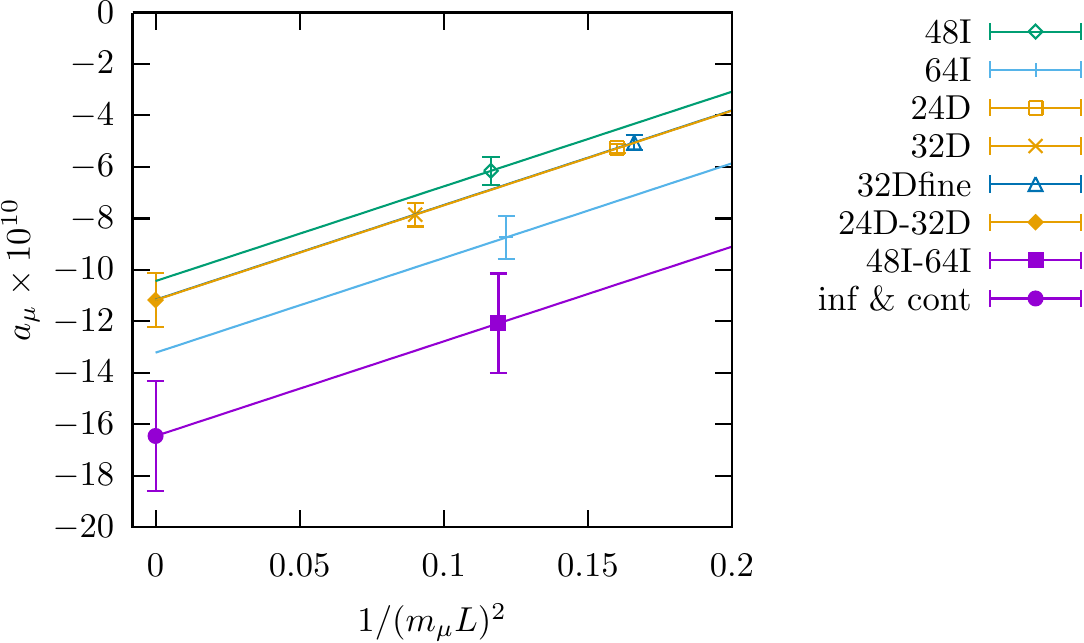}
\includegraphics[width=0.3\textwidth]{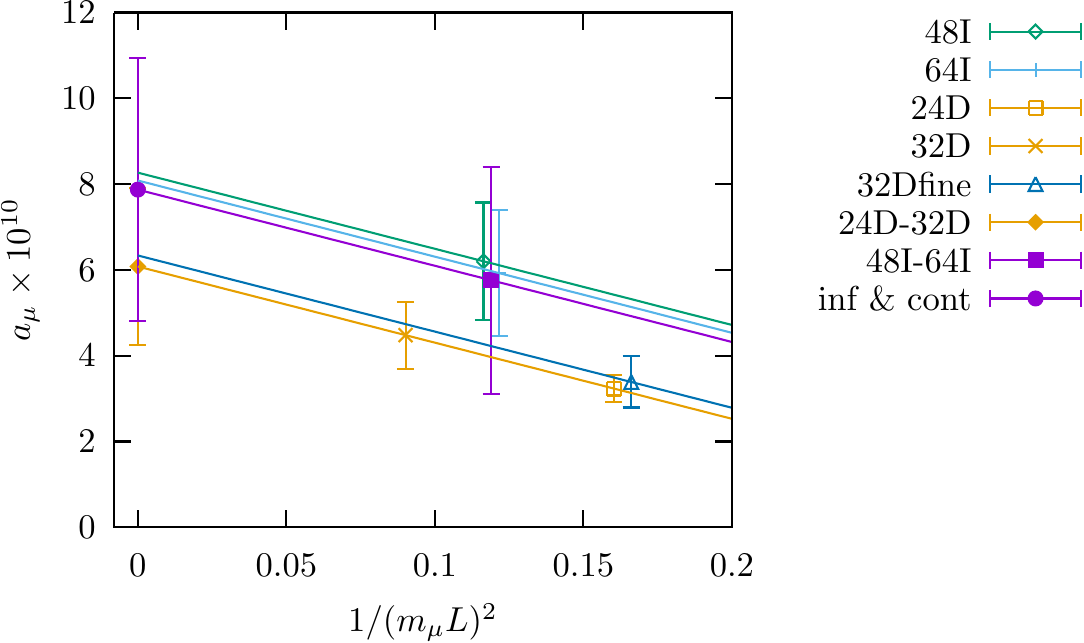}
\caption{Fit plots for \textit{fit-plus-form-2L-2a-2ad-4ad} with the hybrid continuum limit. Connected (left), disconnected (middle), and total (right).}
\label{fig:fit-plus-form-2L-2a-2ad-4ad_hybrid}
\end{figure*}

\begin{table*}
\begin{tabular}{@{}cccccc@{}}
\toprule
 & $a_\mu$ & $b_2$ & $c_1^\text{I}$ & $c_1^\text{D}-c_1^\text{I}$ & $c_2^\text{D}$ \tabularnewline
\midrule
con               & 26.51(2.53) & 4.20(0.57) & 0.63(0.19) & 0.14(0.16) & 0.51(0.15) \tabularnewline
discon            & -18.17(2.29) & 4.30(0.70) & 0.99(0.26) & -0.11(0.12) & 0.59(0.18) \tabularnewline
sum               & 8.34(3.34) \tabularnewline
sum-sys ($a^2$)   & 0.20(0.45) \tabularnewline
tot               & 8.70(3.38) & 4.33(2.32) & -0.13(1.04) & 0.64(0.69) & 0.30(0.63) \tabularnewline
tot-sys ($a^2$)   & 0.20(0.45) \tabularnewline
\bottomrule
\end{tabular}
\caption{Fit results for \textit{fit-plus-form-3L-2a-2ad-4ad} with the hybrid continuum limit.}
\label{tab:fit-plus-form-3L-2a-2ad-4ad_hybrid}
\end{table*}

\begin{figure*}
\centering
\includegraphics[width=0.3\textwidth]{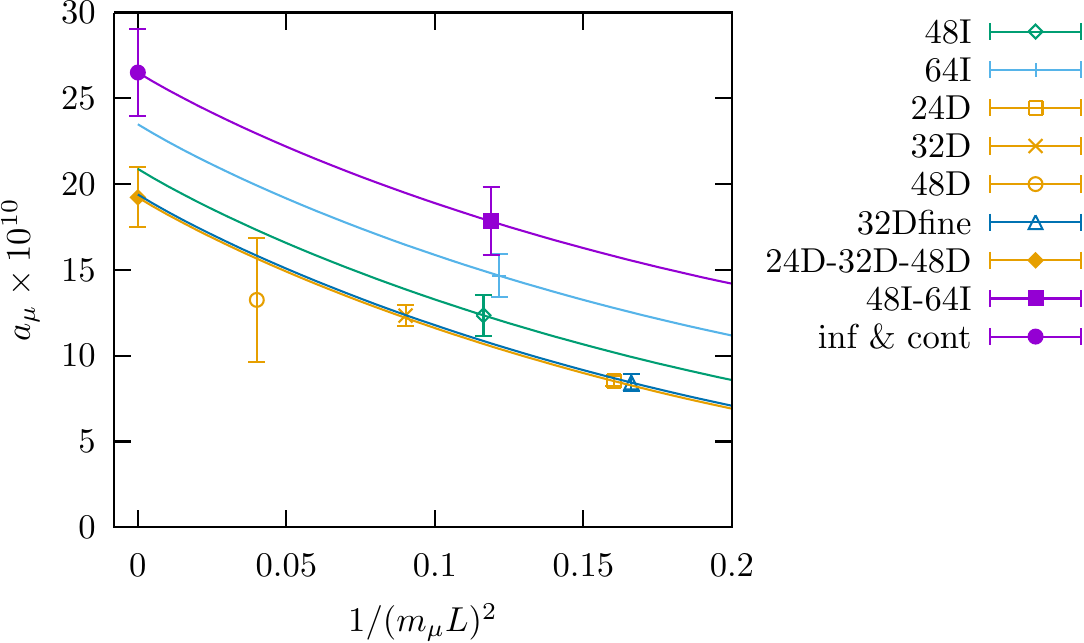}
\includegraphics[width=0.3\textwidth]{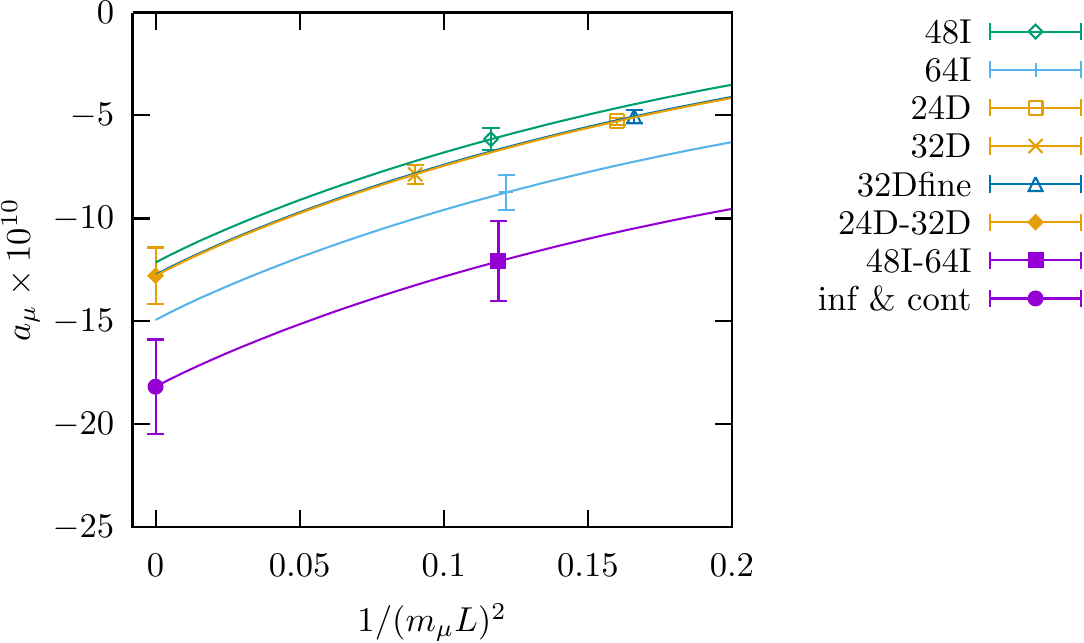}
\includegraphics[width=0.3\textwidth]{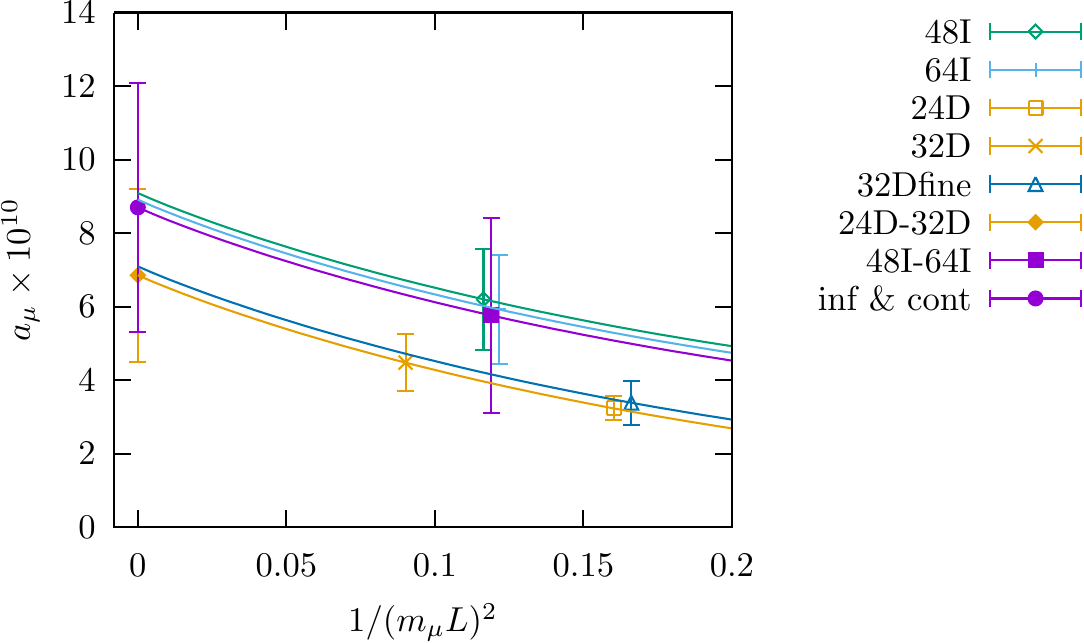}
\caption{Fit plots for \textit{fit-plus-form-3L-2a-2ad-4ad} with the hybrid continuum limit. Connected (left), disconnected (middle), and total (right).}
\label{fig:fit-plus-form-3L-2a-2ad-4ad_hybrid}
\end{figure*}

\nclearpage

\begin{table*}
\begin{tabular}{@{}cccccc@{}}
\toprule
 & $a_\mu$ & $b_2$ & $c_1^\text{I}$ & $c_2$ & $c_1^\text{D}-c_1^\text{I}$ \tabularnewline
\midrule
con               & 25.04(2.50) & 2.13(0.33) & 0.97(0.25) & 0.58(0.17) & -0.07(0.15) \tabularnewline
discon            & -17.16(2.35) & 2.14(0.41) & 1.40(0.34) & 0.69(0.20) & -0.38(0.13) \tabularnewline
sum               & 7.88(3.32) \tabularnewline
sum-sys ($a^2$)   & 0.23(0.51) \tabularnewline
tot               & 8.03(3.33) & 2.21(1.38) & 0.03(1.43) & 0.34(0.75) & 0.55(0.75) \tabularnewline
tot-sys ($a^2$)   & 0.23(0.51) \tabularnewline
\bottomrule
\end{tabular}
\caption{Fit results for \textit{fit-plus-form-2L-2a-4a-2ad} with the hybrid continuum limit.}
\label{tab:fit-plus-form-2L-2a-4a-2ad_hybrid}
\end{table*}

\begin{figure*}
\centering
\includegraphics[width=0.3\textwidth]{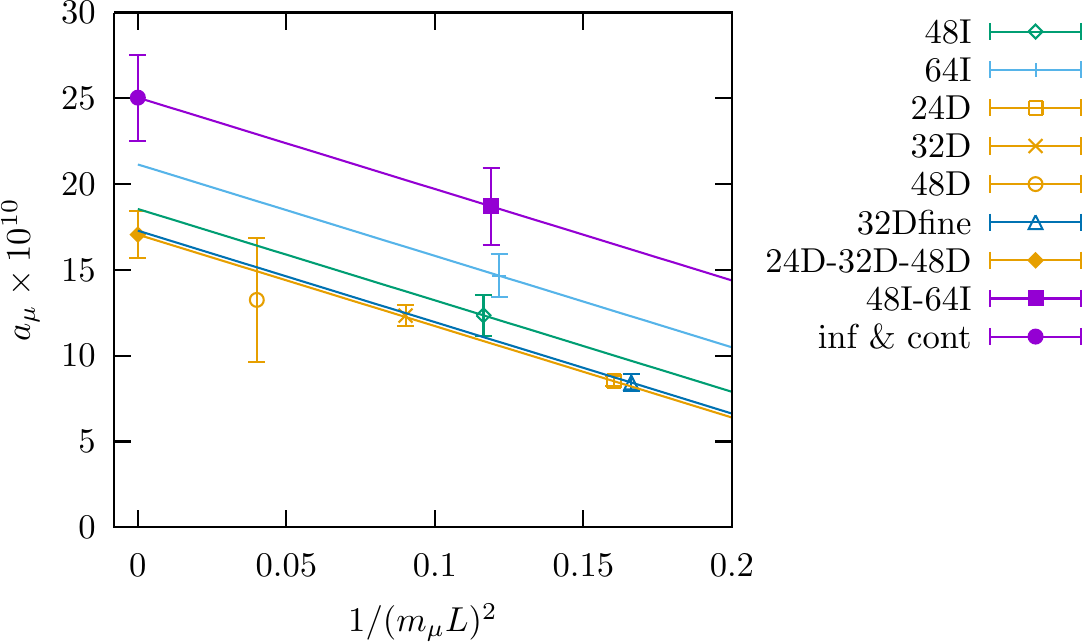}
\includegraphics[width=0.3\textwidth]{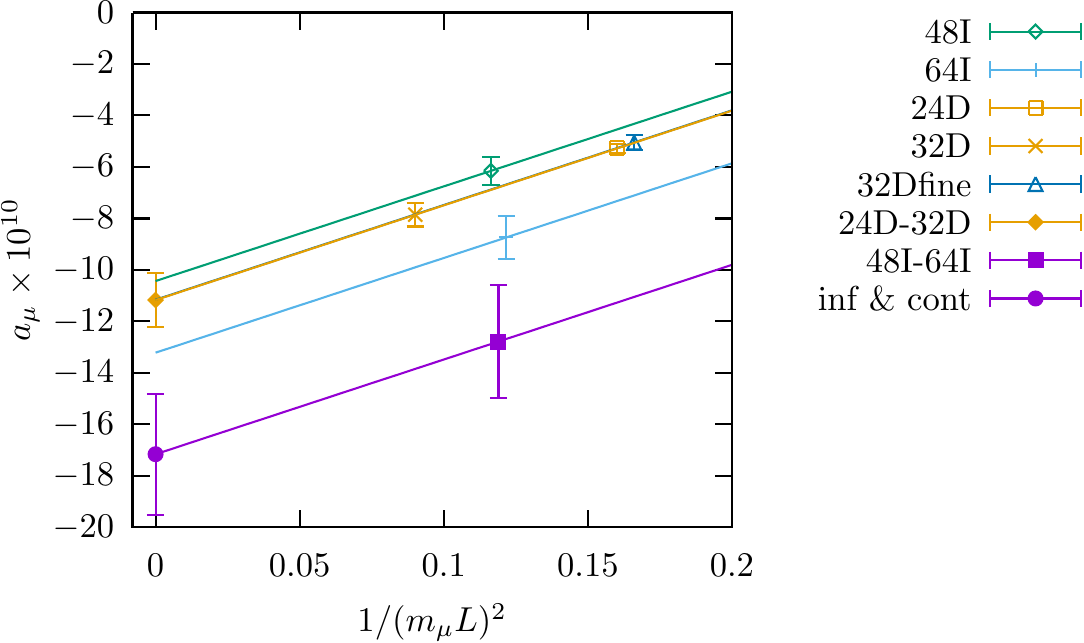}
\includegraphics[width=0.3\textwidth]{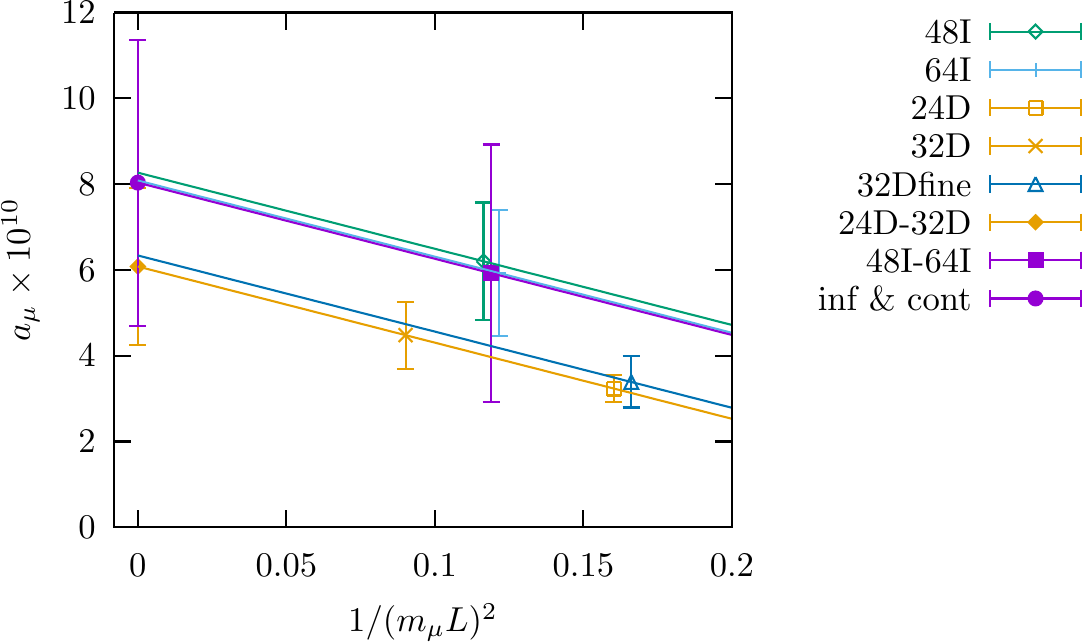}
\caption{Fit plots for \textit{fit-plus-form-2L-2a-4a-2ad} with the hybrid continuum limit. Connected (left), disconnected (middle), and total (right).}
\label{fig:fit-plus-form-2L-2a-4a-2ad_hybrid}
\end{figure*}

\begin{table*}
\begin{tabular}{@{}cccccc@{}}
\toprule
 & $a_\mu$ & $b_2$ & $c_1$ & $c_2^\text{I}$ & $c_2^\text{D}-c_2^\text{I}$ \tabularnewline
\midrule
con               & 24.71(2.43) & 2.16(0.35) & 0.87(0.25) & 0.37(0.51) & 0.20(0.41) \tabularnewline
discon            & -16.01(1.98) & 2.30(0.43) & 0.89(0.29) & -0.46(0.43) & 1.06(0.39) \tabularnewline
sum               & 8.70(3.03) \tabularnewline
sum-sys ($a^2$)   & 0.20(0.44) \tabularnewline
tot               & 8.82(3.04) & 2.01(1.27) & 0.79(0.88) & 1.80(1.78) & -1.32(1.63) \tabularnewline
tot-sys ($a^2$)   & 0.20(0.44) \tabularnewline
\bottomrule
\end{tabular}
\caption{Fit results for \textit{fit-plus-form-2L-2a-4a-4ad} with the hybrid continuum limit.}
\label{tab:fit-plus-form-2L-2a-4a-4ad_hybrid}
\end{table*}

\begin{figure*}
\centering
\includegraphics[width=0.3\textwidth]{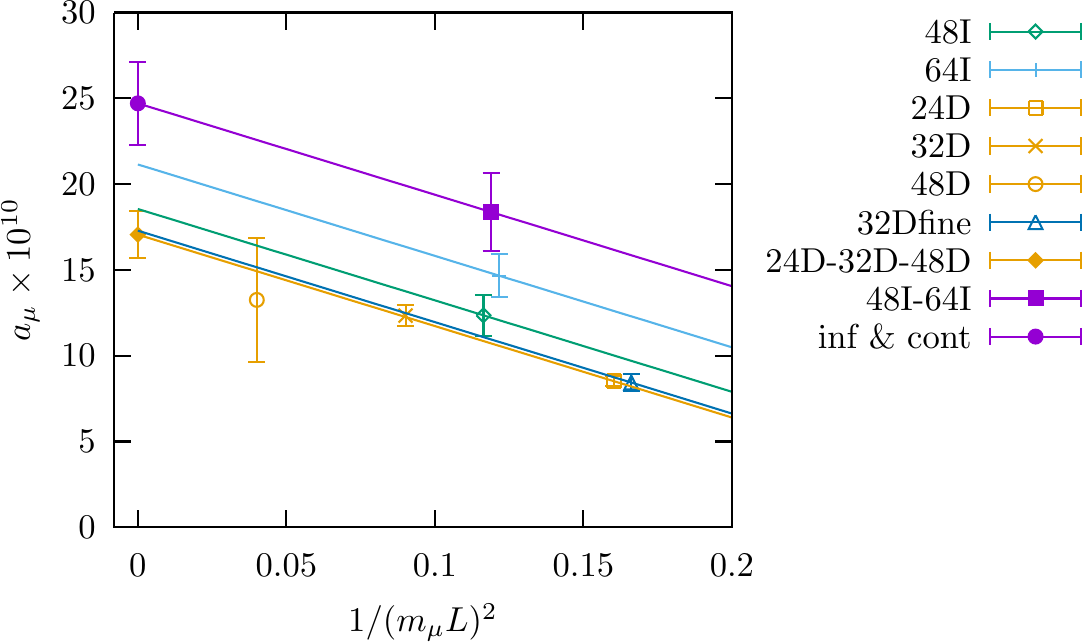}
\includegraphics[width=0.3\textwidth]{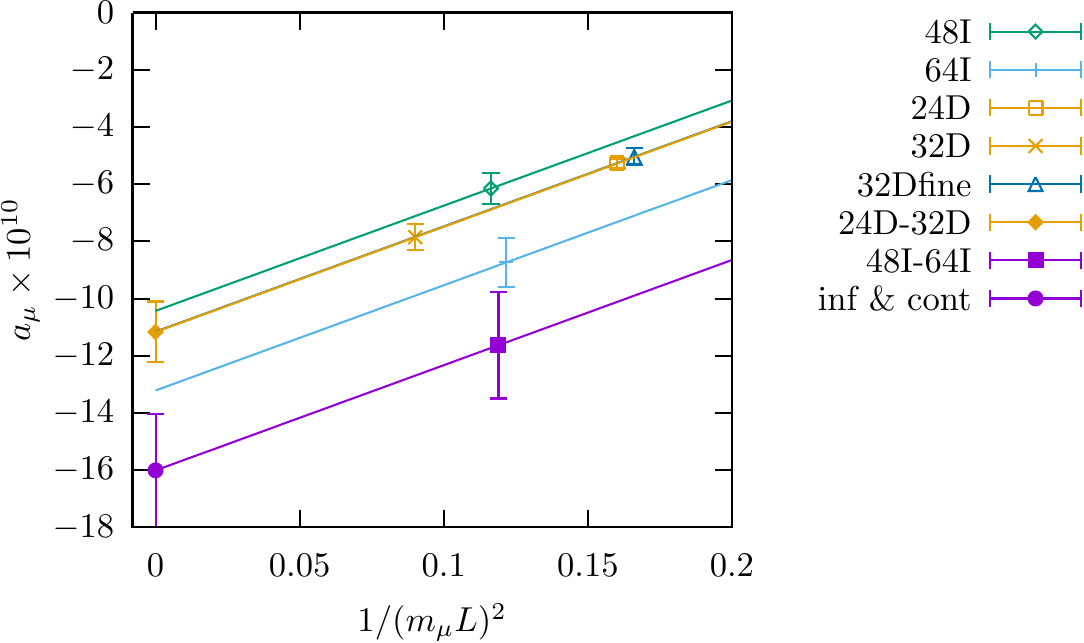}
\includegraphics[width=0.3\textwidth]{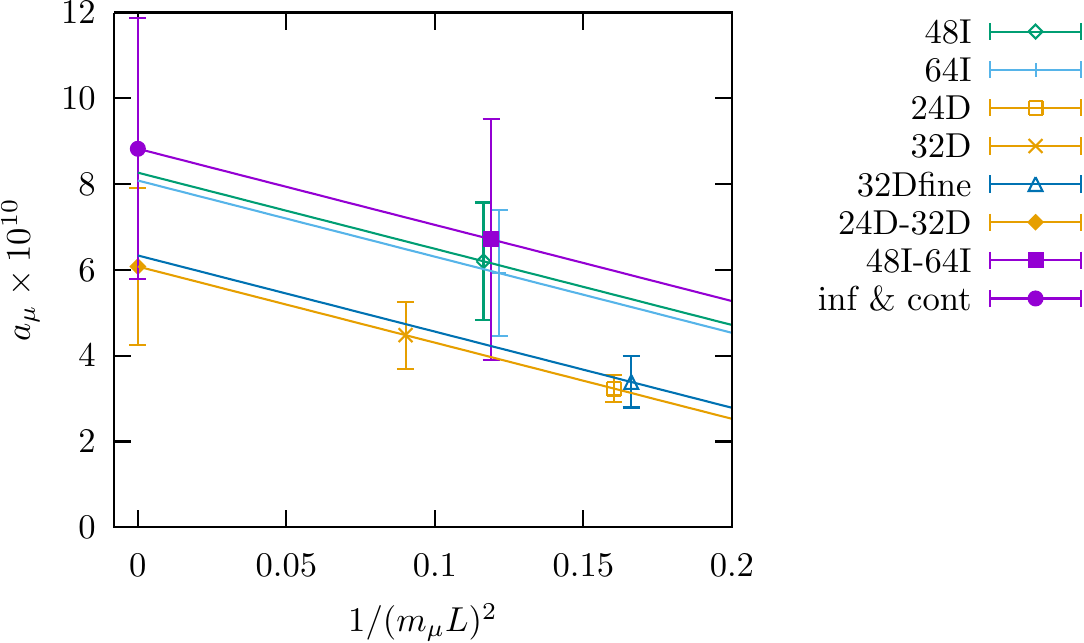}
\caption{Fit plots for \textit{fit-plus-form-2L-2a-4a-4ad} with the hybrid continuum limit. Connected (left), disconnected (middle), and total (right).}
\label{fig:fit-plus-form-2L-2a-4a-4ad_hybrid}
\end{figure*}

\nclearpage

\begin{table*}
\begin{tabular}{@{}cccccc@{}}
\toprule
 & $a_\mu$ & $b_2$ & $c_1^\text{I}$ & $c_1^\text{D}-c_1^\text{I}$ & $c_2^\text{D}$ \tabularnewline
\midrule
con               & 24.39(2.36) & 2.18(0.33) & 0.65(0.19) & 0.12(0.16) & 0.47(0.15) \tabularnewline
discon            & -16.70(2.21) & 2.20(0.40) & 1.02(0.25) & -0.13(0.12) & 0.56(0.17) \tabularnewline
sum               & 7.69(3.14) \tabularnewline
sum-sys ($a^2$)   & 0.21(0.47) \tabularnewline
tot               & 7.85(3.16) & 2.26(1.35) & -0.14(1.09) & 0.62(0.70) & 0.25(0.63) \tabularnewline
tot-sys ($a^2$)   & 0.21(0.47) \tabularnewline
\bottomrule
\end{tabular}
\caption{Fit results for \textit{fit-plus-form-2L-2a-2ad-4ad-lna} with the hybrid continuum limit.}
\label{tab:fit-plus-form-2L-2a-2ad-4ad-lna_hybrid}
\end{table*}

\begin{figure*}
\centering
\includegraphics[width=0.3\textwidth]{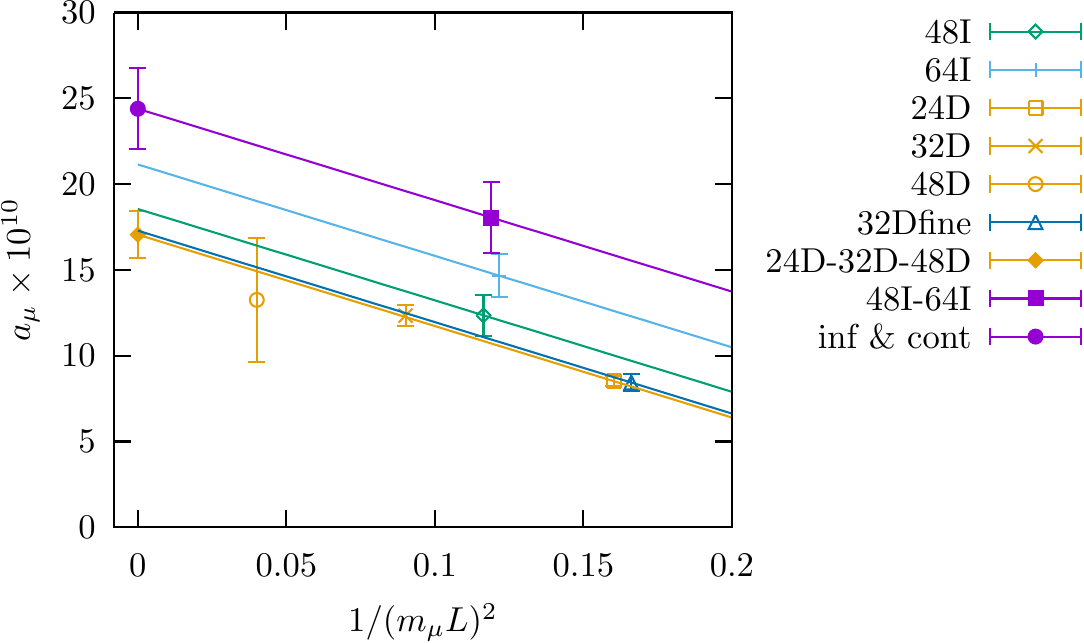}
\includegraphics[width=0.3\textwidth]{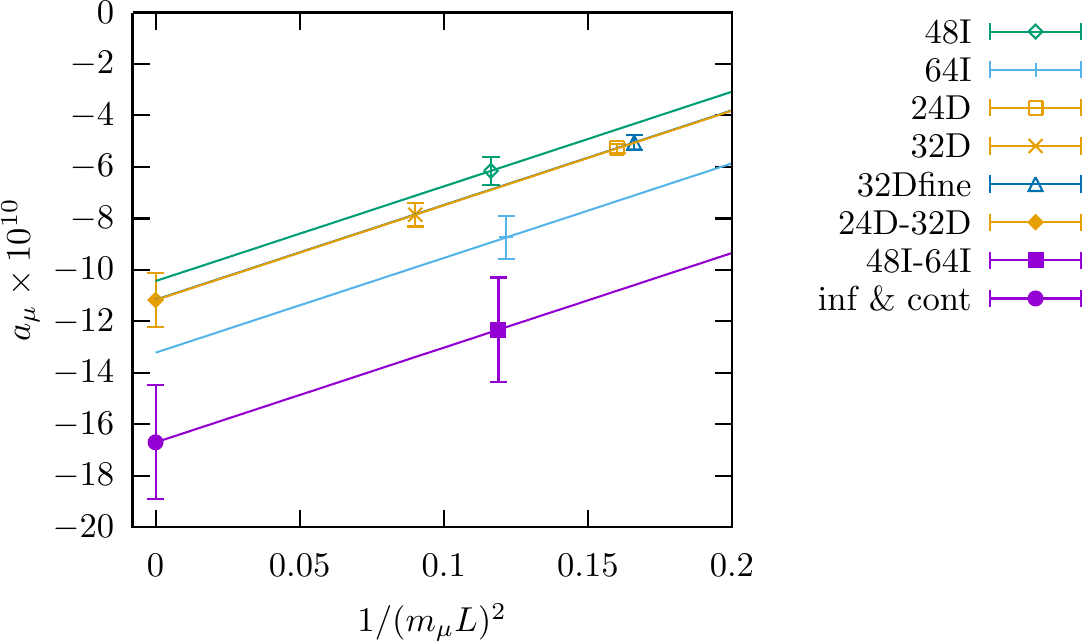}
\includegraphics[width=0.3\textwidth]{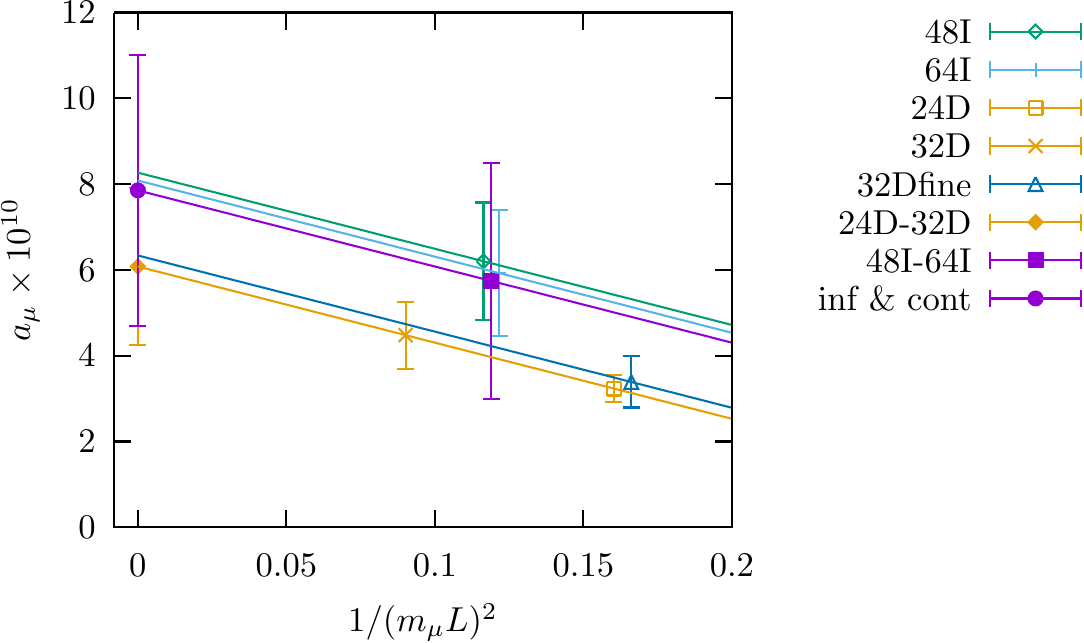}
\caption{Fit plots for \textit{fit-plus-form-2L-2a-2ad-4ad-lna} with the hybrid continuum limit. Connected (left), disconnected (middle), and total (right).}
\label{fig:fit-plus-form-2L-2a-2ad-4ad-lna_hybrid}
\end{figure*}

\begin{table*}
\begin{tabular}{@{}cccccc@{}}
\toprule
 & $a_\mu$ & $b_2$ & $c_1^\text{I}$ & $c_1^\text{D}-c_1^\text{I}$ & $c_2^\text{D}$ \tabularnewline
\midrule
con               & 26.18(2.76) & 2.64(0.24) & 1.00(0.28) & 0.16(0.26) & 0.75(0.23) \tabularnewline
discon            & -17.84(2.59) & 2.71(0.28) & 1.55(0.36) & -0.22(0.20) & 0.88(0.26) \tabularnewline
sum               & 8.34(3.65) \tabularnewline
sum-sys ($a^2$)   & 0.26(0.57) \tabularnewline
tot               & 8.38(3.65) & 2.59(0.98) & -0.19(1.65) & 0.94(1.11) & 0.42(1.00) \tabularnewline
tot-sys ($a^2$)   & 0.26(0.57) \tabularnewline
\bottomrule
\end{tabular}
\caption{Fit results for \textit{fit-plus-form-2L-2a-2ad-4ad-cross} with the hybrid continuum limit.}
\label{tab:fit-plus-form-2L-2a-2ad-4ad-cross_hybrid}
\end{table*}

\begin{figure*}
\centering
\includegraphics[width=0.3\textwidth]{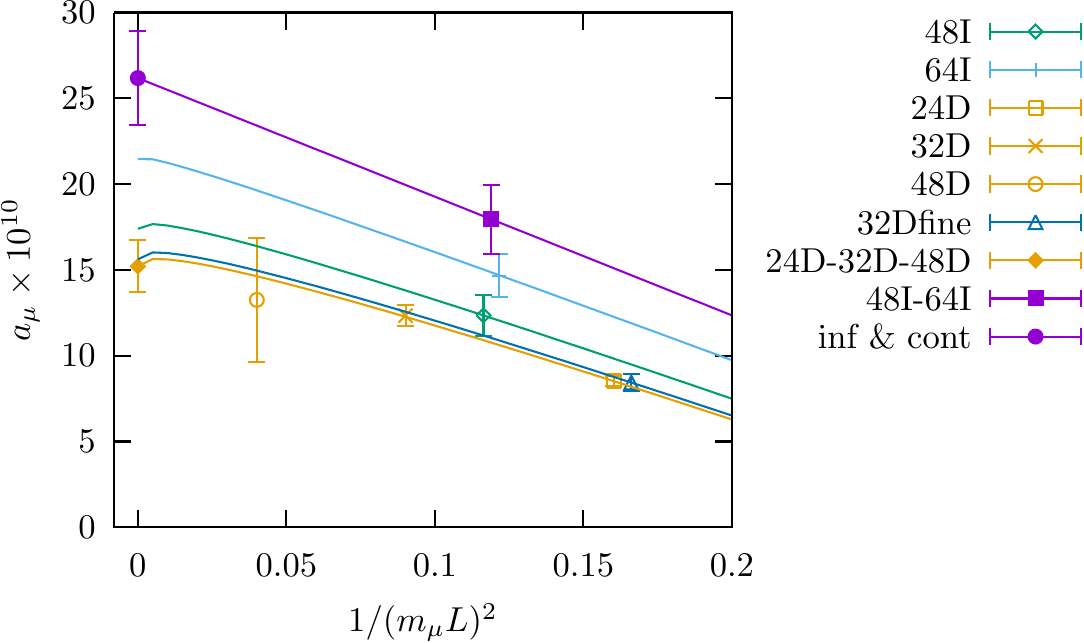}
\includegraphics[width=0.3\textwidth]{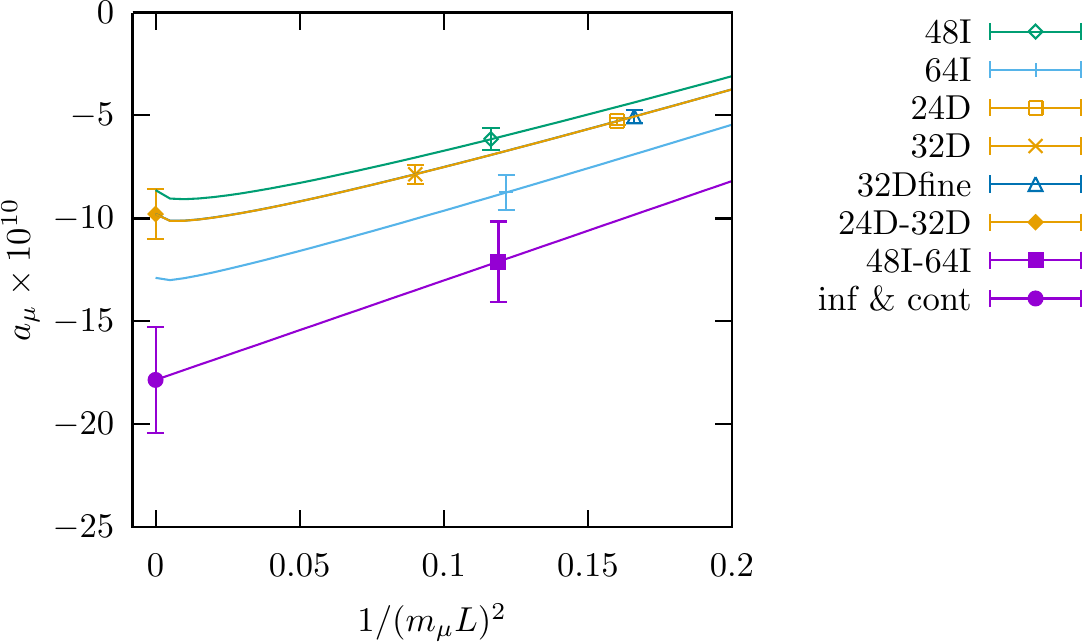}
\includegraphics[width=0.3\textwidth]{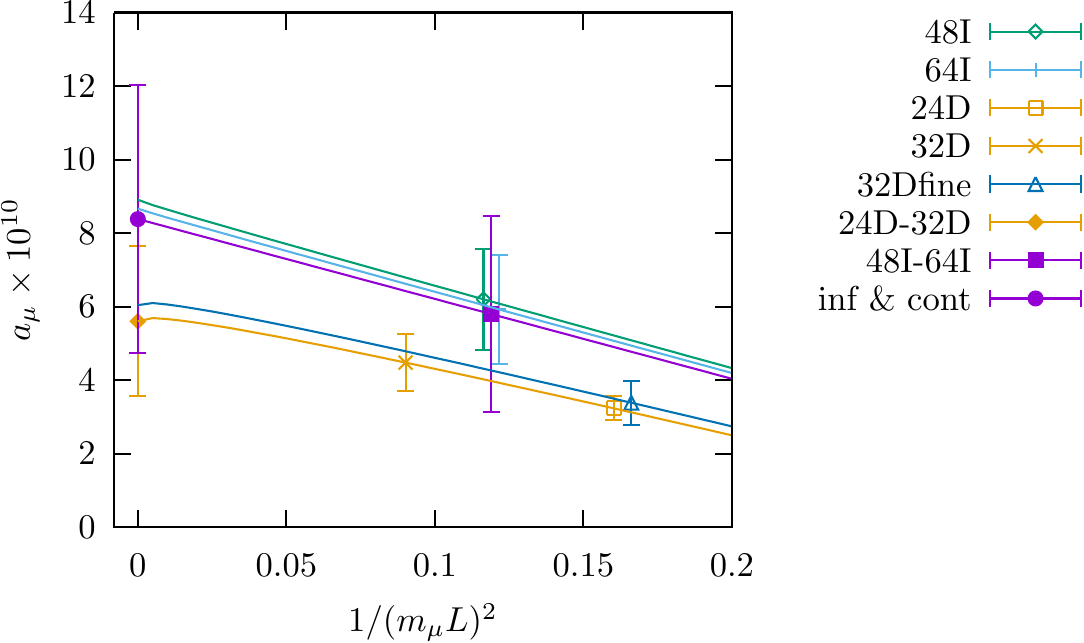}
\caption{Fit plots for \textit{fit-plus-form-2L-2a-2ad-4ad-cross} with the hybrid continuum limit. Connected (left), disconnected (middle), and total (right).}
\label{fig:fit-plus-form-2L-2a-2ad-4ad-cross_hybrid}
\end{figure*}

\nclearpage

\begin{table*}
\begin{tabular}{@{}cccccc@{}}
\toprule
 & $a_\mu$ & $b_2$ & $c_1^\text{I}$ & $c_1^\text{D}-c_1^\text{I}$ & $c_2^\text{D}$ \tabularnewline
\midrule
con               & 28.59(3.65) & 3.12(0.29) & 0.96(0.25) & 0.15(0.26) & 0.72(0.23) \tabularnewline
discon            & -19.94(3.49) & 3.29(0.32) & 1.50(0.31) & -0.20(0.22) & 0.87(0.25) \tabularnewline
sum               & 8.65(4.88) \tabularnewline
sum-sys ($a^2$)   & 0.33(0.73) \tabularnewline
tot               & 8.95(4.59) & 2.92(1.13) & -0.15(1.56) & 0.86(1.06) & 0.39(0.98) \tabularnewline
tot-sys ($a^2$)   & 0.31(0.70) \tabularnewline
\bottomrule
\end{tabular}
\caption{Fit results for \textit{fit-product-form-2L-2a-2ad-4ad} with the hybrid continuum limit.}
\label{tab:fit-product-form-2L-2a-2ad-4ad_hybrid}
\end{table*}

\begin{figure*}
\centering
\includegraphics[width=0.3\textwidth]{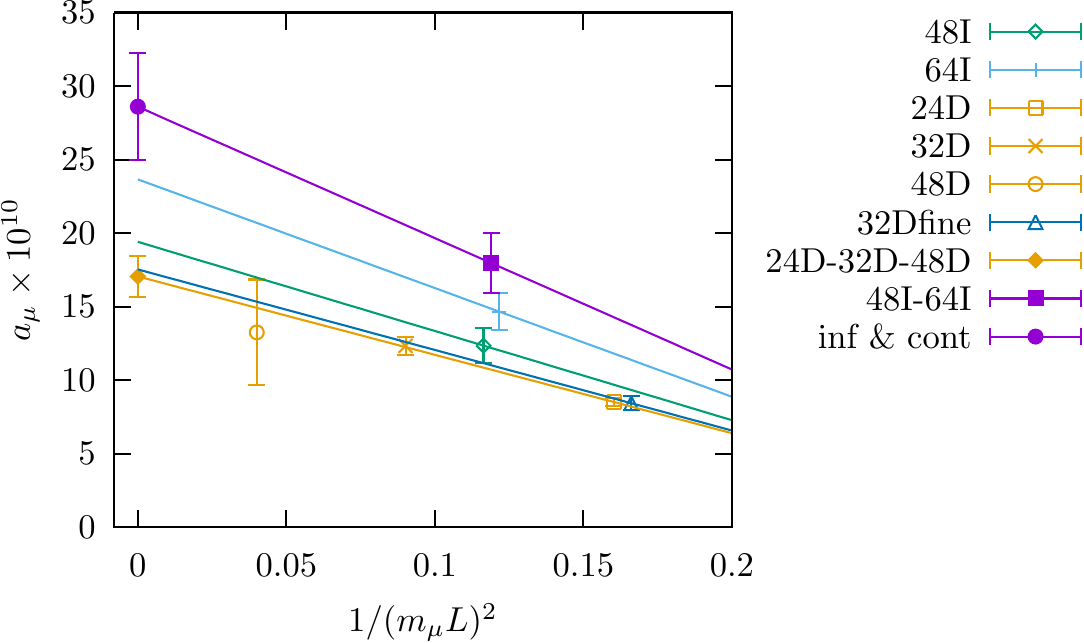}
\includegraphics[width=0.3\textwidth]{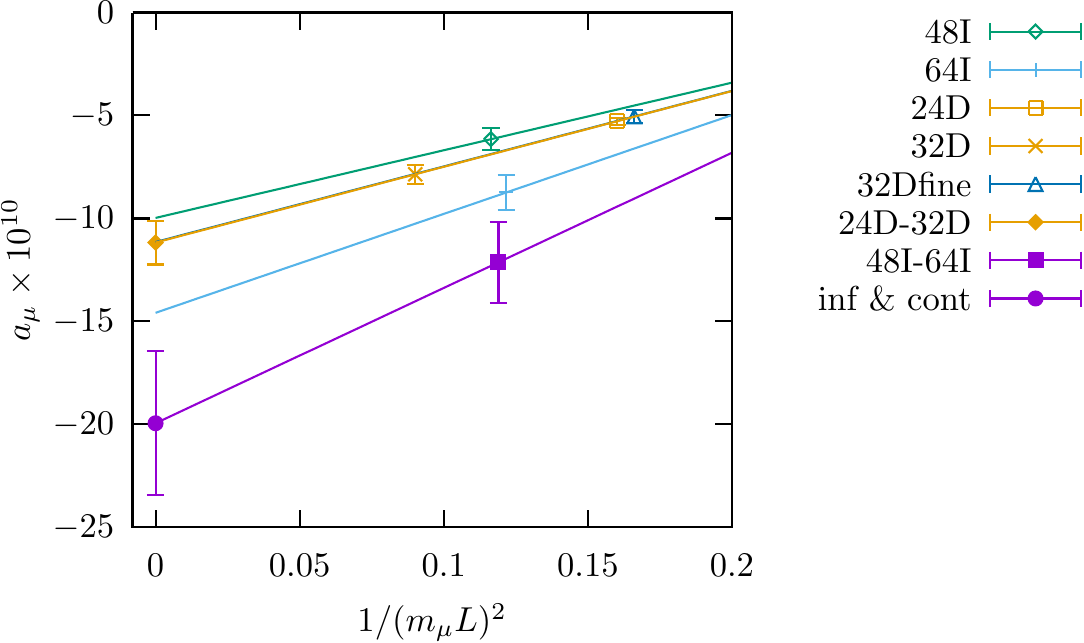}
\includegraphics[width=0.3\textwidth]{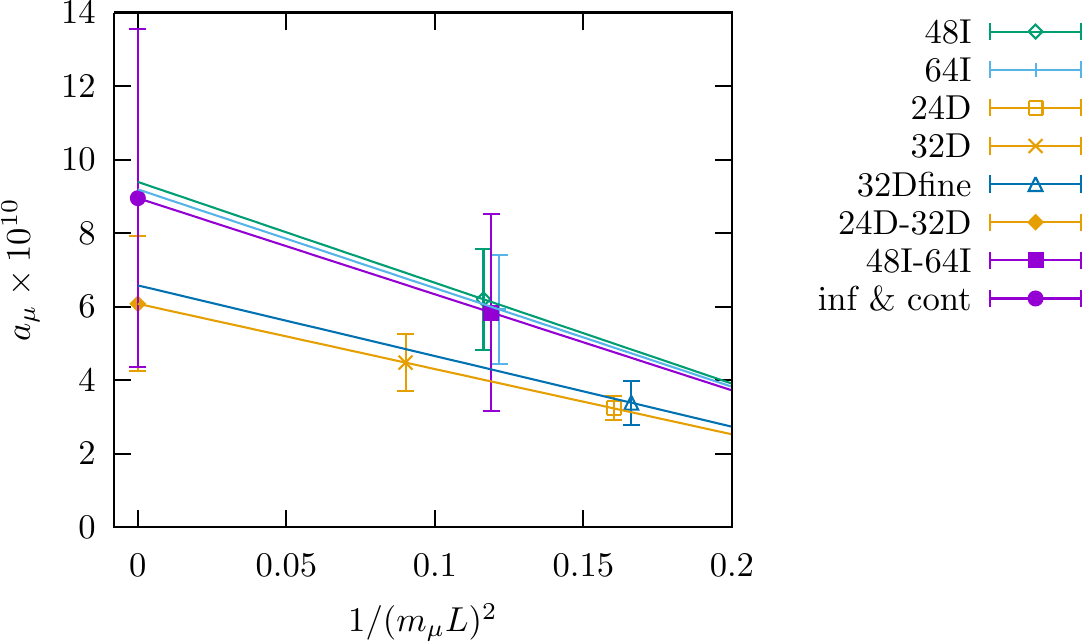}
\caption{Fit plots for \textit{fit-product-form-2L-2a-2ad-4ad} with the hybrid continuum limit. Connected (left), disconnected (middle), and total (right).}
\label{fig:fit-product-form-2L-2a-2ad-4ad_hybrid}
\end{figure*}

\begin{table*}
\begin{tabular}{@{}cccccc@{}}
\toprule
 & $a_\mu$ & $b_2$ & $c_1^\text{I}$ & $c_1^\text{D}-c_1^\text{I}$ & $c_2^\text{D}$ \tabularnewline
\midrule
con               & 32.78(4.42) & 5.79(0.49) & 0.96(0.25) & 0.20(0.26) & 0.76(0.22) \tabularnewline
discon            & -23.19(4.20) & 6.11(0.52) & 1.50(0.31) & -0.16(0.21) & 0.90(0.24) \tabularnewline
sum               & 9.59(5.91) \tabularnewline
sum-sys ($a^2$)   & 0.37(0.83) \tabularnewline
tot               & 10.23(5.51) & 5.50(1.90) & -0.14(1.55) & 0.90(1.04) & 0.43(0.94) \tabularnewline
tot-sys ($a^2$)   & 0.36(0.80) \tabularnewline
\bottomrule
\end{tabular}
\caption{Fit results for \textit{fit-product-form-3L-2a-2ad-4ad} with the hybrid continuum limit.}
\label{tab:fit-product-form-3L-2a-2ad-4ad_hybrid}
\end{table*}

\begin{figure*}
\centering
\includegraphics[width=0.3\textwidth]{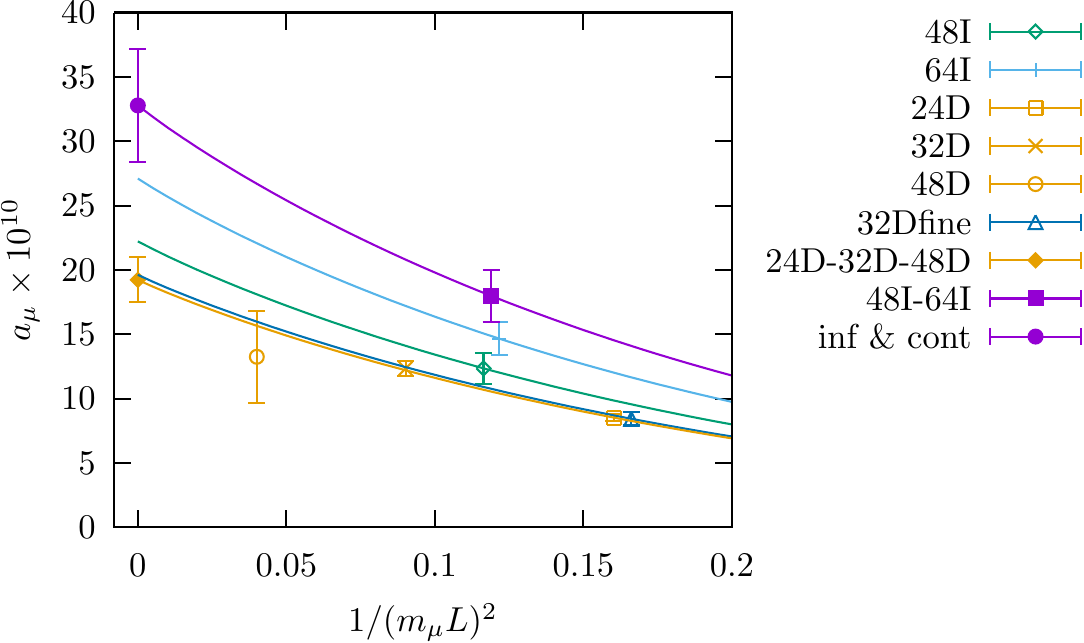}
\includegraphics[width=0.3\textwidth]{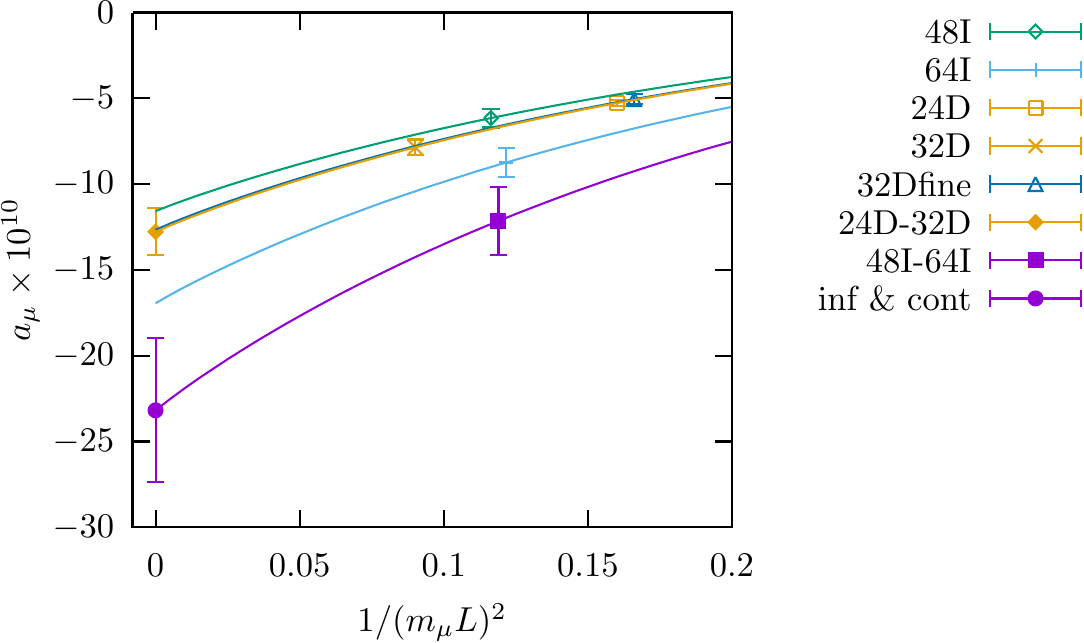}
\includegraphics[width=0.3\textwidth]{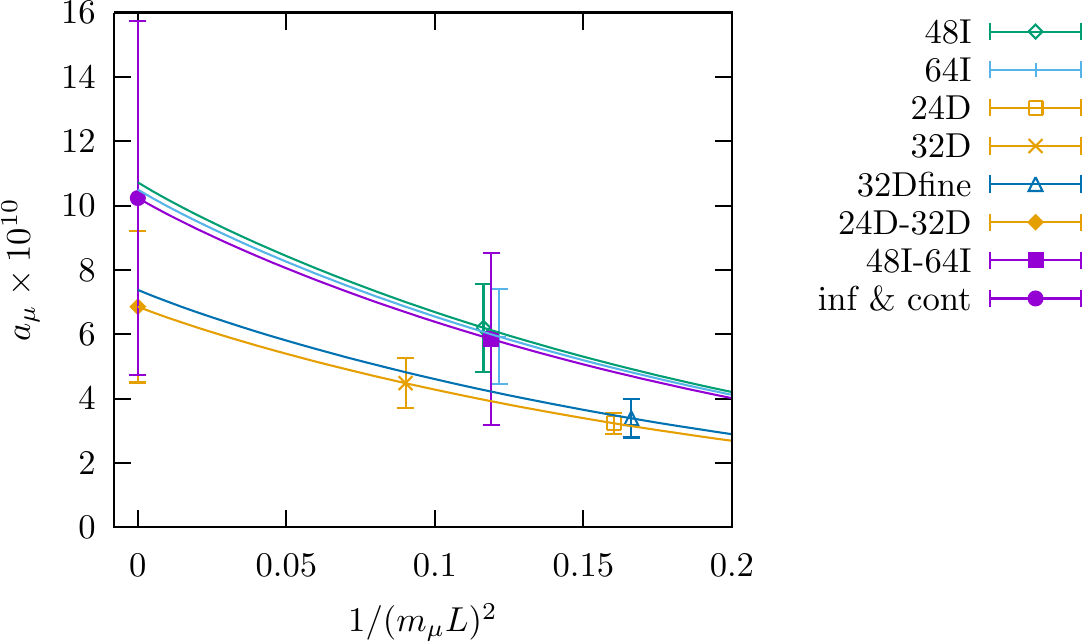}
\caption{Fit plots for \textit{fit-product-form-3L-2a-2ad-4ad} with the hybrid continuum limit. Connected (left), disconnected (middle), and total (right).}
\label{fig:fit-product-form-3L-2a-2ad-4ad_hybrid}
\end{figure*}

\nclearpage

\begin{table*}
\begin{tabular}{@{}cccccc@{}}
\toprule
 & $a_\mu$ & $b_2$ & $c_1^\text{I}$ & $c_2$ & $c_1^\text{D}-c_1^\text{I}$ \tabularnewline
\midrule
con               & 29.99(4.01) & 3.12(0.29) & 1.32(0.30) & 0.78(0.23) & -0.12(0.21) \tabularnewline
discon            & -21.12(3.88) & 3.29(0.32) & 1.89(0.37) & 0.93(0.25) & -0.50(0.18) \tabularnewline
sum               & 8.87(5.38) \tabularnewline
sum-sys ($a^2$)   & 0.37(0.82) \tabularnewline
tot               & 9.19(5.04) & 2.92(1.13) & 0.08(1.92) & 0.43(1.05) & 0.69(1.04) \tabularnewline
tot-sys ($a^2$)   & 0.35(0.79) \tabularnewline
\bottomrule
\end{tabular}
\caption{Fit results for \textit{fit-product-form-2L-2a-4a-2ad} with the hybrid continuum limit.}
\label{tab:fit-product-form-2L-2a-4a-2ad_hybrid}
\end{table*}

\begin{figure*}
\centering
\includegraphics[width=0.3\textwidth]{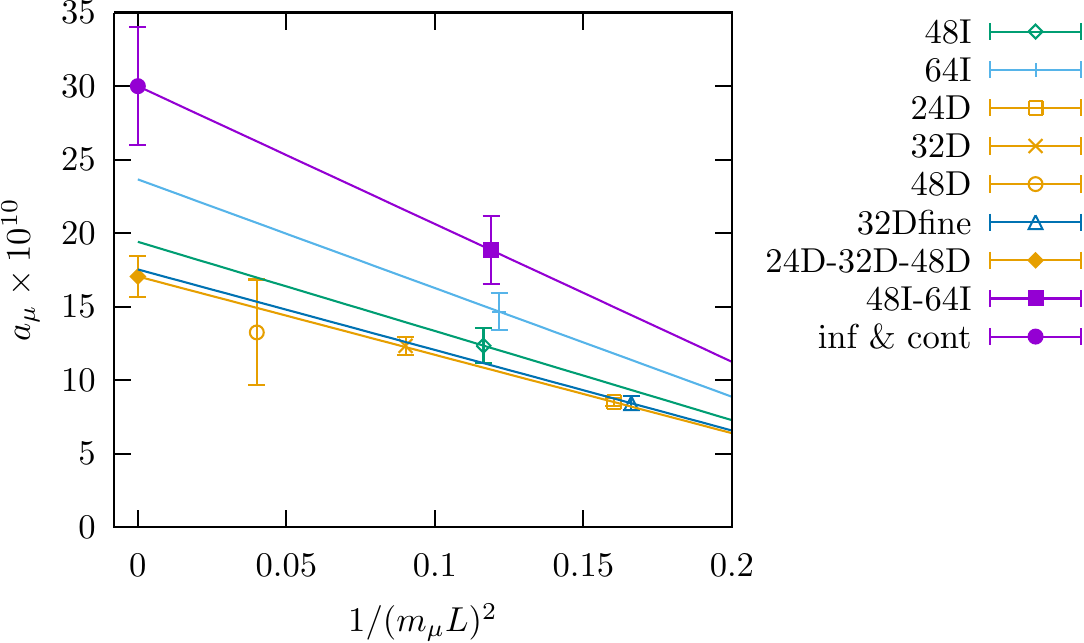}
\includegraphics[width=0.3\textwidth]{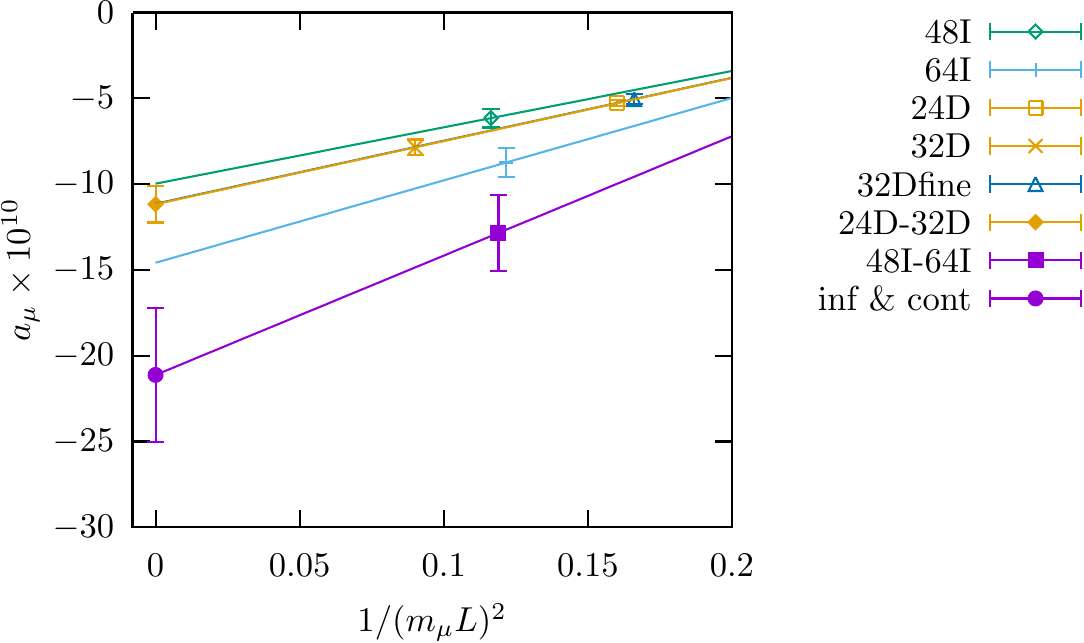}
\includegraphics[width=0.3\textwidth]{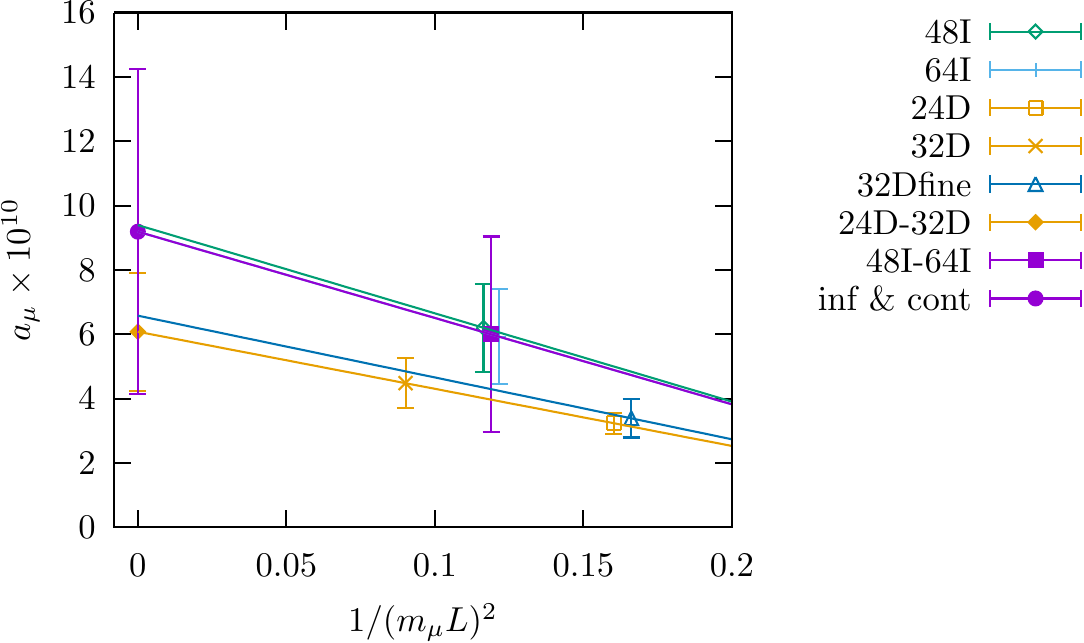}
\caption{Fit plots for \textit{fit-product-form-2L-2a-4a-2ad} with the hybrid continuum limit. Connected (left), disconnected (middle), and total (right).}
\label{fig:fit-product-form-2L-2a-4a-2ad_hybrid}
\end{figure*}

\begin{table*}
\begin{tabular}{@{}cccccc@{}}
\toprule
 & $a_\mu$ & $b_2$ & $c_1$ & $c_2^\text{I}$ & $c_2^\text{D}-c_2^\text{I}$ \tabularnewline
\midrule
con               & 29.36(3.89) & 3.12(0.29) & 1.16(0.32) & 0.44(0.74) & -0.31(0.57) \tabularnewline
discon            & -19.23(3.24) & 3.29(0.32) & 1.23(0.36) & -0.62(0.69) & -1.45(0.57) \tabularnewline
sum               & 10.13(4.88) \tabularnewline
sum-sys ($a^2$)   & 0.32(0.72) \tabularnewline
tot               & 10.31(4.68) & 2.92(1.13) & 1.00(1.12) & 2.20(2.36) & 1.60(2.14) \tabularnewline
tot-sys ($a^2$)   & 0.31(0.69) \tabularnewline
\bottomrule
\end{tabular}
\caption{Fit results for \textit{fit-product-form-2L-2a-4a-4ad} with the hybrid continuum limit.}
\label{tab:fit-product-form-2L-2a-4a-4ad_hybrid}
\end{table*}

\begin{figure*}
\centering
\includegraphics[width=0.3\textwidth]{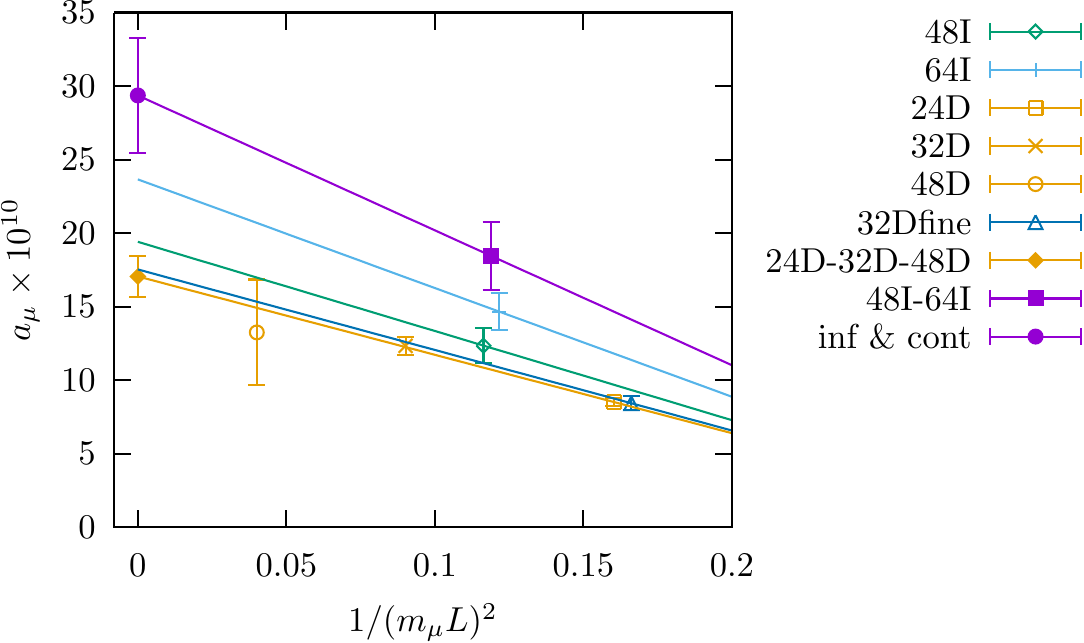}
\includegraphics[width=0.3\textwidth]{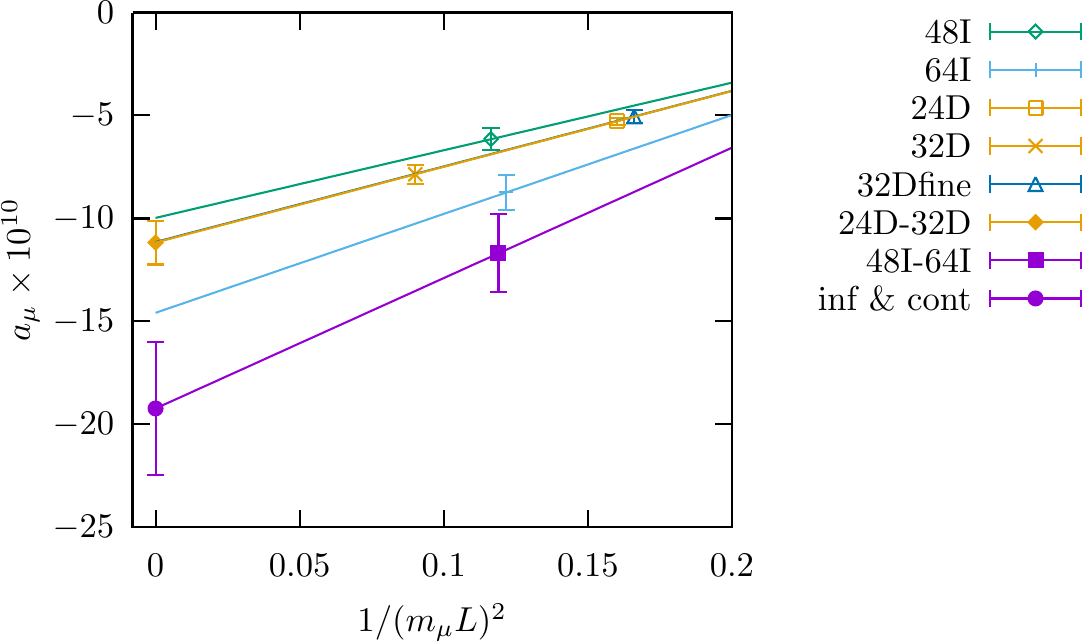}
\includegraphics[width=0.3\textwidth]{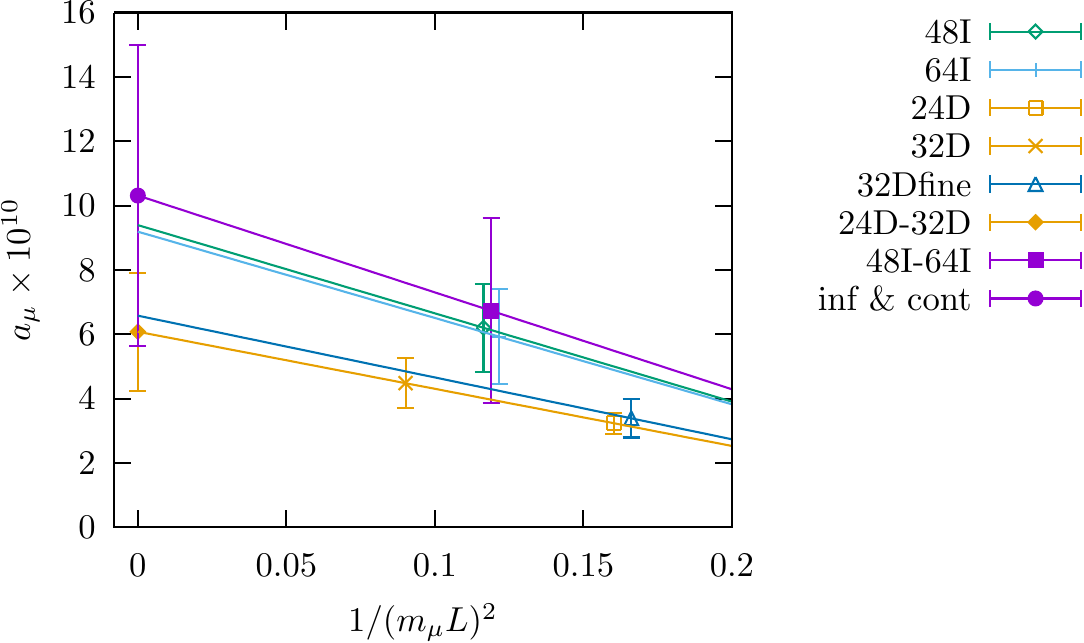}
\caption{Fit plots for \textit{fit-product-form-2L-2a-4a-4ad} with the hybrid continuum limit. Connected (left), disconnected (middle), and total (right).}
\label{fig:fit-product-form-2L-2a-4a-4ad_hybrid}
\end{figure*}

\nclearpage

\begin{table*}
\begin{tabular}{@{}cccccc@{}}
\toprule
 & $a_\mu$ & $b_2$ & $c_1^\text{I}$ & $c_1^\text{D}-c_1^\text{I}$ & $c_2^\text{D}$ \tabularnewline
\midrule
con               & 28.97(3.76) & 3.12(0.29) & 0.89(0.23) & 0.14(0.24) & 0.63(0.21) \tabularnewline
discon            & -20.35(3.64) & 3.29(0.32) & 1.38(0.28) & -0.17(0.20) & 0.77(0.22) \tabularnewline
sum               & 8.62(5.05) \tabularnewline
sum-sys ($a^2$)   & 0.34(0.76) \tabularnewline
tot               & 8.93(4.74) & 2.92(1.13) & -0.14(1.47) & 0.77(0.98) & 0.32(0.90) \tabularnewline
tot-sys ($a^2$)   & 0.33(0.73) \tabularnewline
\bottomrule
\end{tabular}
\caption{Fit results for \textit{fit-product-form-2L-2a-2ad-4ad-lna} with the hybrid continuum limit.}
\label{tab:fit-product-form-2L-2a-2ad-4ad-lna_hybrid}
\end{table*}

\begin{figure*}
\centering
\includegraphics[width=0.3\textwidth]{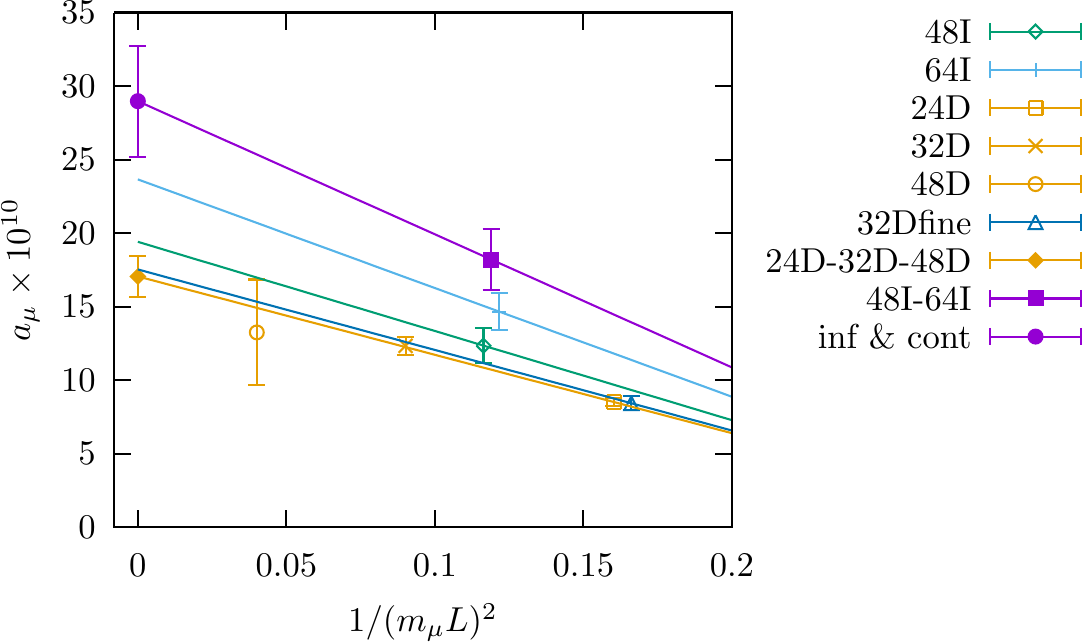}
\includegraphics[width=0.3\textwidth]{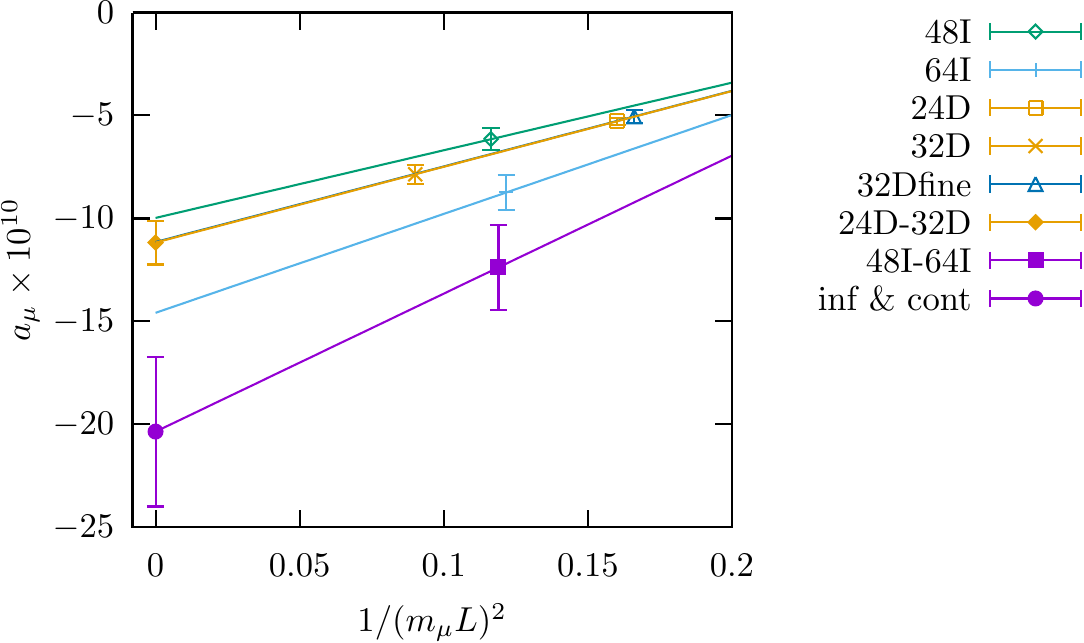}
\includegraphics[width=0.3\textwidth]{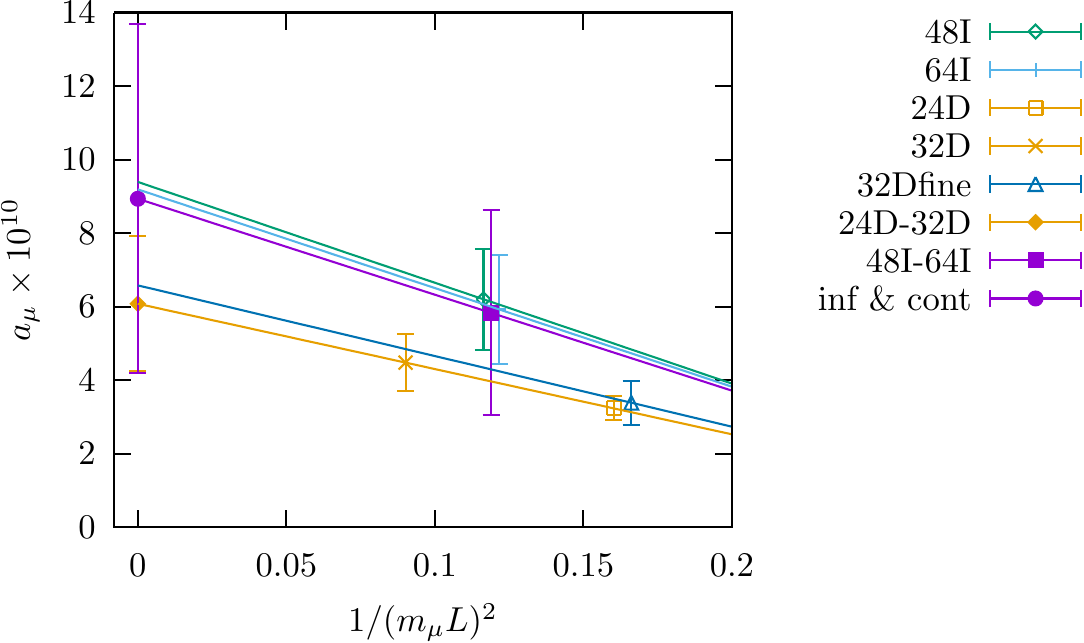}
\caption{Fit plots for \textit{fit-product-form-2L-2a-2ad-4ad-lna} with the hybrid continuum limit. Connected (left), disconnected (middle), and total (right).}
\label{fig:fit-product-form-2L-2a-2ad-4ad-lna_hybrid}
\end{figure*}

\begin{table*}
\begin{tabular}{@{}cccccc@{}}
\toprule
 & $a_\mu$ & $b_2$ & $c_1^\text{I}$ & $c_1^\text{D}-c_1^\text{I}$ & $c_2^\text{D}$ \tabularnewline
\midrule
con               & 31.86(4.75) & 3.61(0.25) & 1.51(0.39) & 0.12(0.44) & 1.03(0.39) \tabularnewline
discon            & -22.06(4.47) & 3.78(0.27) & 2.28(0.48) & -0.38(0.38) & 1.26(0.41) \tabularnewline
sum               & 9.80(6.27) \tabularnewline
sum-sys ($a^2$)   & 0.43(0.97) \tabularnewline
tot               & 9.69(5.70) & 3.29(1.01) & -0.17(2.35) & 1.17(1.72) & 0.51(1.57) \tabularnewline
tot-sys ($a^2$)   & 0.41(0.96) \tabularnewline
\bottomrule
\end{tabular}
\caption{Fit results for \textit{fit-product-form-2L-2a-2ad-4ad-cross} with the hybrid continuum limit.}
\label{tab:fit-product-form-2L-2a-2ad-4ad-cross_hybrid}
\end{table*}

\begin{figure*}
\centering
\includegraphics[width=0.3\textwidth]{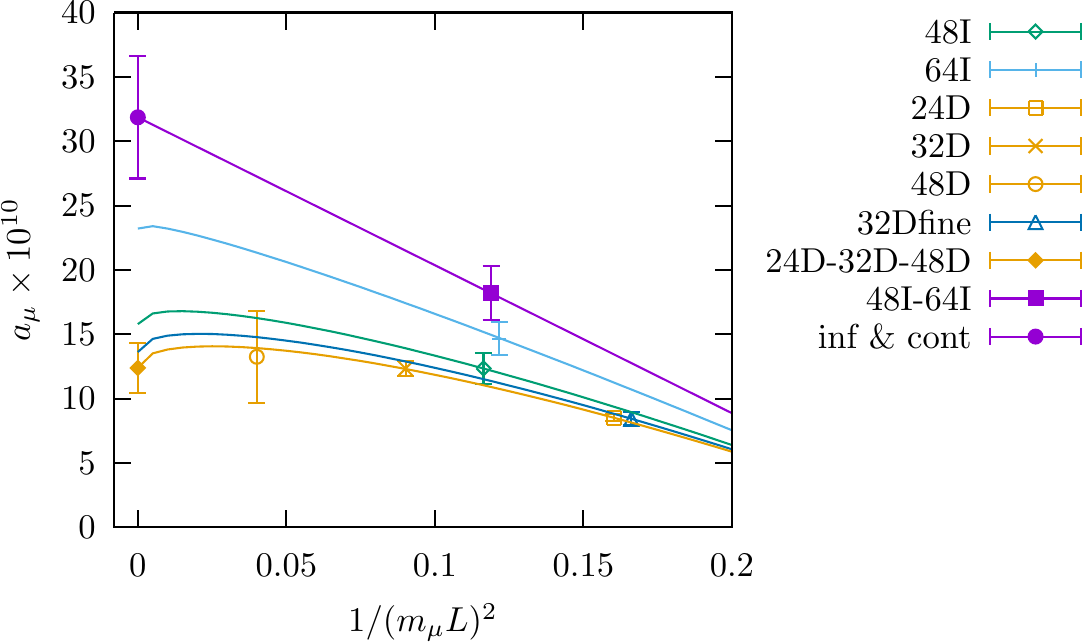}
\includegraphics[width=0.3\textwidth]{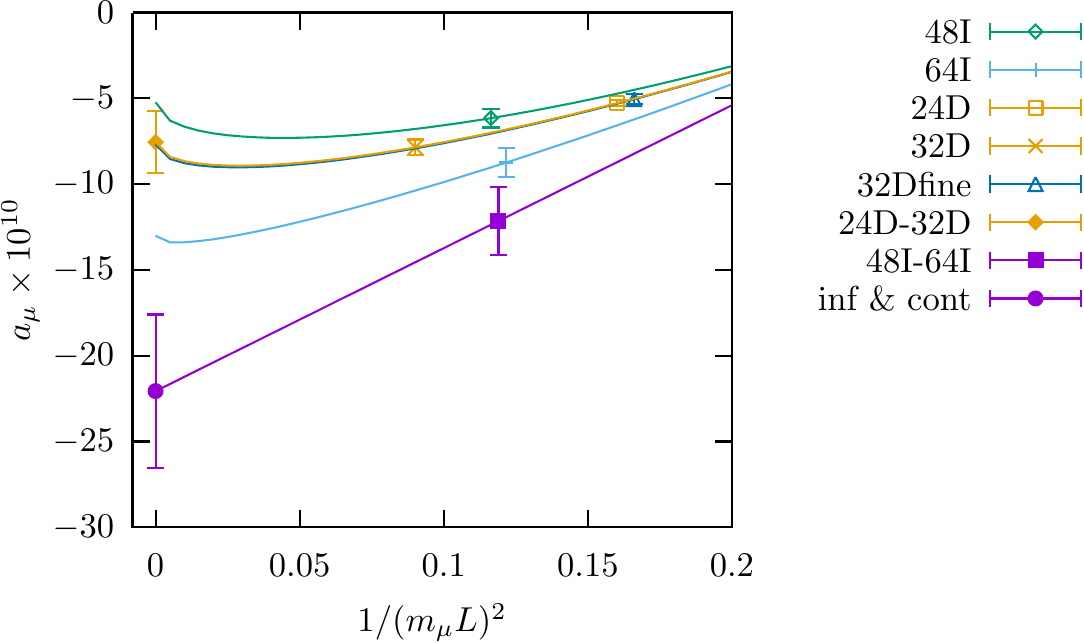}
\includegraphics[width=0.3\textwidth]{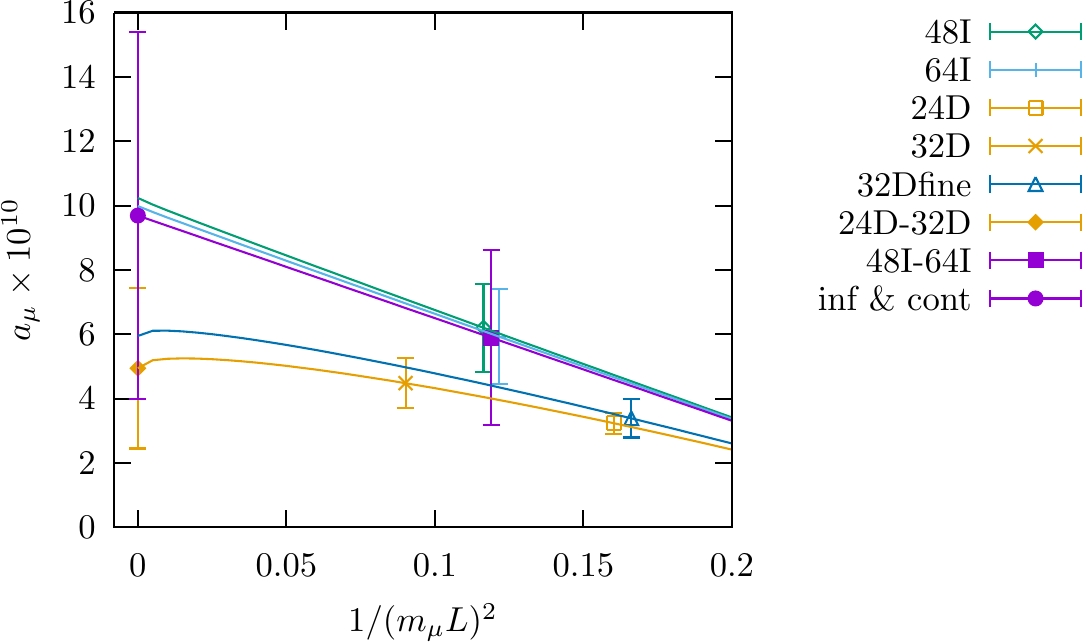}
\caption{Fit plots for \textit{fit-product-form-2L-2a-2ad-4ad-cross} with the hybrid continuum limit. Connected (left), disconnected (middle), and total (right).}
\label{fig:fit-product-form-2L-2a-2ad-4ad-cross_hybrid}
\end{figure*}

\nclearpage

%% file: src-nohybrid.tex
\begin{table*}
\begin{tabular}{@{}cccccc@{}}
\toprule
 & $a_\mu$ & $b_2$ & $c_1^\text{I}$ & $c_1^\text{D}-c_1^\text{I}$ & $c_2^\text{D}$ \tabularnewline
\midrule
con               & 23.76(3.96) & 2.24(0.45) & 0.66(0.45) & 0.13(0.20) & 0.51(0.26) \tabularnewline
discon            & -16.45(2.13) & 2.24(0.40) & 1.10(0.28) & -0.15(0.14) & 0.63(0.19) \tabularnewline
sum               & 7.31(4.23) \tabularnewline
sum-sys ($a^2$)   & 0.00(0.00) \tabularnewline
tot               & 7.47(4.24) & 2.38(1.67) & -0.32(2.01) & 0.72(0.97) & 0.22(0.99) \tabularnewline
tot-sys ($a^2$)   & -0.00(0.00) \tabularnewline
\bottomrule
\end{tabular}
\caption{Fit results for \textit{fit-plus-form-2L-2a-2ad-4ad} without the hybrid continuum limit.}
\label{tab:fit-plus-form-2L-2a-2ad-4ad_nohybrid}
\end{table*}

\begin{figure*}
\centering
\includegraphics[width=0.3\textwidth]{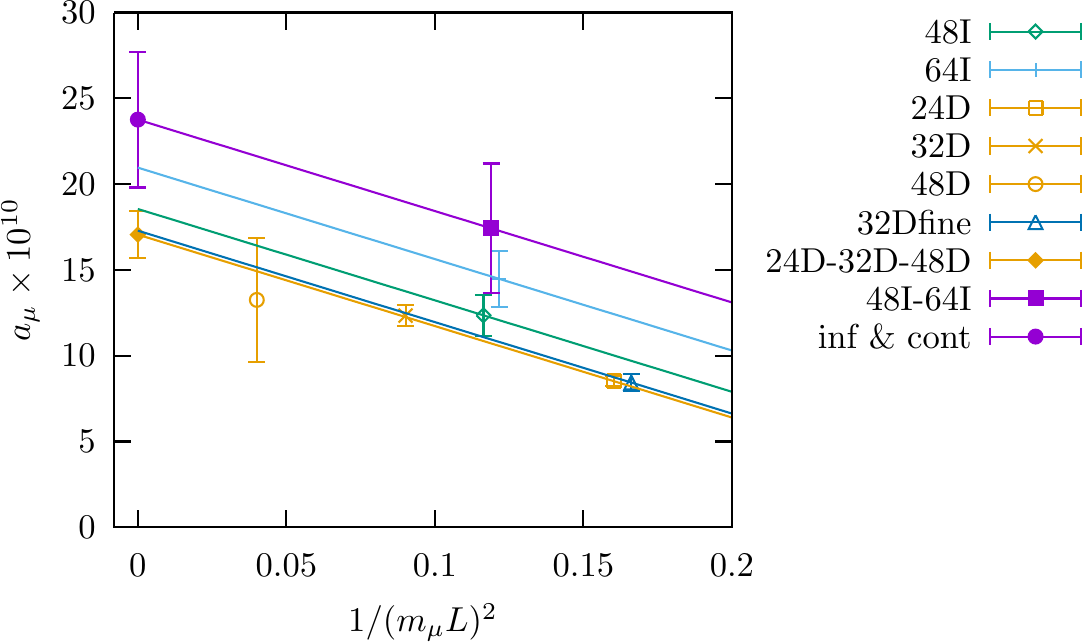}
\includegraphics[width=0.3\textwidth]{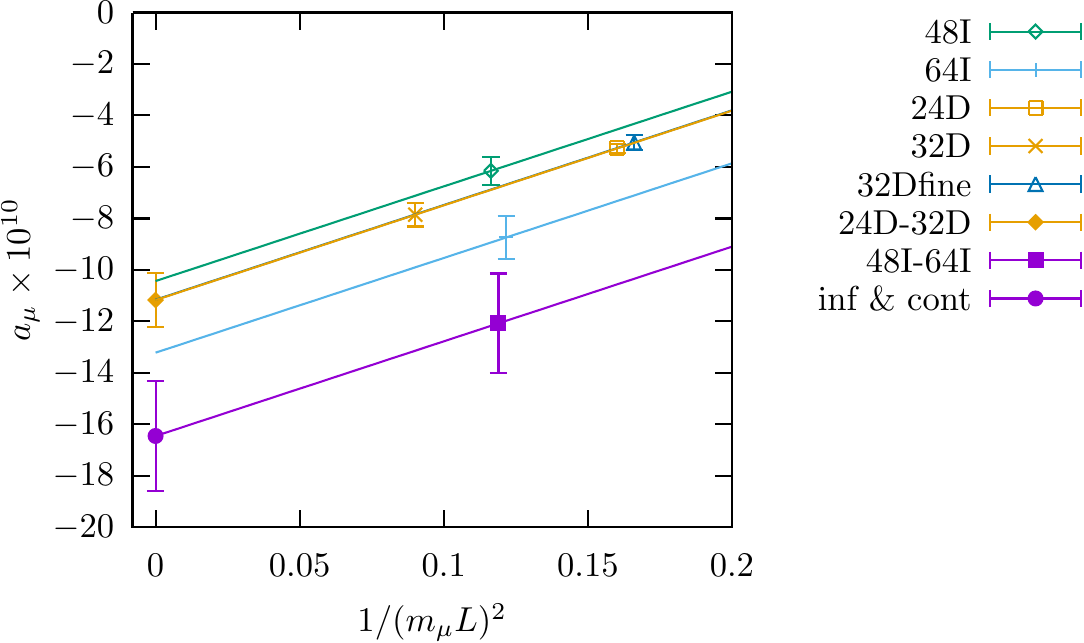}
\includegraphics[width=0.3\textwidth]{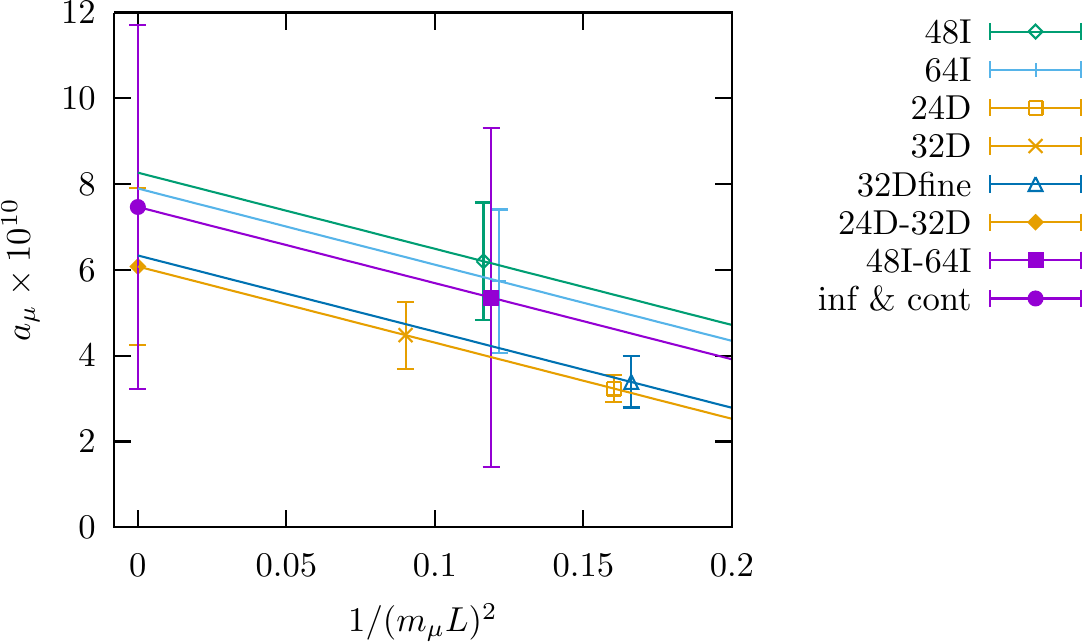}
\caption{Fit plots for \textit{fit-plus-form-2L-2a-2ad-4ad} without the hybrid continuum limit. Connected (left), disconnected (middle), and total (right).}
\label{fig:fit-plus-form-2L-2a-2ad-4ad_nohybrid}
\end{figure*}

\begin{table*}
\begin{tabular}{@{}cccccc@{}}
\toprule
 & $a_\mu$ & $b_2$ & $c_1^\text{I}$ & $c_1^\text{D}-c_1^\text{I}$ & $c_2^\text{D}$ \tabularnewline
\midrule
con               & 26.10(4.10) & 4.27(0.78) & 0.60(0.41) & 0.15(0.18) & 0.49(0.24) \tabularnewline
discon            & -18.17(2.29) & 4.30(0.70) & 0.99(0.26) & -0.11(0.12) & 0.59(0.18) \tabularnewline
sum               & 7.93(4.44) \tabularnewline
sum-sys ($a^2$)   & 0.00(0.00) \tabularnewline
tot               & 8.30(4.48) & 4.54(2.88) & -0.29(1.80) & 0.67(0.87) & 0.22(0.89) \tabularnewline
tot-sys ($a^2$)   & 0.00(0.00) \tabularnewline
\bottomrule
\end{tabular}
\caption{Fit results for \textit{fit-plus-form-3L-2a-2ad-4ad} without the hybrid continuum limit.}
\label{tab:fit-plus-form-3L-2a-2ad-4ad_nohybrid}
\end{table*}

\begin{figure*}
\centering
\includegraphics[width=0.3\textwidth]{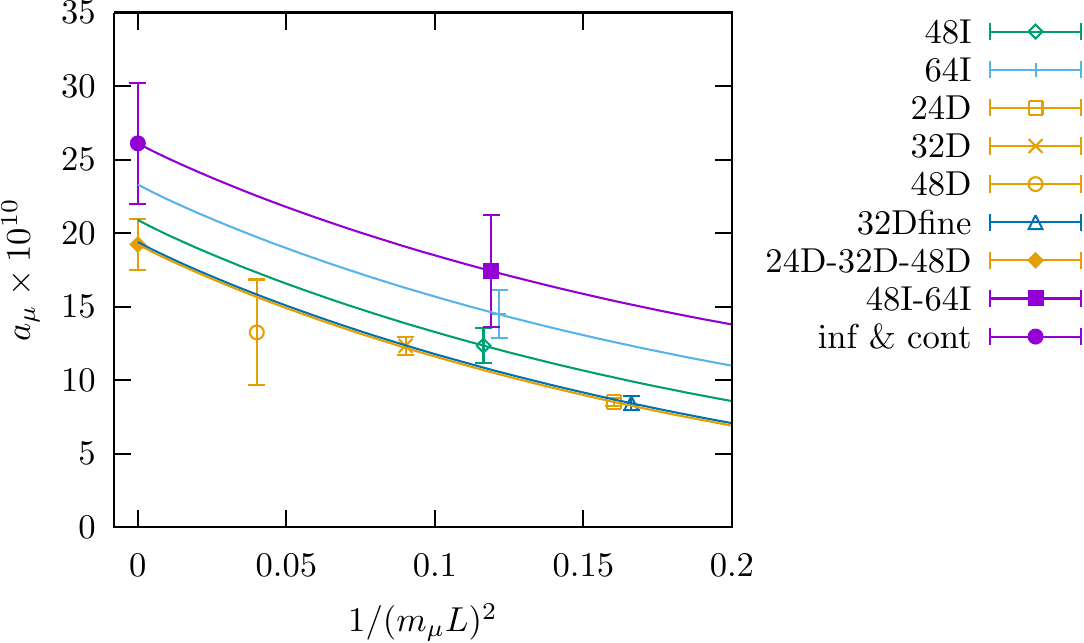}
\includegraphics[width=0.3\textwidth]{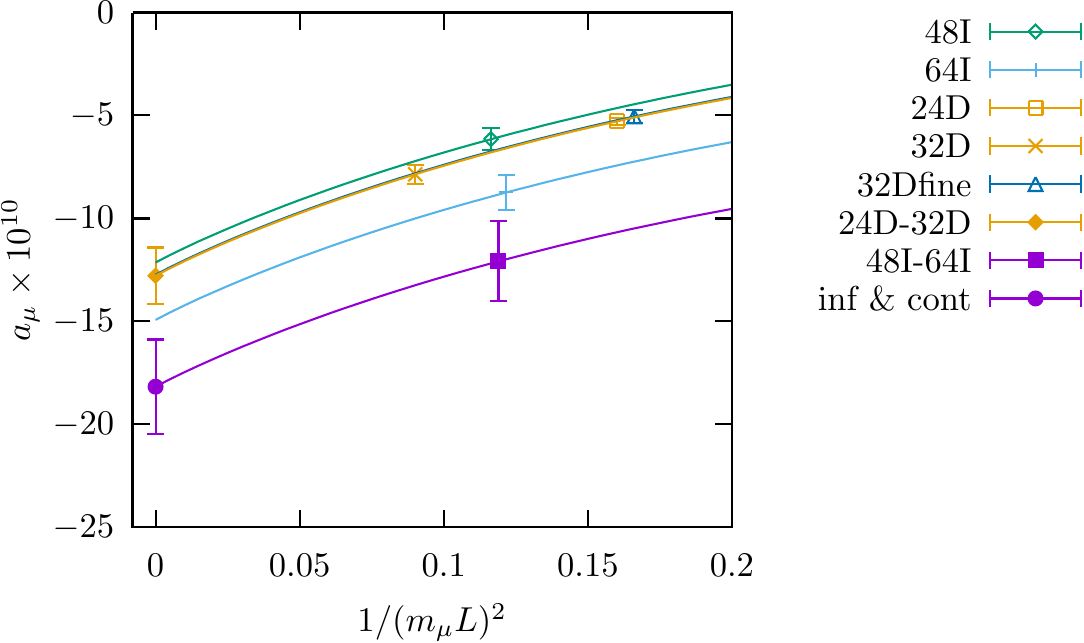}
\includegraphics[width=0.3\textwidth]{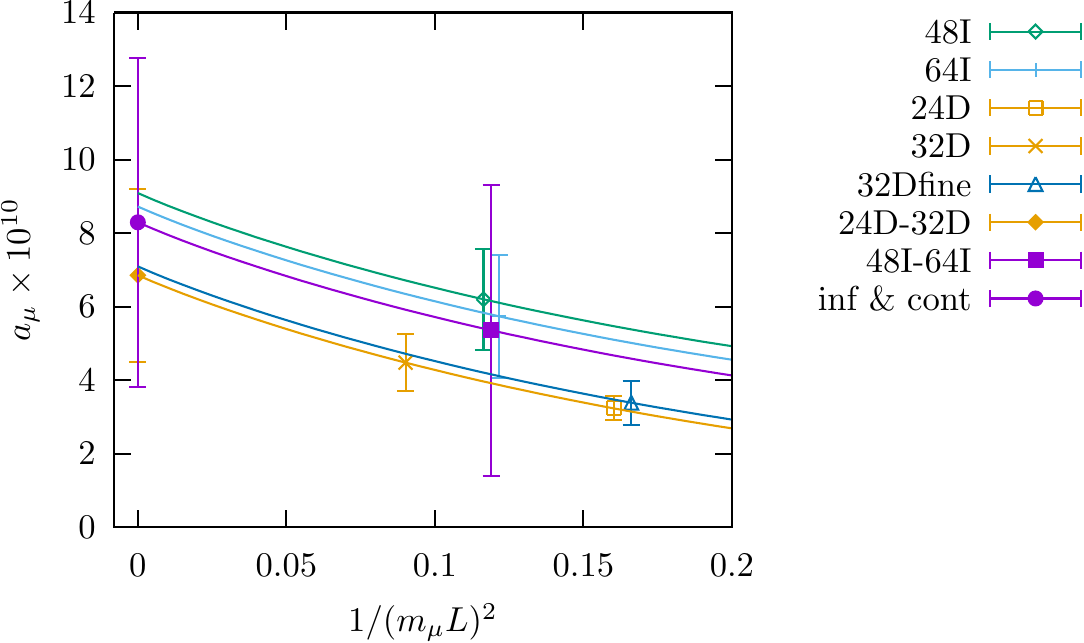}
\caption{Fit plots for \textit{fit-plus-form-3L-2a-2ad-4ad} without the hybrid continuum limit. Connected (left), disconnected (middle), and total (right).}
\label{fig:fit-plus-form-3L-2a-2ad-4ad_nohybrid}
\end{figure*}

\nclearpage

\begin{table*}
\begin{tabular}{@{}cccccc@{}}
\toprule
 & $a_\mu$ & $b_2$ & $c_1^\text{I}$ & $c_2$ & $c_1^\text{D}-c_1^\text{I}$ \tabularnewline
\midrule
con               & 24.58(4.43) & 2.17(0.46) & 0.92(0.54) & 0.56(0.27) & -0.06(0.22) \tabularnewline
discon            & -17.16(2.35) & 2.14(0.41) & 1.40(0.34) & 0.69(0.20) & -0.38(0.13) \tabularnewline
sum               & 7.42(4.70) \tabularnewline
sum-sys ($a^2$)   & 0.00(0.00) \tabularnewline
tot               & 7.58(4.71) & 2.34(1.77) & -0.19(2.50) & 0.24(1.09) & 0.63(1.18) \tabularnewline
tot-sys ($a^2$)   & -0.00(0.00) \tabularnewline
\bottomrule
\end{tabular}
\caption{Fit results for \textit{fit-plus-form-2L-2a-4a-2ad} without the hybrid continuum limit.}
\label{tab:fit-plus-form-2L-2a-4a-2ad_nohybrid}
\end{table*}

\begin{figure*}
\centering
\includegraphics[width=0.3\textwidth]{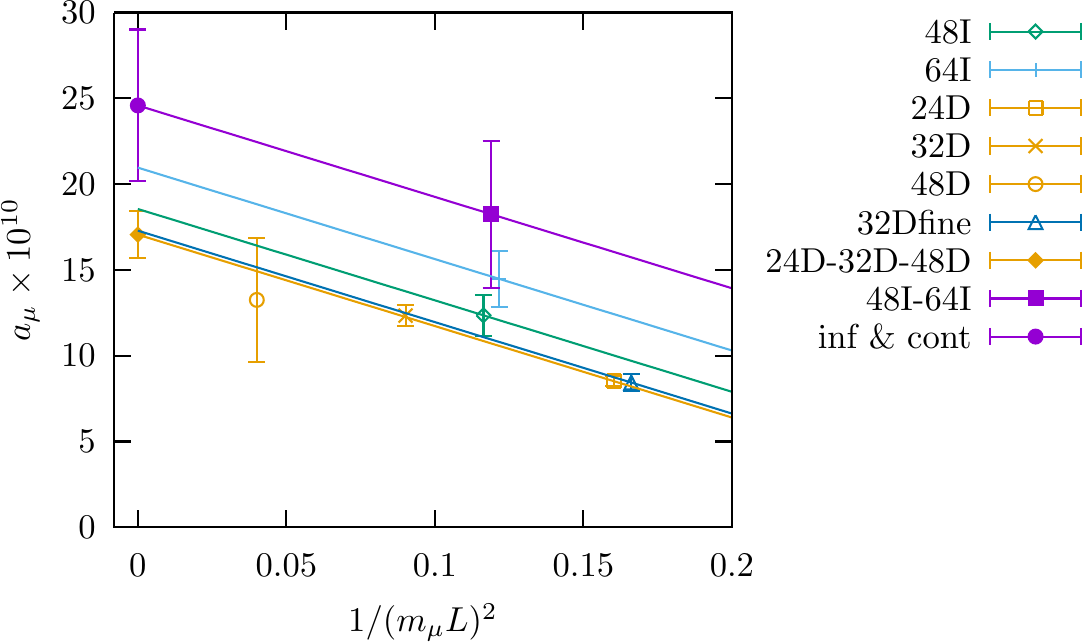}
\includegraphics[width=0.3\textwidth]{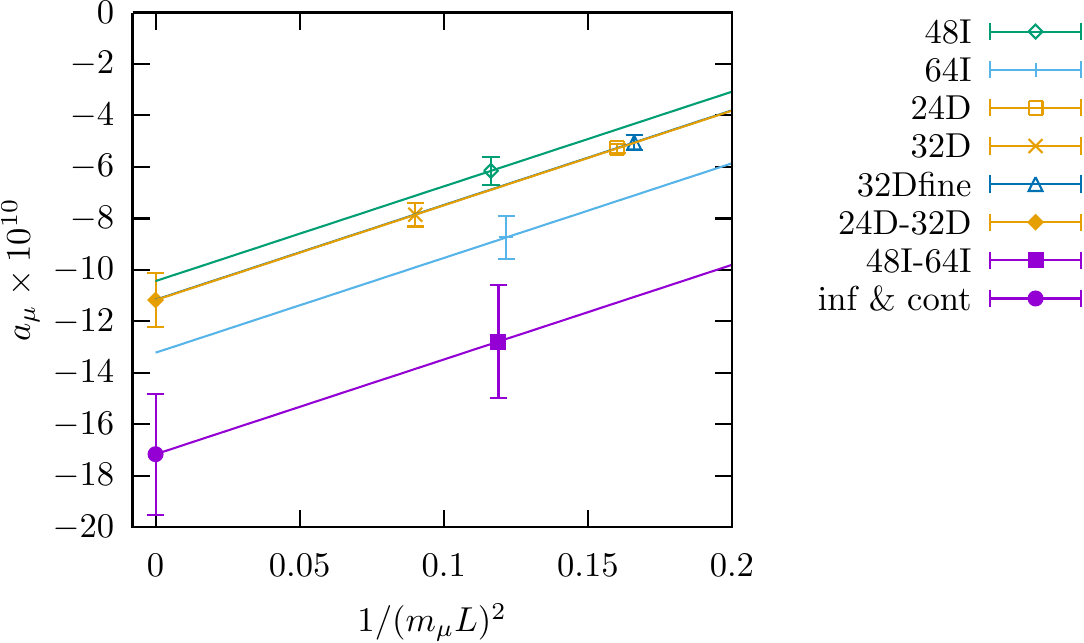}
\includegraphics[width=0.3\textwidth]{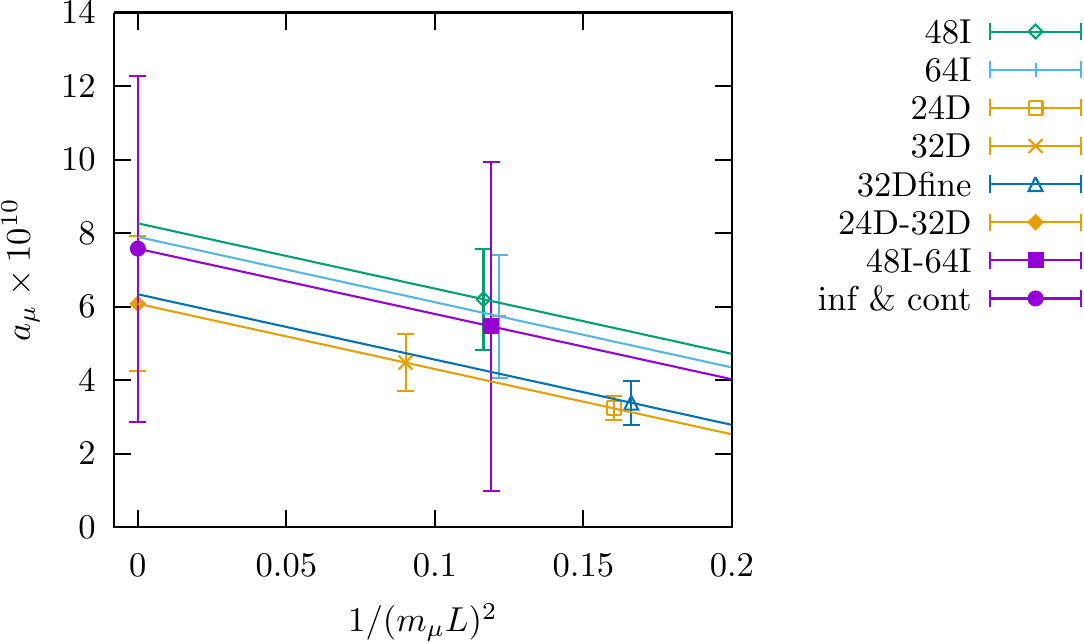}
\caption{Fit plots for \textit{fit-plus-form-2L-2a-4a-2ad} without the hybrid continuum limit. Connected (left), disconnected (middle), and total (right).}
\label{fig:fit-plus-form-2L-2a-4a-2ad_nohybrid}
\end{figure*}

\begin{table*}
\begin{tabular}{@{}cccccc@{}}
\toprule
 & $a_\mu$ & $b_2$ & $c_1$ & $c_2^\text{I}$ & $c_2^\text{D}-c_2^\text{I}$ \tabularnewline
\midrule
con               & 24.31(3.69) & 2.19(0.44) & 0.83(0.34) & 0.38(0.56) & 0.17(0.60) \tabularnewline
discon            & -16.01(1.98) & 2.30(0.43) & 0.89(0.29) & -0.46(0.43) & 1.06(0.39) \tabularnewline
sum               & 8.30(3.87) \tabularnewline
sum-sys ($a^2$)   & 0.00(0.00) \tabularnewline
tot               & 8.43(3.88) & 2.11(1.45) & 0.69(1.12) & 1.89(2.26) & -1.48(2.48) \tabularnewline
tot-sys ($a^2$)   & -0.00(0.00) \tabularnewline
\bottomrule
\end{tabular}
\caption{Fit results for \textit{fit-plus-form-2L-2a-4a-4ad} without the hybrid continuum limit.}
\label{tab:fit-plus-form-2L-2a-4a-4ad_nohybrid}
\end{table*}

\begin{figure*}
\centering
\includegraphics[width=0.3\textwidth]{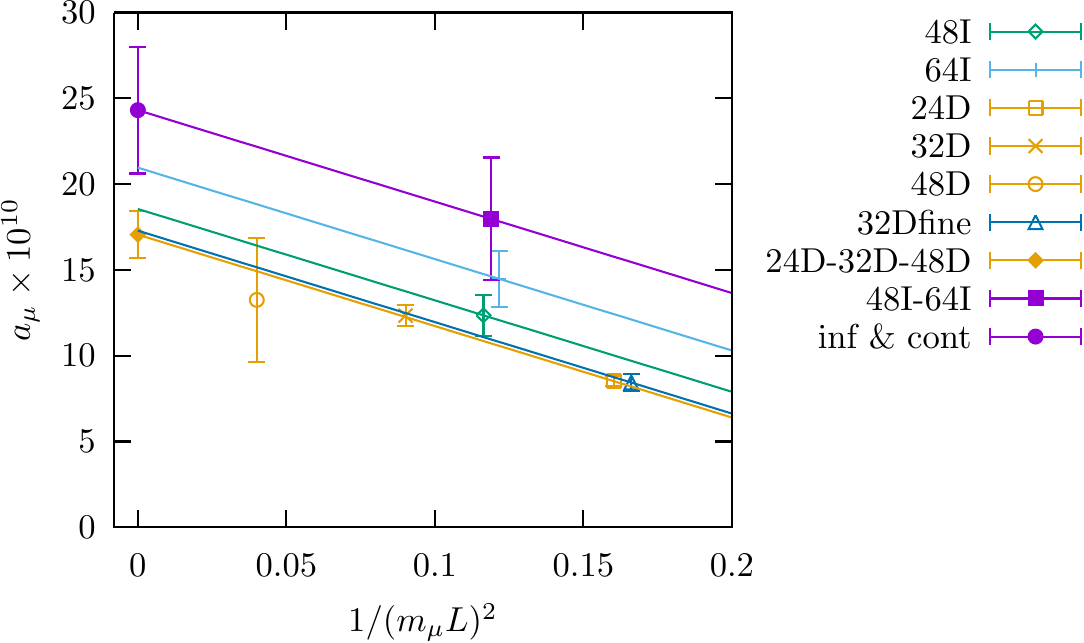}
\includegraphics[width=0.3\textwidth]{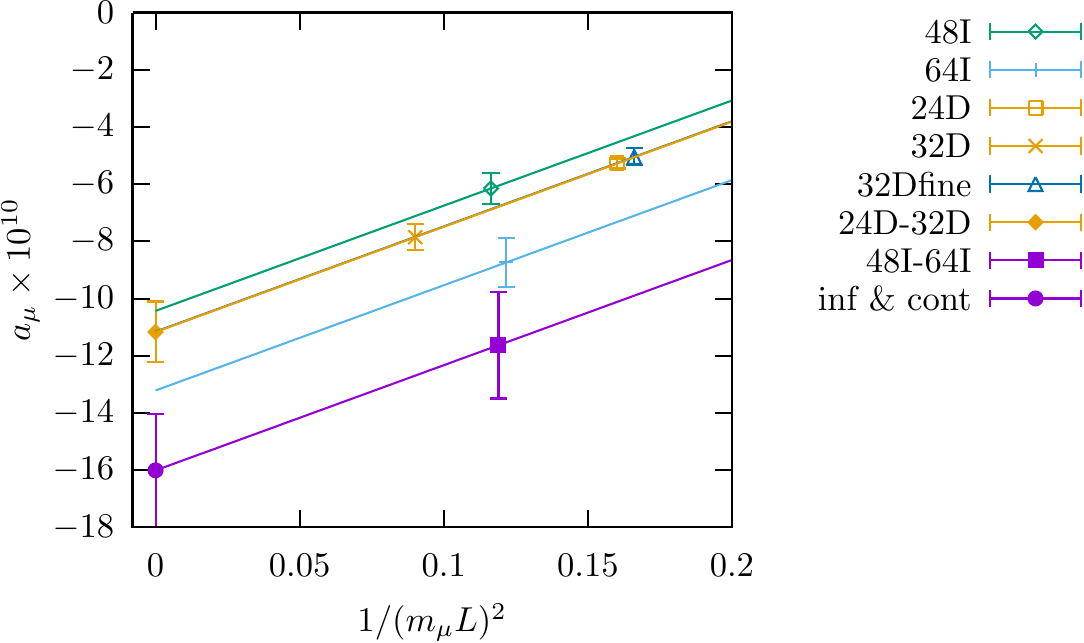}
\includegraphics[width=0.3\textwidth]{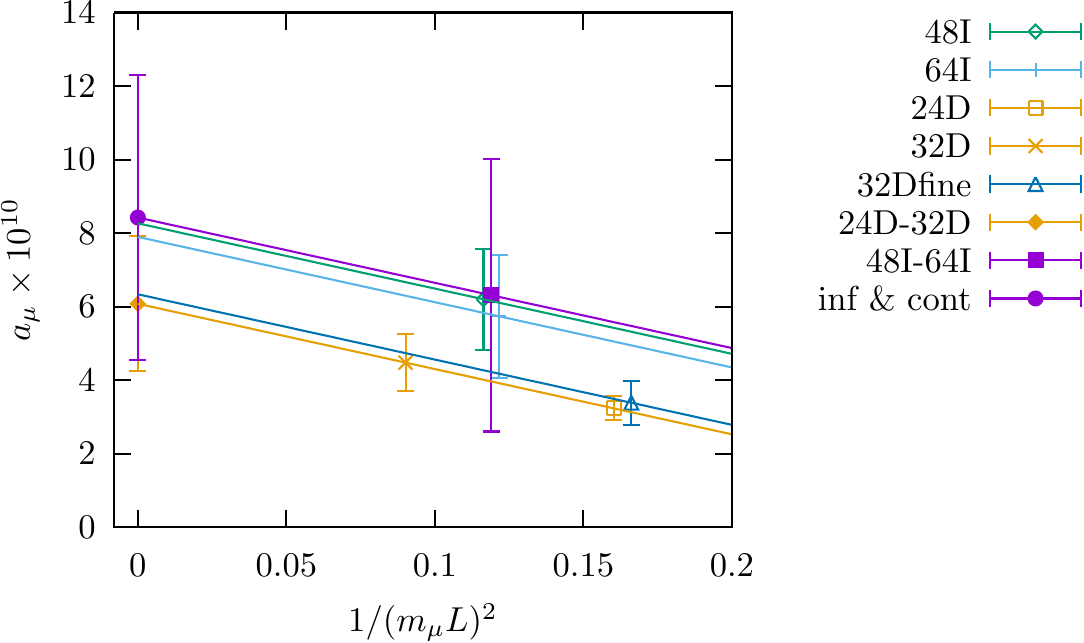}
\caption{Fit plots for \textit{fit-plus-form-2L-2a-4a-4ad} without the hybrid continuum limit. Connected (left), disconnected (middle), and total (right).}
\label{fig:fit-plus-form-2L-2a-4a-4ad_nohybrid}
\end{figure*}

\nclearpage

\begin{table*}
\begin{tabular}{@{}cccccc@{}}
\toprule
 & $a_\mu$ & $b_2$ & $c_1^\text{I}$ & $c_1^\text{D}-c_1^\text{I}$ & $c_2^\text{D}$ \tabularnewline
\midrule
con               & 23.97(4.13) & 2.22(0.45) & 0.61(0.41) & 0.12(0.18) & 0.45(0.23) \tabularnewline
discon            & -16.70(2.21) & 2.20(0.40) & 1.02(0.25) & -0.13(0.12) & 0.56(0.17) \tabularnewline
sum               & 7.27(4.40) \tabularnewline
sum-sys ($a^2$)   & 0.00(0.00) \tabularnewline
tot               & 7.43(4.41) & 2.39(1.72) & -0.30(1.91) & 0.65(0.90) & 0.17(0.91) \tabularnewline
tot-sys ($a^2$)   & 0.00(0.00) \tabularnewline
\bottomrule
\end{tabular}
\caption{Fit results for \textit{fit-plus-form-2L-2a-2ad-4ad-lna} without the hybrid continuum limit.}
\label{tab:fit-plus-form-2L-2a-2ad-4ad-lna_nohybrid}
\end{table*}

\begin{figure*}
\centering
\includegraphics[width=0.3\textwidth]{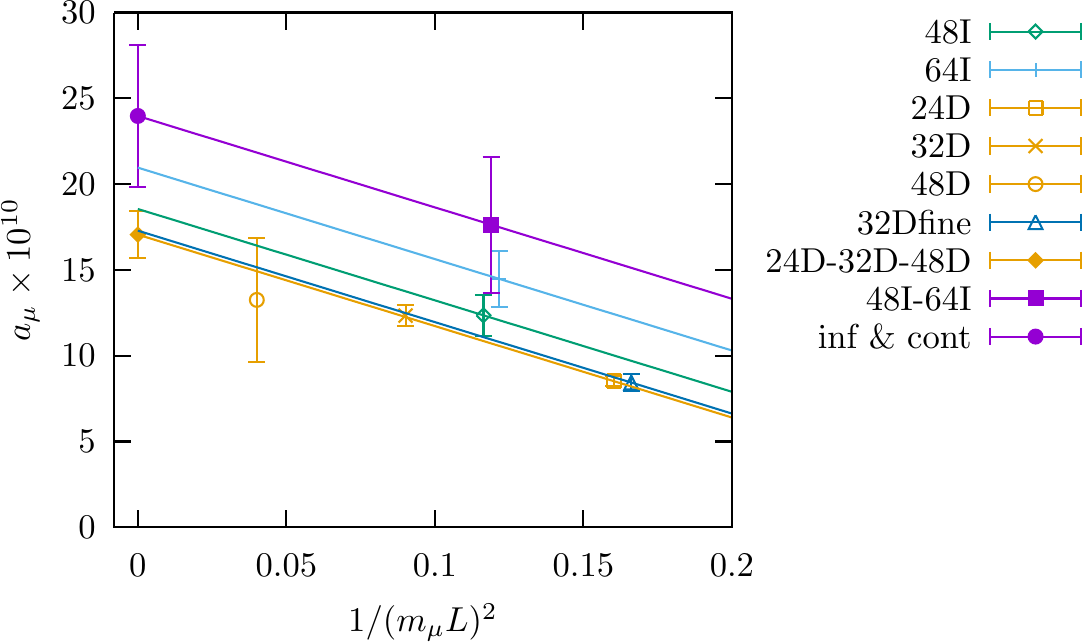}
\includegraphics[width=0.3\textwidth]{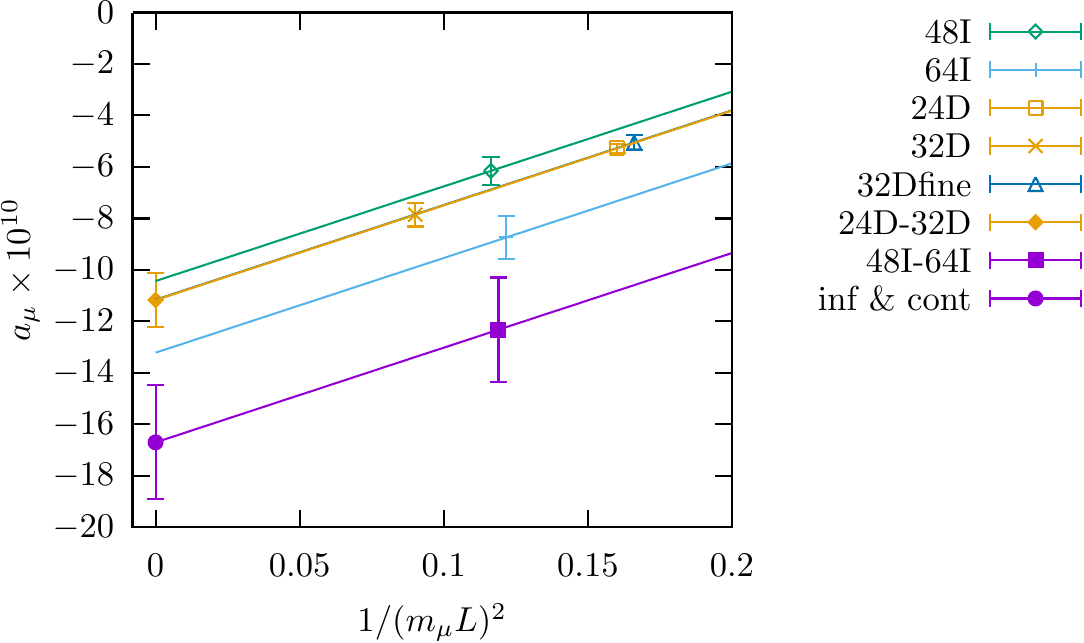}
\includegraphics[width=0.3\textwidth]{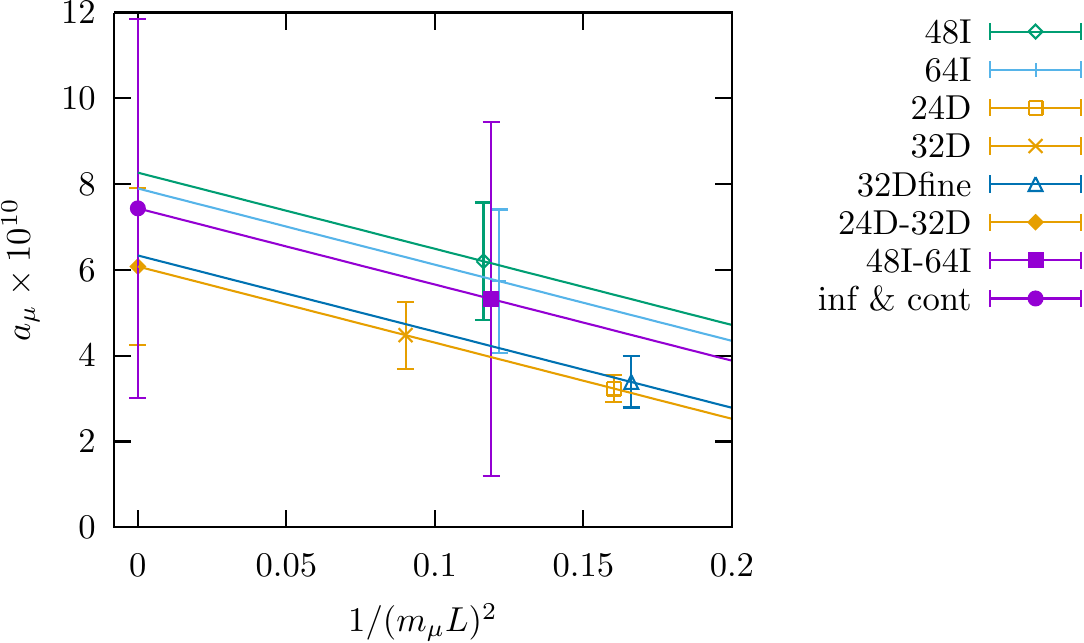}
\caption{Fit plots for \textit{fit-plus-form-2L-2a-2ad-4ad-lna} without the hybrid continuum limit. Connected (left), disconnected (middle), and total (right).}
\label{fig:fit-plus-form-2L-2a-2ad-4ad-lna_nohybrid}
\end{figure*}

\begin{table*}
\begin{tabular}{@{}cccccc@{}}
\toprule
 & $a_\mu$ & $b_2$ & $c_1^\text{I}$ & $c_1^\text{D}-c_1^\text{I}$ & $c_2^\text{D}$ \tabularnewline
\midrule
con               & 25.67(4.92) & 2.66(0.28) & 0.95(0.60) & 0.16(0.29) & 0.72(0.35) \tabularnewline
discon            & -17.84(2.59) & 2.71(0.28) & 1.55(0.36) & -0.22(0.20) & 0.88(0.26) \tabularnewline
sum               & 7.83(5.20) \tabularnewline
sum-sys ($a^2$)   & 0.00(0.00) \tabularnewline
tot               & 7.87(5.20) & 2.65(1.13) & -0.44(2.94) & 1.01(1.49) & 0.30(1.43) \tabularnewline
tot-sys ($a^2$)   & 0.00(0.00) \tabularnewline
\bottomrule
\end{tabular}
\caption{Fit results for \textit{fit-plus-form-2L-2a-2ad-4ad-cross} without the hybrid continuum limit.}
\label{tab:fit-plus-form-2L-2a-2ad-4ad-cross_nohybrid}
\end{table*}

\begin{figure*}
\centering
\includegraphics[width=0.3\textwidth]{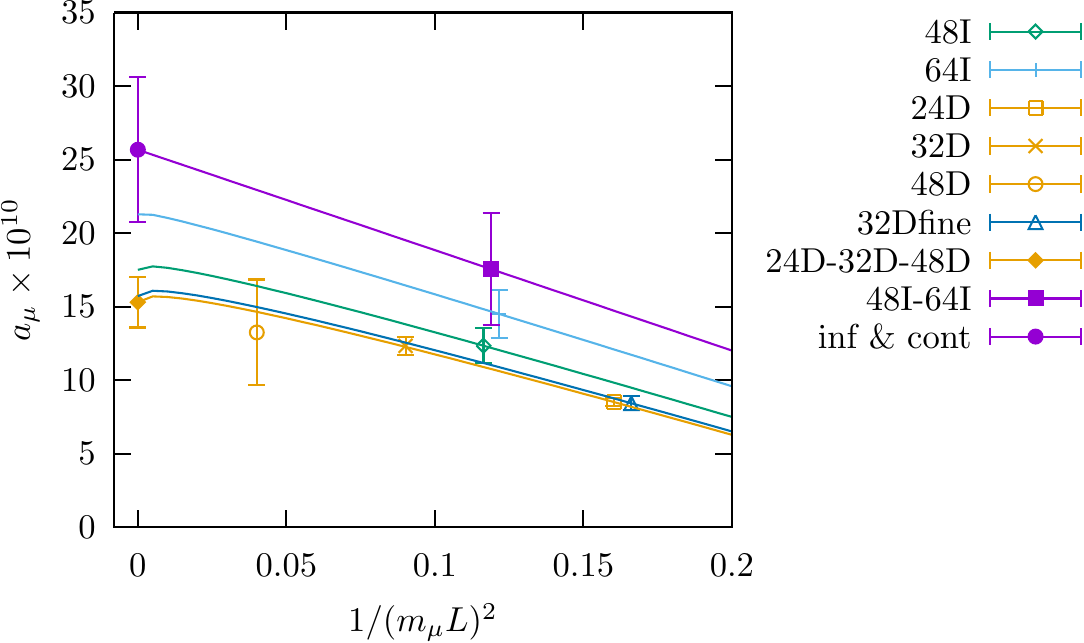}
\includegraphics[width=0.3\textwidth]{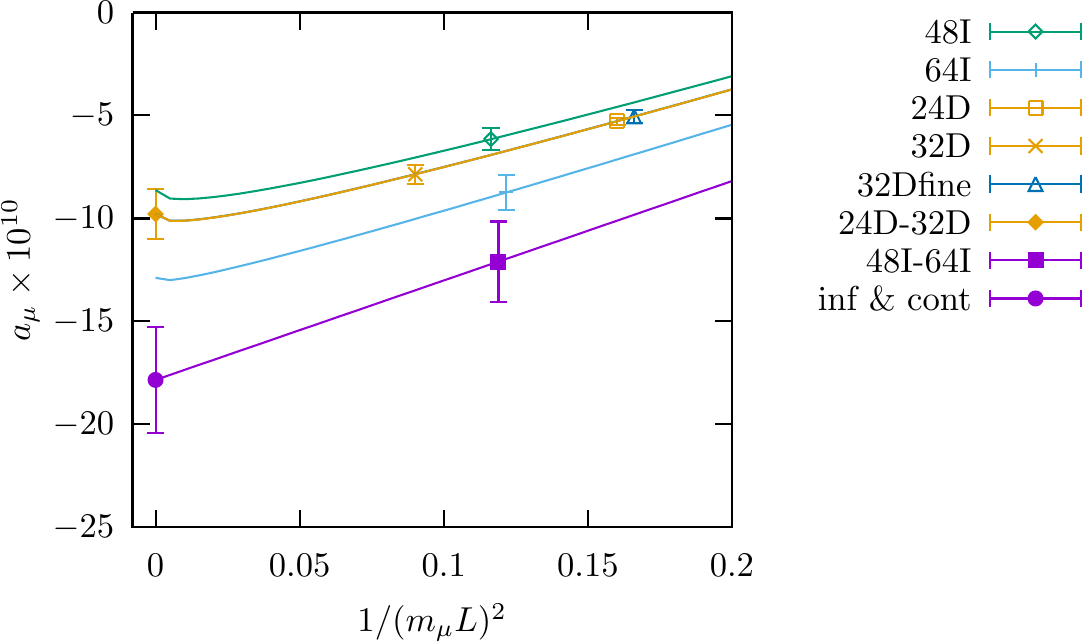}
\includegraphics[width=0.3\textwidth]{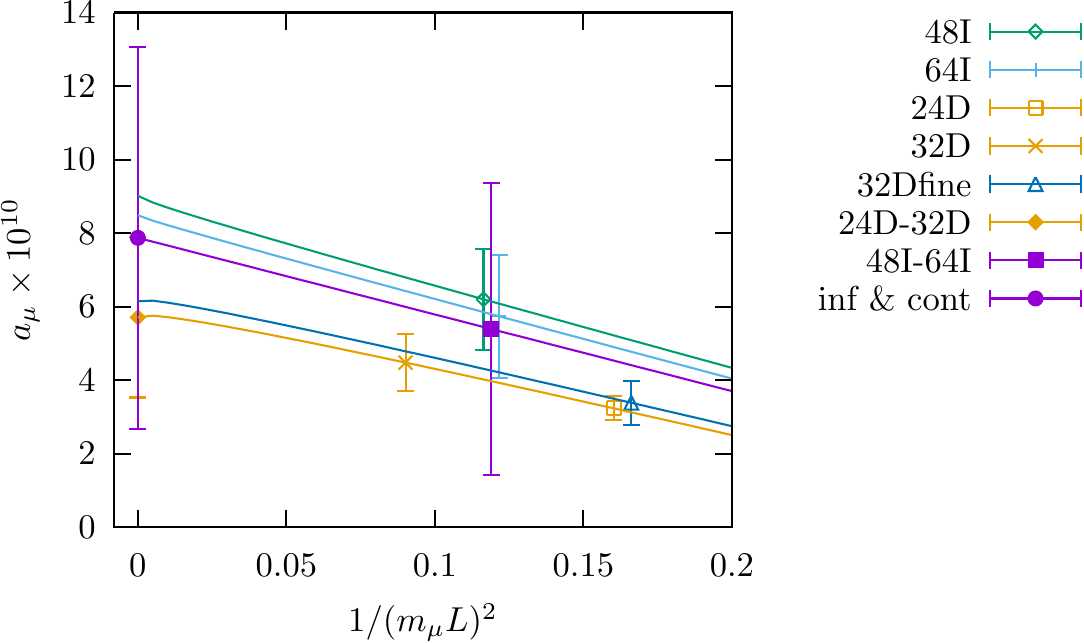}
\caption{Fit plots for \textit{fit-plus-form-2L-2a-2ad-4ad-cross} without the hybrid continuum limit. Connected (left), disconnected (middle), and total (right).}
\label{fig:fit-plus-form-2L-2a-2ad-4ad-cross_nohybrid}
\end{figure*}

\nclearpage

\begin{table*}
\begin{tabular}{@{}cccccc@{}}
\toprule
 & $a_\mu$ & $b_2$ & $c_1^\text{I}$ & $c_1^\text{D}-c_1^\text{I}$ & $c_2^\text{D}$ \tabularnewline
\midrule
con               & 27.94(6.31) & 3.12(0.29) & 0.91(0.55) & 0.16(0.29) & 0.69(0.33) \tabularnewline
discon            & -19.94(3.49) & 3.29(0.32) & 1.50(0.31) & -0.20(0.22) & 0.87(0.25) \tabularnewline
sum               & 8.00(6.75) \tabularnewline
sum-sys ($a^2$)   & 0.00(0.00) \tabularnewline
tot               & 8.33(6.40) & 2.92(1.13) & -0.38(2.81) & 0.93(1.45) & 0.27(1.40) \tabularnewline
tot-sys ($a^2$)   & 0.00(0.00) \tabularnewline
\bottomrule
\end{tabular}
\caption{Fit results for \textit{fit-product-form-2L-2a-2ad-4ad} without the hybrid continuum limit.}
\label{tab:fit-product-form-2L-2a-2ad-4ad_nohybrid}
\end{table*}

\begin{figure*}
\centering
\includegraphics[width=0.3\textwidth]{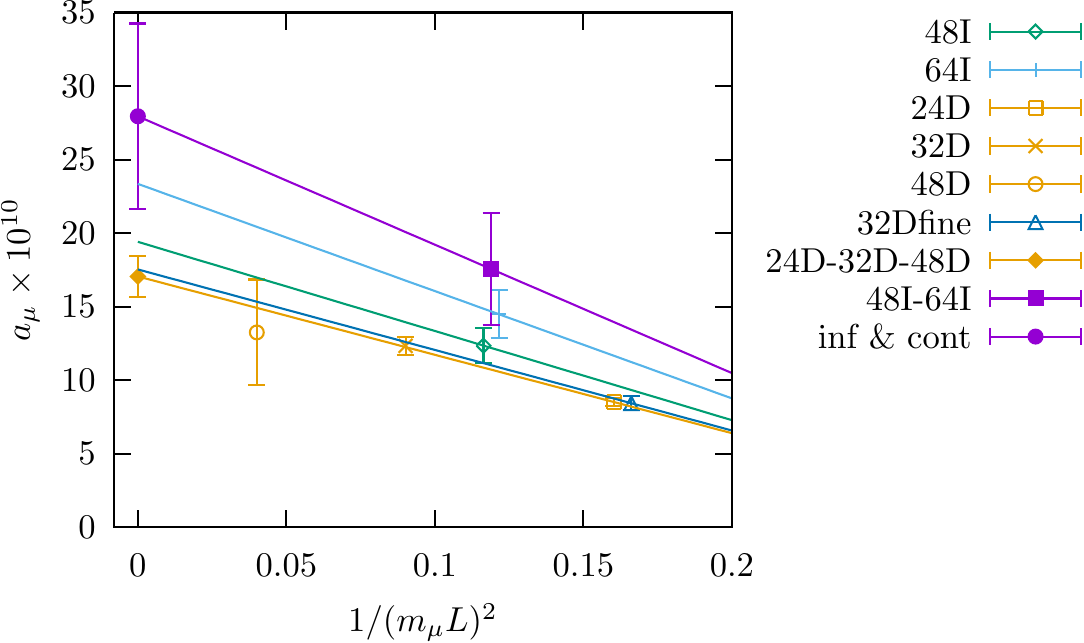}
\includegraphics[width=0.3\textwidth]{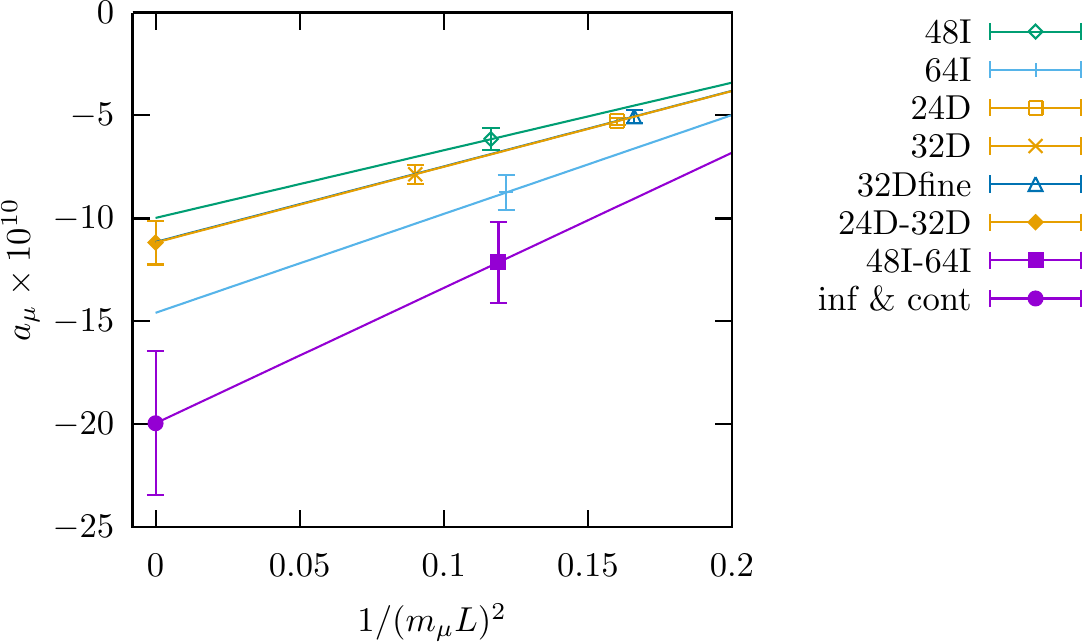}
\includegraphics[width=0.3\textwidth]{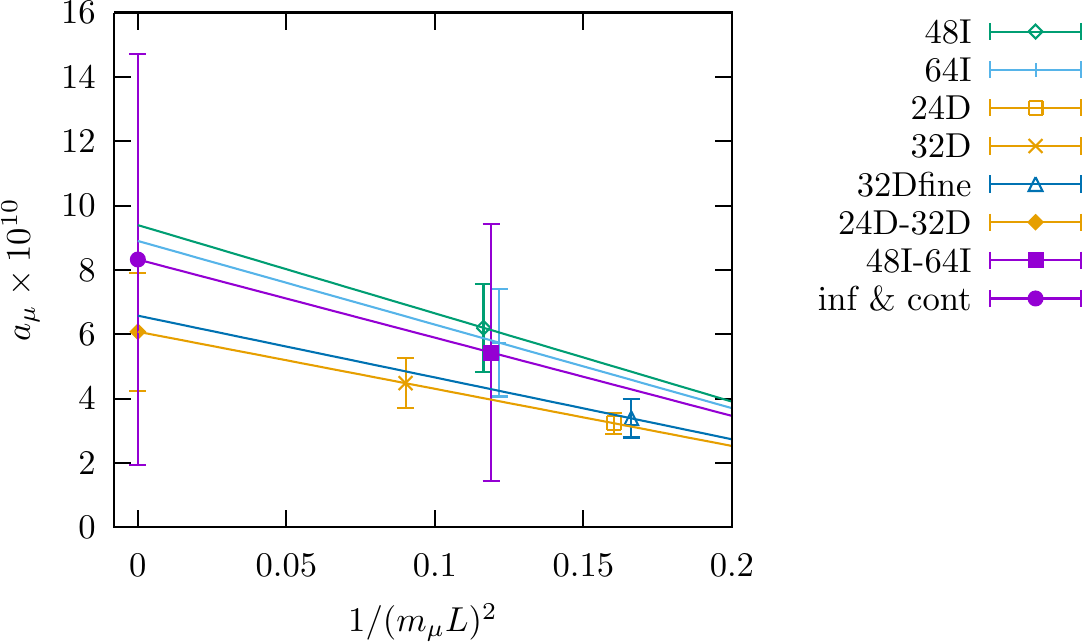}
\caption{Fit plots for \textit{fit-product-form-2L-2a-2ad-4ad} without the hybrid continuum limit. Connected (left), disconnected (middle), and total (right).}
\label{fig:fit-product-form-2L-2a-2ad-4ad_nohybrid}
\end{figure*}

\begin{table*}
\begin{tabular}{@{}cccccc@{}}
\toprule
 & $a_\mu$ & $b_2$ & $c_1^\text{I}$ & $c_1^\text{D}-c_1^\text{I}$ & $c_2^\text{D}$ \tabularnewline
\midrule
con               & 32.03(7.36) & 5.79(0.49) & 0.92(0.55) & 0.20(0.29) & 0.73(0.32) \tabularnewline
discon            & -23.19(4.20) & 6.11(0.52) & 1.50(0.31) & -0.16(0.21) & 0.90(0.24) \tabularnewline
sum               & 8.85(7.95) \tabularnewline
sum-sys ($a^2$)   & 0.00(0.00) \tabularnewline
tot               & 9.52(7.47) & 5.50(1.90) & -0.38(2.80) & 0.97(1.45) & 0.31(1.36) \tabularnewline
tot-sys ($a^2$)   & 0.00(0.00) \tabularnewline
\bottomrule
\end{tabular}
\caption{Fit results for \textit{fit-product-form-3L-2a-2ad-4ad} without the hybrid continuum limit.}
\label{tab:fit-product-form-3L-2a-2ad-4ad_nohybrid}
\end{table*}

\begin{figure*}
\centering
\includegraphics[width=0.3\textwidth]{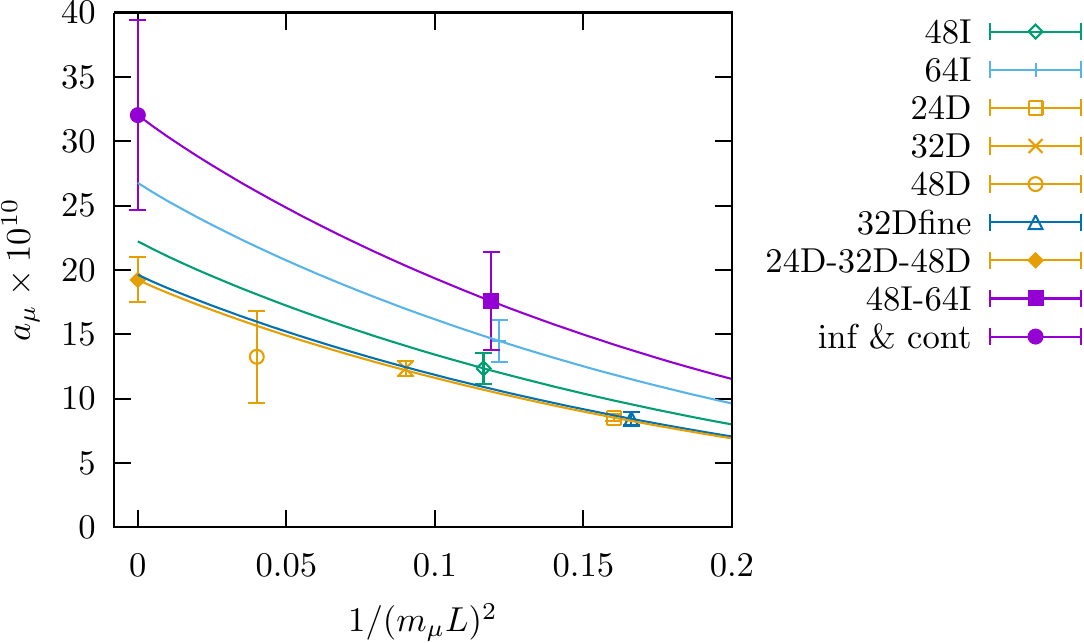}
\includegraphics[width=0.3\textwidth]{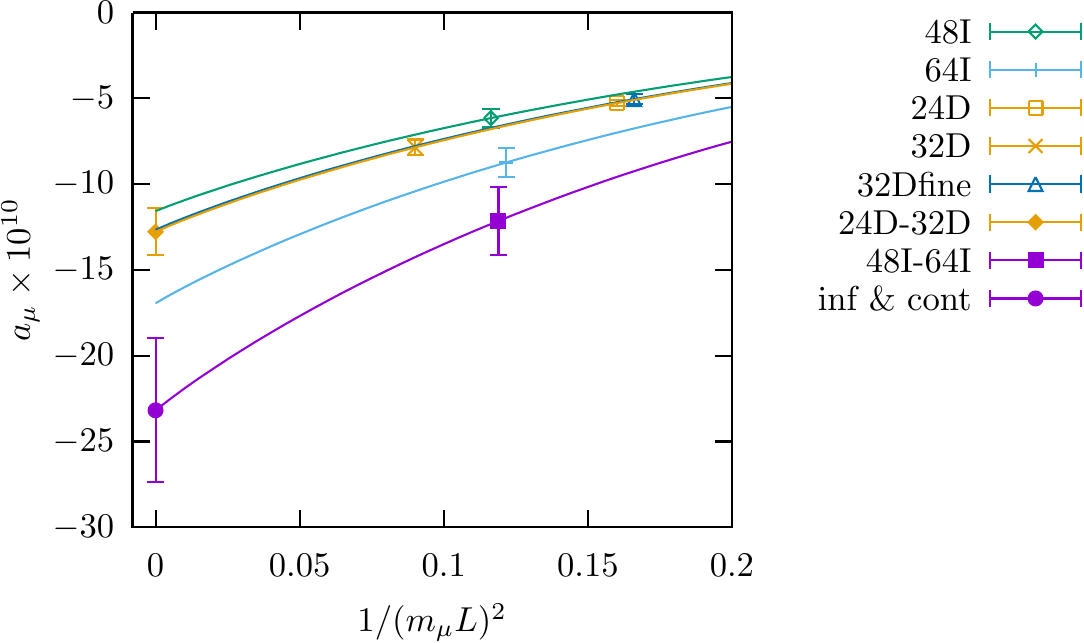}
\includegraphics[width=0.3\textwidth]{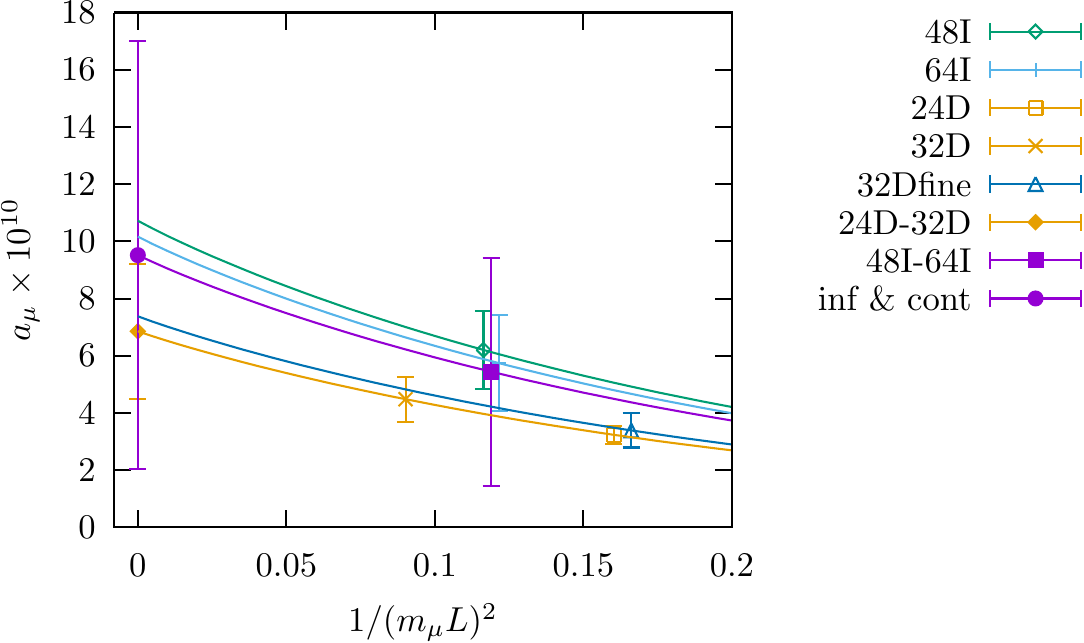}
\caption{Fit plots for \textit{fit-product-form-3L-2a-2ad-4ad} without the hybrid continuum limit. Connected (left), disconnected (middle), and total (right).}
\label{fig:fit-product-form-3L-2a-2ad-4ad_nohybrid}
\end{figure*}

\nclearpage

\begin{table*}
\begin{tabular}{@{}cccccc@{}}
\toprule
 & $a_\mu$ & $b_2$ & $c_1^\text{I}$ & $c_2$ & $c_1^\text{D}-c_1^\text{I}$ \tabularnewline
\midrule
con               & 29.25(7.08) & 3.12(0.29) & 1.26(0.65) & 0.75(0.34) & -0.10(0.30) \tabularnewline
discon            & -21.12(3.88) & 3.29(0.32) & 1.89(0.37) & 0.93(0.25) & -0.50(0.18) \tabularnewline
sum               & 8.13(7.54) \tabularnewline
sum-sys ($a^2$)   & 0.00(0.00) \tabularnewline
tot               & 8.48(7.14) & 2.92(1.13) & -0.22(3.47) & 0.30(1.53) & 0.81(1.72) \tabularnewline
tot-sys ($a^2$)   & -0.00(0.00) \tabularnewline
\bottomrule
\end{tabular}
\caption{Fit results for \textit{fit-product-form-2L-2a-4a-2ad} without the hybrid continuum limit.}
\label{tab:fit-product-form-2L-2a-4a-2ad_nohybrid}
\end{table*}

\begin{figure*}
\centering
\includegraphics[width=0.3\textwidth]{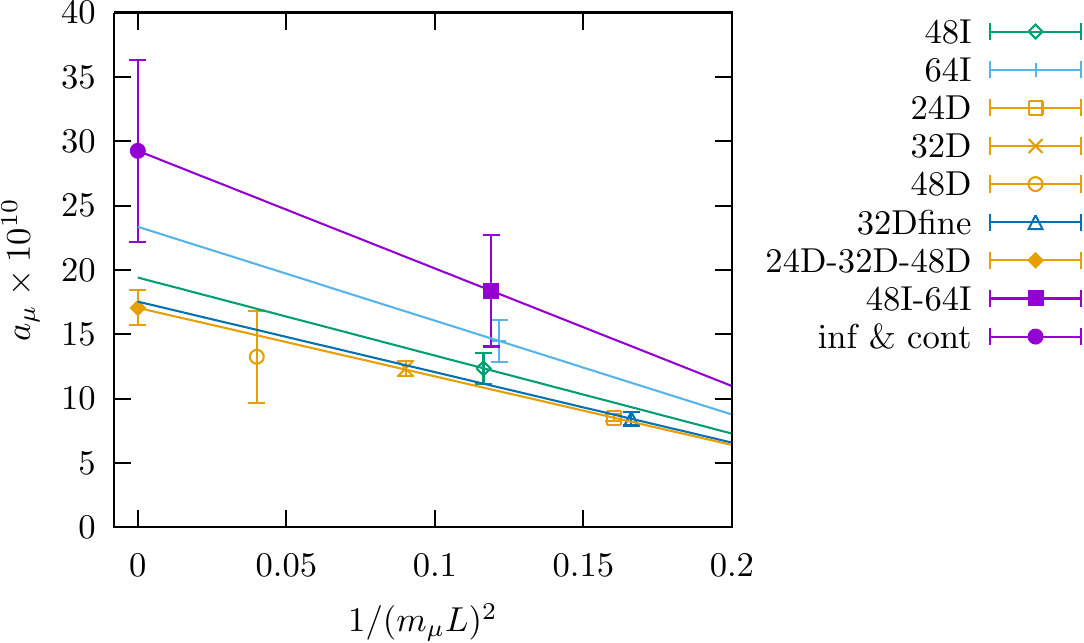}
\includegraphics[width=0.3\textwidth]{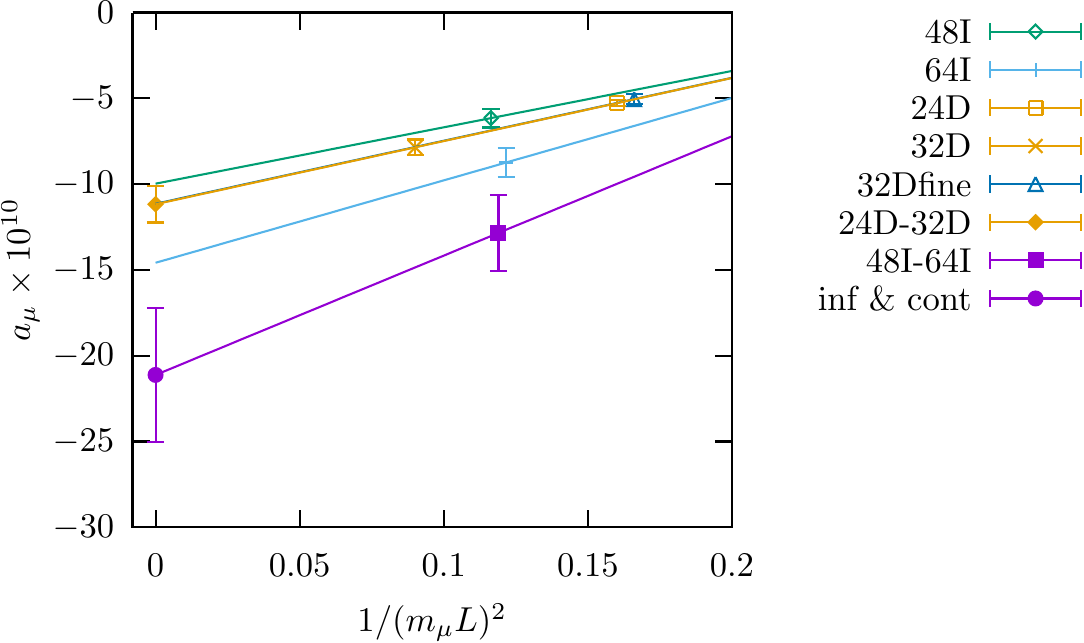}
\includegraphics[width=0.3\textwidth]{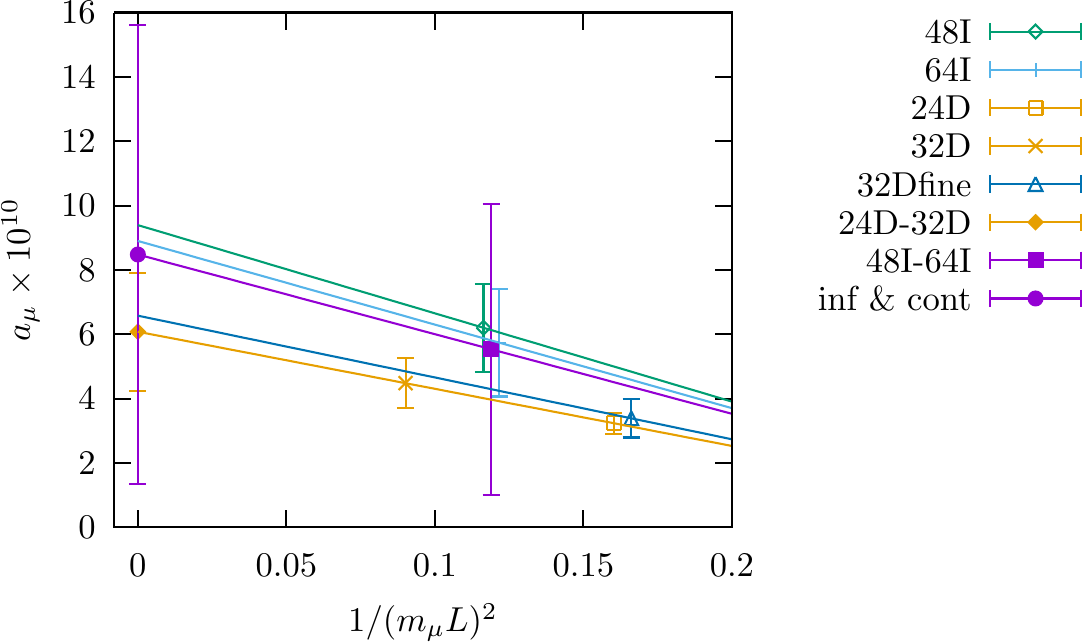}
\caption{Fit plots for \textit{fit-product-form-2L-2a-4a-2ad} without the hybrid continuum limit. Connected (left), disconnected (middle), and total (right).}
\label{fig:fit-product-form-2L-2a-4a-2ad_nohybrid}
\end{figure*}

\begin{table*}
\begin{tabular}{@{}cccccc@{}}
\toprule
 & $a_\mu$ & $b_2$ & $c_1$ & $c_2^\text{I}$ & $c_2^\text{D}-c_2^\text{I}$ \tabularnewline
\midrule
con               & 28.72(5.91) & 3.12(0.29) & 1.12(0.43) & 0.45(0.81) & -0.27(0.81) \tabularnewline
discon            & -19.23(3.24) & 3.29(0.32) & 1.23(0.36) & -0.62(0.69) & -1.45(0.57) \tabularnewline
sum               & 9.49(6.21) \tabularnewline
sum-sys ($a^2$)   & 0.00(0.00) \tabularnewline
tot               & 9.69(5.95) & 2.92(1.13) & 0.88(1.44) & 2.35(3.14) & 1.84(3.43) \tabularnewline
tot-sys ($a^2$)   & -0.00(0.00) \tabularnewline
\bottomrule
\end{tabular}
\caption{Fit results for \textit{fit-product-form-2L-2a-4a-4ad} without the hybrid continuum limit.}
\label{tab:fit-product-form-2L-2a-4a-4ad_nohybrid}
\end{table*}

\begin{figure*}
\centering
\includegraphics[width=0.3\textwidth]{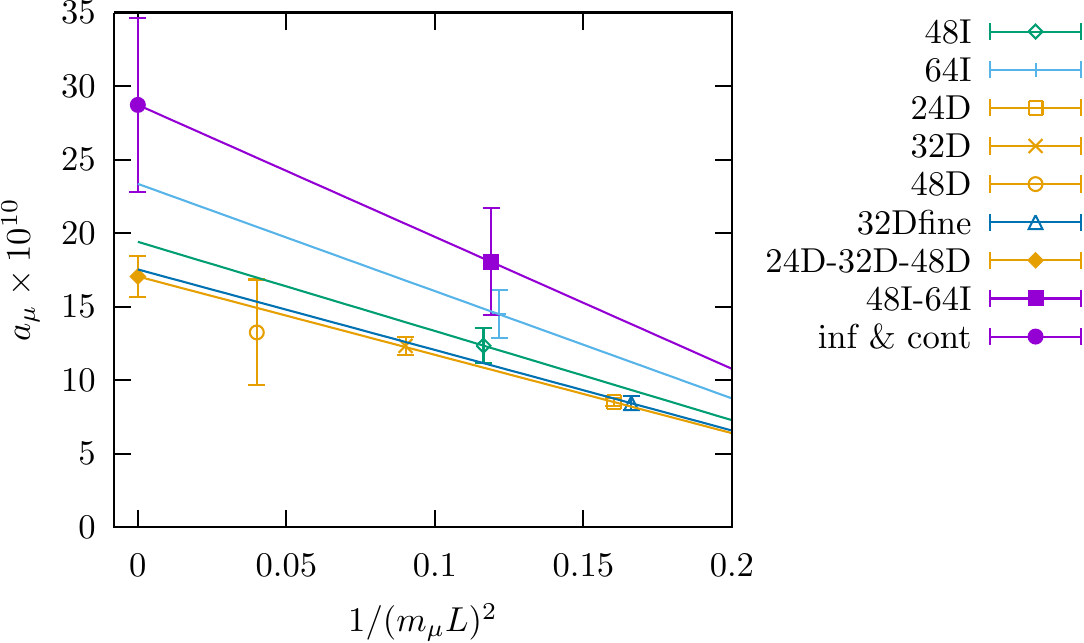}
\includegraphics[width=0.3\textwidth]{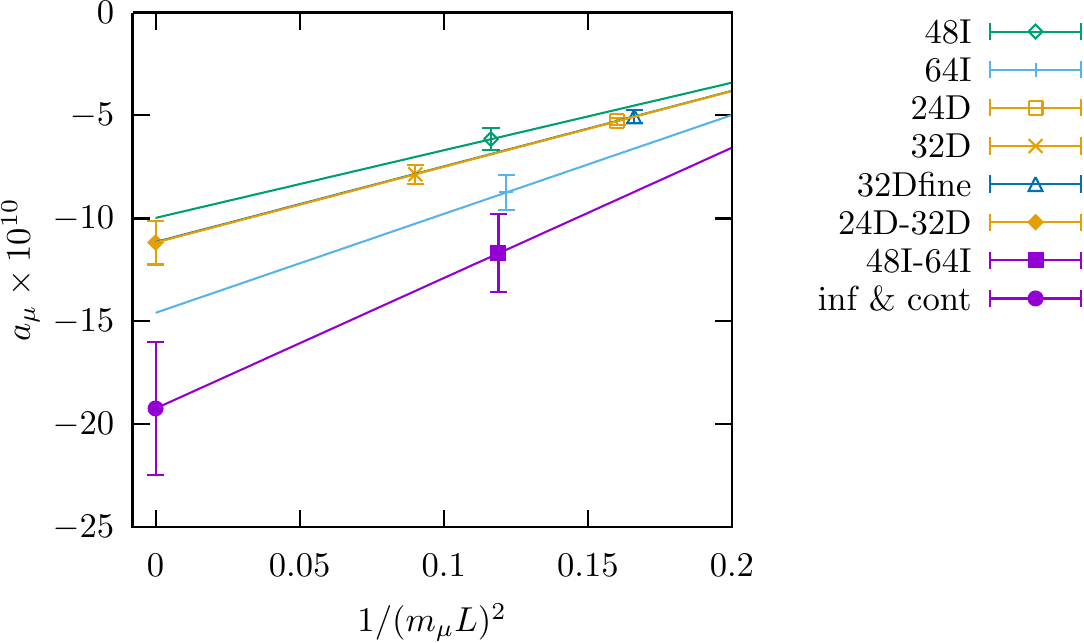}
\includegraphics[width=0.3\textwidth]{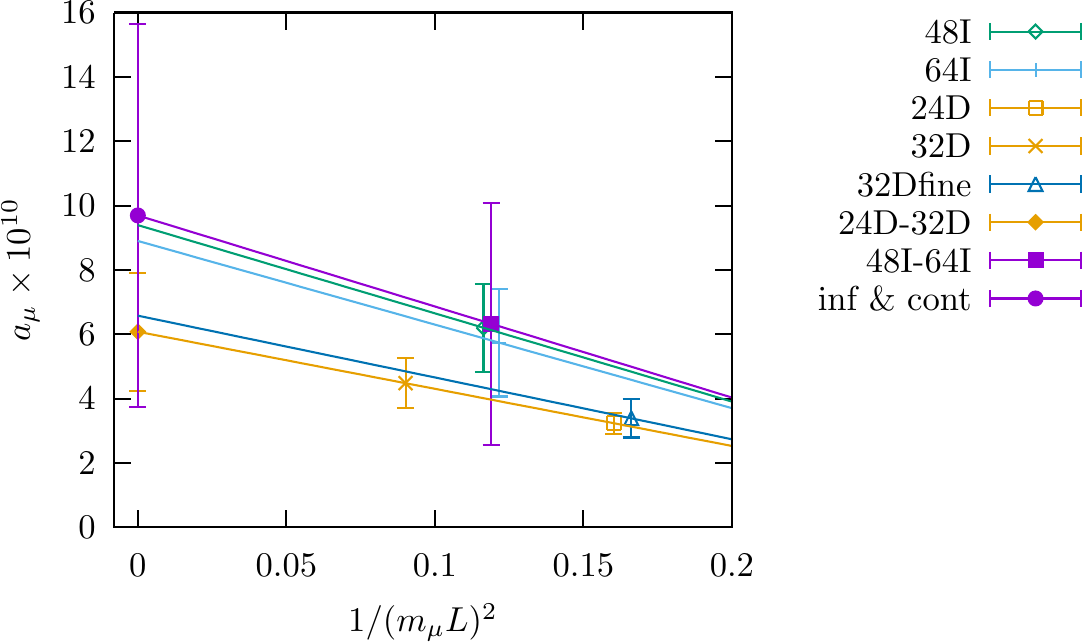}
\caption{Fit plots for \textit{fit-product-form-2L-2a-4a-4ad} without the hybrid continuum limit. Connected (left), disconnected (middle), and total (right).}
\label{fig:fit-product-form-2L-2a-4a-4ad_nohybrid}
\end{figure*}

\nclearpage

\begin{table*}
\begin{tabular}{@{}cccccc@{}}
\toprule
 & $a_\mu$ & $b_2$ & $c_1^\text{I}$ & $c_1^\text{D}-c_1^\text{I}$ & $c_2^\text{D}$ \tabularnewline
\midrule
con               & 28.29(6.59) & 3.12(0.29) & 0.85(0.51) & 0.14(0.26) & 0.60(0.29) \tabularnewline
discon            & -20.35(3.64) & 3.29(0.32) & 1.38(0.28) & -0.17(0.20) & 0.77(0.22) \tabularnewline
sum               & 7.94(7.04) \tabularnewline
sum-sys ($a^2$)   & 0.00(0.00) \tabularnewline
tot               & 8.28(6.67) & 2.92(1.13) & -0.36(2.68) & 0.84(1.35) & 0.21(1.30) \tabularnewline
tot-sys ($a^2$)   & 0.00(0.00) \tabularnewline
\bottomrule
\end{tabular}
\caption{Fit results for \textit{fit-product-form-2L-2a-2ad-4ad-lna} without the hybrid continuum limit.}
\label{tab:fit-product-form-2L-2a-2ad-4ad-lna_nohybrid}
\end{table*}

\begin{figure*}
\centering
\includegraphics[width=0.3\textwidth]{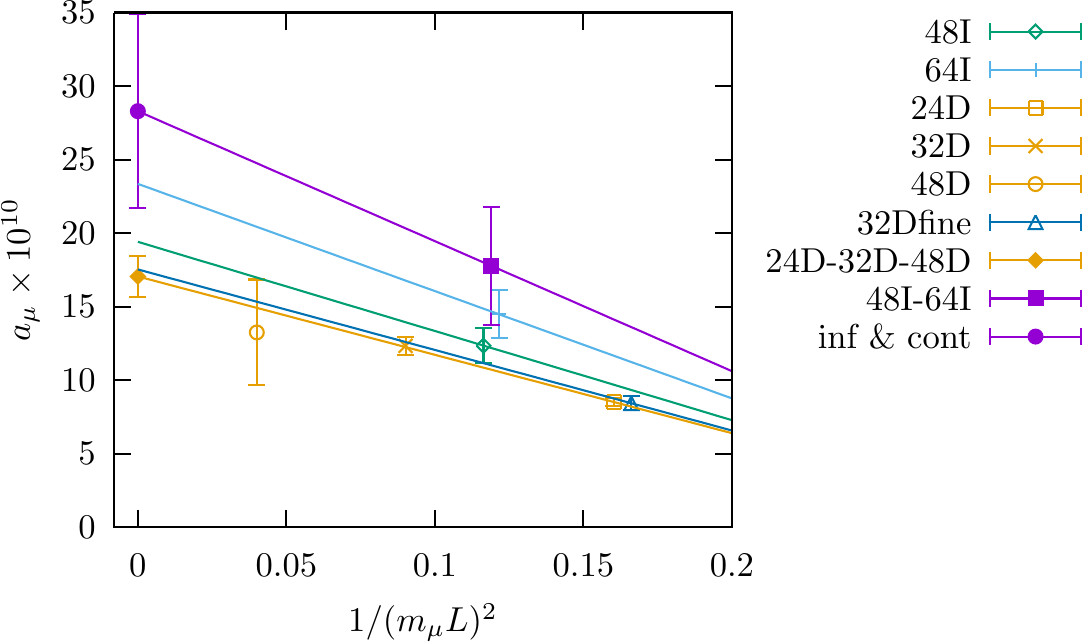}
\includegraphics[width=0.3\textwidth]{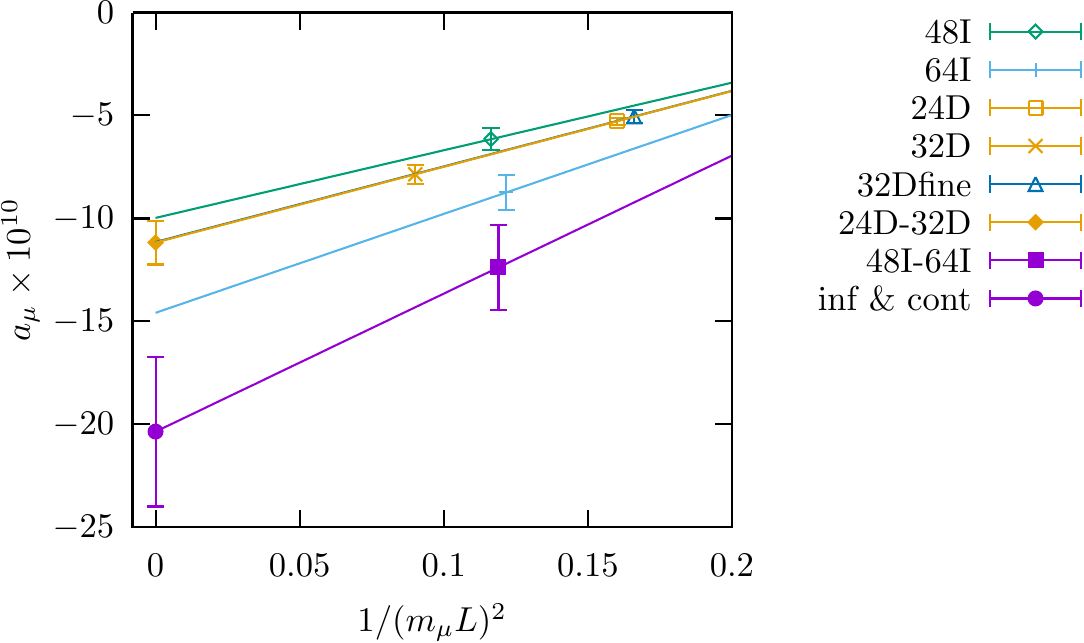}
\includegraphics[width=0.3\textwidth]{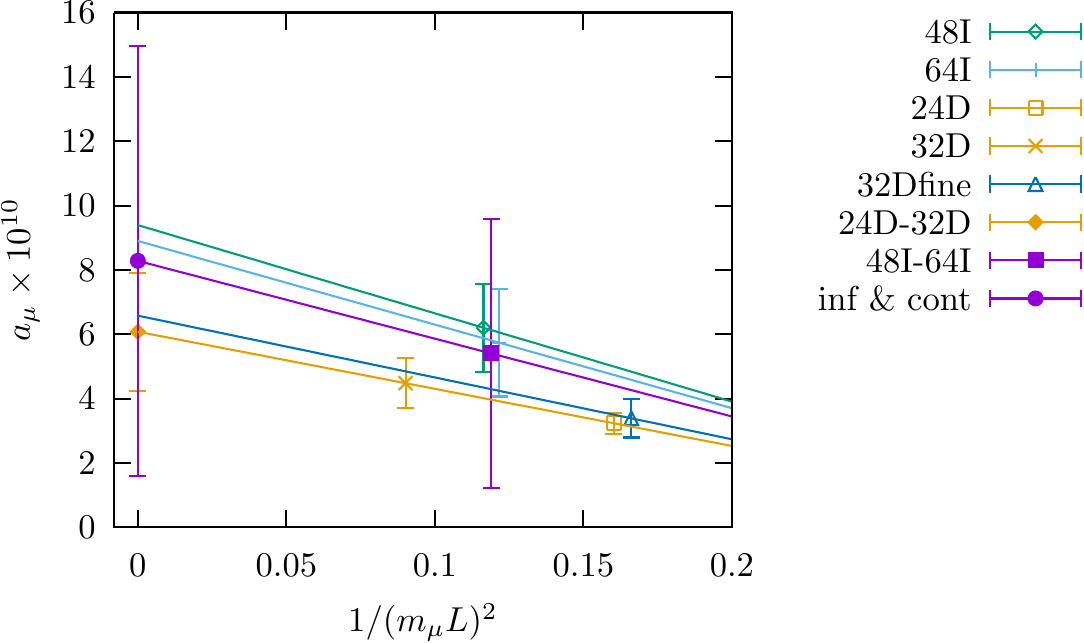}
\caption{Fit plots for \textit{fit-product-form-2L-2a-2ad-4ad-lna} without the hybrid continuum limit. Connected (left), disconnected (middle), and total (right).}
\label{fig:fit-product-form-2L-2a-2ad-4ad-lna_nohybrid}
\end{figure*}

\begin{table*}
\begin{tabular}{@{}cccccc@{}}
\toprule
 & $a_\mu$ & $b_2$ & $c_1^\text{I}$ & $c_1^\text{D}-c_1^\text{I}$ & $c_2^\text{D}$ \tabularnewline
\midrule
con               & 31.00(8.24) & 3.58(0.31) & 1.44(0.82) & 0.13(0.49) & 0.99(0.51) \tabularnewline
discon            & -22.06(4.47) & 3.78(0.27) & 2.28(0.48) & -0.38(0.38) & 1.26(0.41) \tabularnewline
sum               & 8.94(8.73) \tabularnewline
sum-sys ($a^2$)   & 0.00(0.00) \tabularnewline
tot               & 8.88(8.04) & 3.22(1.18) & -0.53(4.25) & 1.29(2.33) & 0.34(2.14) \tabularnewline
tot-sys ($a^2$)   & 0.00(0.00) \tabularnewline
\bottomrule
\end{tabular}
\caption{Fit results for \textit{fit-product-form-2L-2a-2ad-4ad-cross} without the hybrid continuum limit.}
\label{tab:fit-product-form-2L-2a-2ad-4ad-cross_nohybrid}
\end{table*}

\begin{figure*}
\centering
\includegraphics[width=0.3\textwidth]{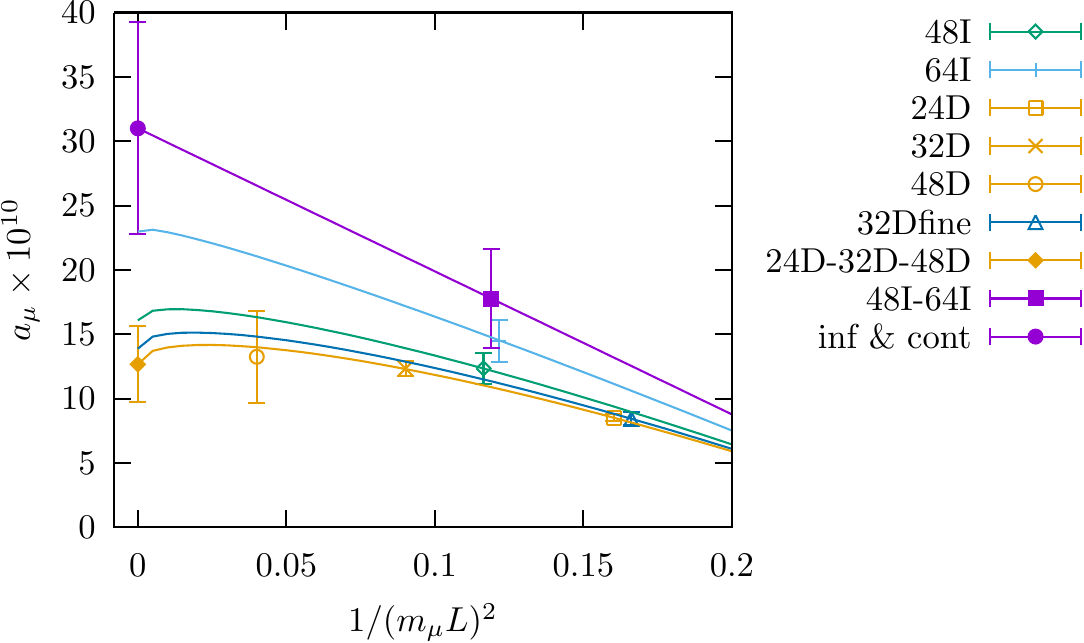}
\includegraphics[width=0.3\textwidth]{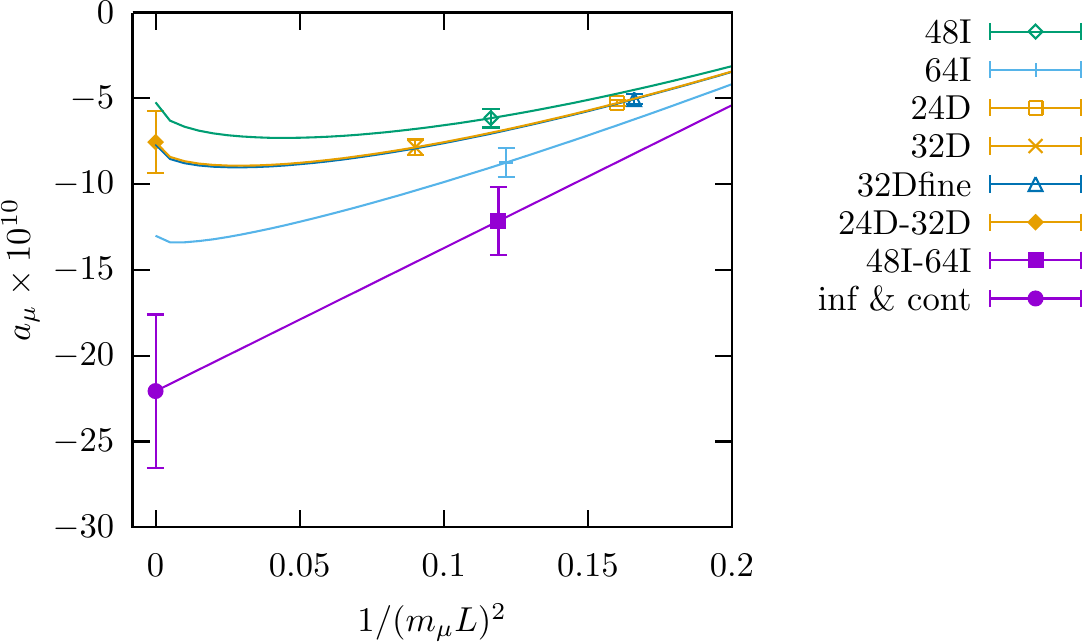}
\includegraphics[width=0.3\textwidth]{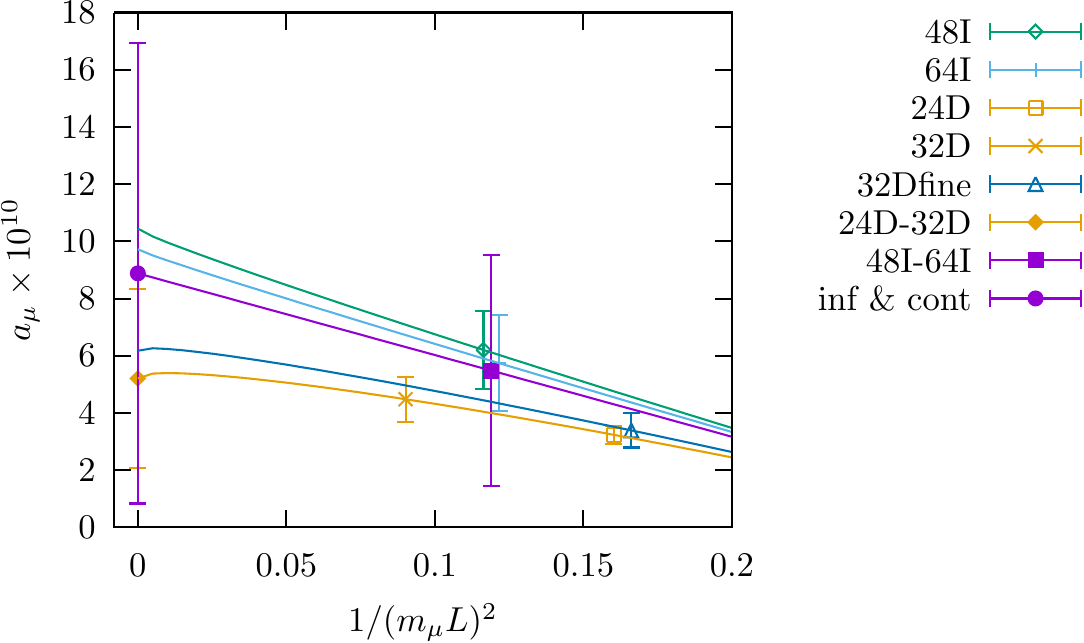}
\caption{Fit plots for \textit{fit-product-form-2L-2a-2ad-4ad-cross} without the hybrid continuum limit. Connected (left), disconnected (middle), and total (right).}
\label{fig:fit-product-form-2L-2a-2ad-4ad-cross_nohybrid}
\end{figure*}

\nclearpage